\documentclass[twocolumn,english,superscriptaddress,floatfix,longbibliography]{revtex4}
\usepackage[T1]{fontenc}
\usepackage[utf8]{inputenc}
\usepackage{xcolor}
\usepackage{pdfcolmk}
\usepackage{bm}
\usepackage{amsmath}
\usepackage{amssymb}
\usepackage{graphicx}
\usepackage{esint}
\PassOptionsToPackage{normalem}{ulem}
\usepackage{ulem}

\makeatletter

\providecommand{\tabularnewline}{\\}
\providecolor{lyxadded}{rgb}{0,0,1}
\providecolor{lyxdeleted}{rgb}{1,0,0}

\@ifundefined{textcolor}{}
{%
 \definecolor{BLACK}{gray}{0}
 \definecolor{WHITE}{gray}{1}
 \definecolor{RED}{rgb}{1,0,0}
 \definecolor{GREEN}{rgb}{0,1,0}
 \definecolor{BLUE}{rgb}{0,0,1}
 \definecolor{CYAN}{cmyk}{1,0,0,0}
 \definecolor{MAGENTA}{cmyk}{0,1,0,0}
 \definecolor{YELLOW}{cmyk}{0,0,1,0}
}

\usepackage{upgreek}
\usepackage{braket}

\usepackage{babel}

\makeatother

\usepackage{babel}
\begin{document}

\title{Non-local hydrodynamic transport and collective excitations in Dirac
fluids}

\author{Egor I. Kiselev}

\affiliation{Institut für Theorie der Kondensierten Materie, Karlsruher Institut
für Technologie, 76131 Karlsruhe, Germany}

\author{Jörg Schmalian}

\affiliation{Institut für Theorie der Kondensierten Materie, Karlsruher Institut
für Technologie, 76131 Karlsruhe, Germany}

\affiliation{Institut für Quantenmaterialien und -technologien, Karlsruher Institut
für Technologie, 76131 Karlsruhe, Germany}
\begin{abstract}
We study the response of a Dirac fluid to electric fields and thermal
gradients at finite wave-numbers and frequencies in the hydrodynamic
regime. We find that non-local transport in the hydrodynamic regime
is governed by infinite set of kinetic modes that describe non-collinear
scattering events in different angular harmonic channels. The scattering
rates of these modes $\tau_{m}^{-1}$ increase as $\left|m\right|$,
where $m$ labels the angular harmonics. In an earlier publication,
we pointed out that this dependence leads to anomalous, Lévy-flight-like
phase space diffusion \cite{Kiselev2019b}. Here, we show how this
surprisingly simple, non-analytic dependence allows us to obtain exact
expressions for the non-local charge and electronic thermal conductivities.
The peculiar dependence of the scattering rates on $m$ also leads
to a non-trivial structure of collective excitations: Besides the
well known plasmon, second sound and diffusive modes, we find non-degenerate
damped modes corresponding to excitations of higher angular harmonics.
We use these results to investigate the transport of a Dirac fluid
through Poiseuille-type geometries of different widths, and to study
the response to surface acoustic waves in graphene-piezoelectric devices. 
\end{abstract}
\maketitle

\section{Introduction}

In many instances transport properties can be described in terms of
a local relationship between forces and currents. Examples are Fourier's
law of heat conduction $\boldsymbol{j}_{\varepsilon}=-\kappa\nabla T$,
Fick's law of diffusion $\boldsymbol{j}_{c}=-D\nabla\mu$, or Ohm's
law of electrical conduction $\boldsymbol{j}_{c}=\sigma\boldsymbol{E}.$
Here the thermal conductivity $\kappa$, the diffusion coefficient
$D$, or the electrical conductivity $\sigma$ establish a relationship
between the value of the forces, such as a temperature gradient or
electric field, and the corresponding current density at the same
location. Such local relations break down when the electron propagation
is almost ballistic. Important examples worked out in particular by
Brian Pippard are the nonlocal current-field relations to describe
the Meissner effect in clean superconductors or the anomalous skin
effect in clean metals\cite{Pippard1947,Pippard1955,Waldram1970}.
However, non-local transport relations are not limited to the ballistic
transport regime. Another example for non-local transport occurs when
hydrodynamic flow of charge or heat sets in. Indeed, hydrodynamic
flow patterns are frequently identified by complex ``non-local''
flow lines. It is therefore necessary to find closed expressions for
the nonlocal heat conductivity $\kappa_{\alpha\beta}\left(\mathbf{r}-\mathbf{r}',t-t'\right)$,
electrical conductivity $\sigma_{\alpha\beta}\left(\mathbf{r}-\mathbf{r}',t-t'\right)$
or even non-local shear viscosities $\eta$$\left(\mathbf{r}-\mathbf{r}',t-t'\right)$
of many-body systems in the hydrodynamic regime. In this regime collisions
between partices are not weak, it merely holds that momentum relaxing
collisions are weak while momentum-conserving collisions are not.
A formulation in terms of non-local transport coefficients allow for
a microscopic description of hydrodynamic flow pattern and goes beyond
the usual description in terms of the linear Navier-Stokes equation.
The latter corresponds to the leading gradient expansion of the theory.
In addition, the inclusion of dynamical phenomena - here expressed
in terms of the dependency on the time difference $t-t'$ between
force and current – allows to determine the system's collective modes.

In this paper we develop the theory of non-local transport in Dirac
systems at charge neutrality in the collision-dominated hydrodynamic
regime and find closed expressions for the frequency and wave-vector-dependent,
charge and electronic thermal conductivities as well as the non-local
viscosity. Remarkably, the calculations of this paper are exact in
the limit of a small graphene fine structure constant $\alpha$ in
the regime of linear response. This is made possible by the peculiar
$\propto\left|m\right|$ dependence of the scattering rates of collinear
zero modes in higher angular momentum channels $m>2$ - a behavior
that was shown to lead to a super-diffusive Lévy-flight-like phase
space dynamics in an earlier work \cite{Kiselev2019b}. Collinear
zero modes do not decay due to the strong collinear scattering that
give rise to rapid equilibration and therefore dominate the long-time
dynamics. We make specific predictions for measurements such as the
velocity shift of surface acoustic waves, determine the flow of charge
and heat in finite geometries, and determine the collective mode spectrum
of the system including plasma waves and second-sound-ike thermal
waves. The dispersion relations of collective modes can be derived
from the poles of transport coefficients, or found from the solutions
of the homogeneous quantum Boltzmann equation. Here, focusing on the
charge neutrality point, we go beyond the phenomenological treatment
of electron-electron interactions of Refs. \cite{Svintsov2018,Torre2019}.
Our detailed analysis reveals a complex structure of damped collective
excitations. These excitations are similar to the so-called ``non-hydrodynamic''
modes that were shown to be relevant for the equilibration of unitary
fermi gases \cite{Brewer2015} and QCD plasmas \cite{Romatschke2016,Romatschke2018,Heller2018}.
In fact, the term non-hydrodynamic is somewhat misleading. What is
meant is that these modes correspond to excitations of high angular
momentum components of the kinetic distribution function, which are
not captured by the Navier-Stokes equations.

Transport in a Dirac fluid is in many respects different from the
archetypical example of the Fermi liquid. One important difference
is that electric currents in a Dirac fluid are not protected by momentum
conservation, and therefore decay even in a perfectly clean system.
Negatively charged electrons and positive holes flowing in opposite
directions sum up to a finite electric current with zero momentum.
Thus, even in the absence of impurities, pristine graphene -- the
prime example of a Dirac fluid -- has a finite conductivity that is
induced by electron-electron interactions \cite{Fritz2008,Kashuba2008}.
On the other hand, the energy current is proportional to the momentum
density, and therefore propagates ballistically \cite{Mueller2009,Foster2009}.
Both phenomena, the interaction induced conductivity and the ballistic
transport of energy, are relevant in the broader context of quantum
criticality \cite{Damle1997,Hartnoll2007,Sheehy2007}. Several experiments
addressed the unique transport properties of graphene at the charge
neutrality point. A violation of the Wiedemann-Franz law was observed
in Ref. \cite{Crossno2016}, indicating the ballistic transport of
energy. The interaction induced resistivity was recently measured
at finite frequencies \cite{Gallagher2019} and showen to be in good
agreement with the theoretical prediction of Ref. \cite{Fritz2008}.
Graphene has become one of the most important host systems for electron
hydrodynamics in general, extensively studied in both experiment \cite{Sulpizio2019,Bandurin2018,Bandurin2016,Berdyugin2019,KrishnaKumar2017}
and theory \cite{Narozhny2015,Briskot2015,Torre2015,Narozhny2017,Xie2019,Klug2018,Narozhny2019,Link2018,Levitov2013,Borgnia2015,Narozhny2019HydroNotes,Narozhny2019a,Danz2020,Kashuba2018,Lucas2018,sun2018}.

An important experimental prerequisite for the realization of hydrodynamic
electron flow is the dominance of electron-electron scattering over
any momentum relaxing scattering mechanism. Besides graphene, materials
such as delafossite metals \cite{Moll2016,Mackenzie2017} and Weyl
semimetals \cite{Gooth2018} show non-local transport patterns and
have been identified as potential candidates for the realization of
hydrodynamic electron flows - a development that boosted experimental
and theoretical work on the subject \cite{Gusev2018,Andreev2011,Principi2015,Alekseev2015,Alekseev2016,Guo2017,Scaffidi2017,Alekseev2018,Moessner2018,Cohen2018,Burmistrov2019,Moessner2019,Zdyrski2019,Alekseev2019,Svintsov2019,Holder2019,Cook2019,Pellegrino2017,Alekseev2018Counterflows,Link2018Out,Matthaiakakis2020}.

\begin{figure}
\centering{}\includegraphics[scale=0.4]{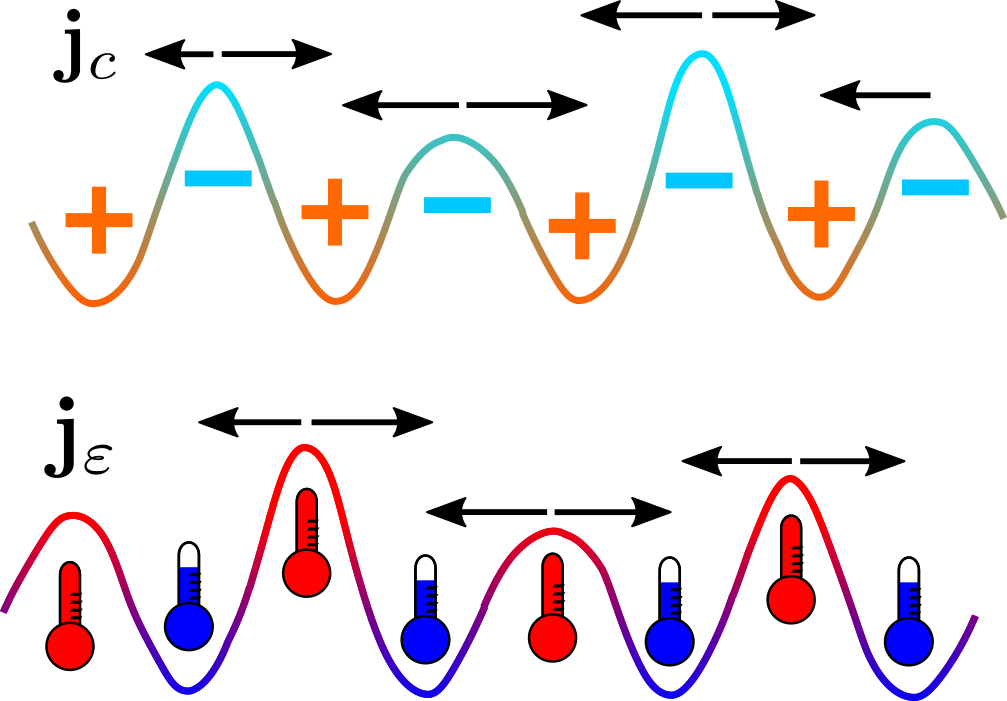}\caption{Charge (upper row) and energy currents (lower row) excited by wavelike
longitudinal electric fields and temperature differences.\label{fig:Charge-and-energy_currents}}
\end{figure}

In a clean system, hydrodynamics prevails when the electron-electron
scattering rate $l_{ee}$ is much smaller than the system size $l_{\mathrm{geo}}$.
The ratio between these two lengths is the Knudsen number ${\rm Kn}=l_{ee}/l_{\mathrm{geo}}$.
In a Poiseuille-like geometry $l_{\mathrm{geo}}$ corresponds to the
width of the sample. The geometry of the system then sets a finite
wavenumber $q\sim2\pi/l_{\mathrm{geo}}$. Therefore, for finite Knudsen
numbers, the wave-vector dependence of transport coefficients determines
the behavior of the fluid. Thinking in real space, this means that
higher-order spatial derivatives have to be included into the equations
of motion of the fluid, and the flow becomes highly non-local. A very
similar situation occurs when the system is subjected to spatially
modulated force fields, e.g. an electric field of the form $E_{\mathbf{q}}=E_{0}e^{i\mathbf{q}\cdot\mathbf{x}}e^{-i\omega t}$
(see Fig. \ref{fig:Charge-and-energy_currents}). The response of
the fluid is then determined by a non-local conductivity tensor $\sigma_{\alpha\beta}\left(\mathbf{q},\omega\right)$.
An important example that is treated in Sec \ref{sec:Surface-acoustic-waves}
are surface acoustic waves (SAWs) in piezoelectric materials, which
produce spatially modulated electric fields and can be used to study
the longitudinal part of the non-local charge conductivity.

\section{main results}

In this paper, we focus on the non-local transport properties and
collective excitations of graphene electrons at the charge neutrality
point - prime example for a Dirac fluid. The quantum Boltzmann method
developed in Ref. \cite{Fritz2008} is used. This method relies on
the fact, that at low temperatures the graphene fine structure constant
$\alpha$ is renormalized to small values. Thermally excited electrons
and holes therefore appear as sharply-defined quasiparticles, whose
transport properties can be studied by means of a kinetic equation.
The solution of this equation is facilitated by the presence of so-called
collinear modes, whose scattering rates are enhanced by a large factor
of $\log\left(1/\alpha\right)$. Here, the velocities of the interacting
particles are parallel to each other. Due to the linear graphene spectrum,
all particles travel at the same speed, regardless of their momentum.
Particles traveling in parallel have a particularly long time to interact
with each other, hence the strong enhancement. Transport in the hydrodynamic
regime, however, is dominated by processes, which have the smallest
scattering rates (for details see Eq. (\ref{eq:Collinear_non_collinear_separation})
and below). Such ``slow'' processes are represented by collinear
zero modes -- functions that set the collinear part of the collision
operator to zero \cite{Fritz2008,Kashuba2008}.

\begin{figure}
\centering{}\includegraphics[scale=0.35]{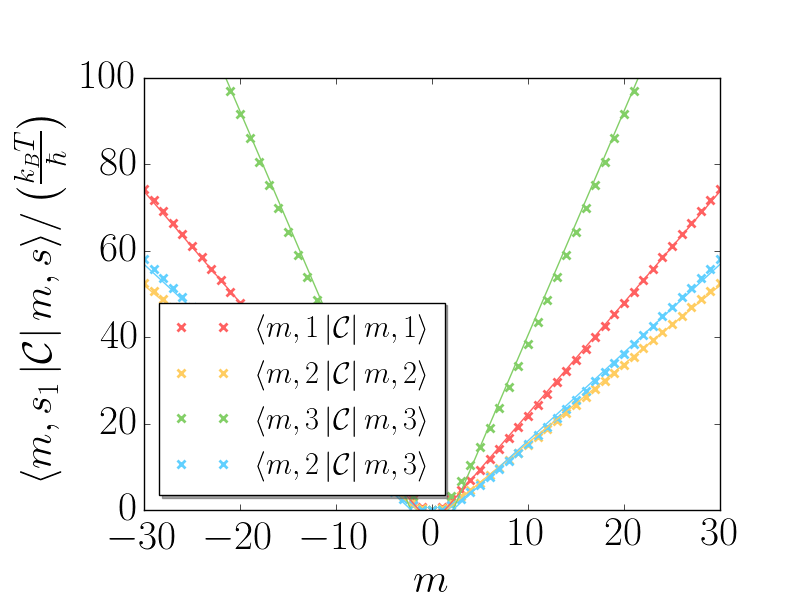}\caption{The matrix elements of the collision operator $\mathcal{C}$ of Eq.
(\ref{eq:Collision_operator}) with respect to the collinear zero
modes $\chi_{\mathbf{k},\lambda}^{\left(m,s\right)}=\lambda^{m}e^{im\theta}\left\{ 1,\lambda,\lambda\beta v\hbar k\right\} $
of Eq. (\ref{eq:Collinear_modes}) grow linearly with increasing angular
harmonic numbers $m$. The linear fits of Eqs. (\ref{eq:linear_scattering_rates_charge}),
(\ref{eq:linear_scattering_rates_energy}) are plotted as solid red
and green lines. The linear behavior of the matrix elements and scattering
rates allows to solve the quantum Boltzmann equation exactly.\label{fig:The-matrix-elements}}
\end{figure}

We solve the kinetic equation by reducing it to a matrix equation
in the space of collinear zero modes $\chi_{\mathbf{k},\lambda}^{\left(m,s\right)}=\lambda^{m}e^{im\theta}\left\{ 1,\lambda,\lambda\beta v\hbar k\right\} $
(see Sec. \ref{subsec:Collinear-zero-modes}). Here, $\theta$ is
the polar angle and $k$ the modulus of the momentum variable $\mathbf{k}$,
$\lambda=\pm1$ is the band index, $m$ labels the angular harmonics
$\exp\left(im\theta\right)$, and $s\in\left\{ 1,2,3\right\} $ labels
the three basis functions written in curly brackets. To an excellent
approximation, it is sufficient to retain only the $s=1$ and $s=3$
modes. These modes describe charge ($c$) and energy ($\varepsilon$)
excitations, respectively. A numerical evaluation of the collision
integral's matrix elements with respect to the modes $\chi_{\mathbf{k},\lambda}^{\left(m,s\right)}$
(see Fig. \ref{fig:The-matrix-elements}) shows, that the relaxation
rates of these modes grow linearly with increasing $m$: 
\begin{equation}
\tau_{\varepsilon/c,m}^{-1}\sim\left|m\right|,\label{eq:summary_scattering_rates}
\end{equation}
for large $m$ (see sections \ref{subsec:Collinear-zero-modes} and
\ref{subsec:Scattering-times}). This unusual behavior allows us to
solve the (linearized) Boltzmann equation exactly in the limit of
a small $\alpha$. The details of this solution are given in Sec.
\ref{subsec:Non-local-transport-coefficients}. 

It is an important feature of graphene at the neutrality point, that
the hydrodynamic modes excited by electric and thermal fields decouple
in linear response, and in the absence of magnetic fields \cite{Mueller2008,Mueller2009}.
The modes are characterized by the distinct scattering, with all of
them following Eq. (\ref{eq:summary_scattering_rates}). Using our
full solution of the Boltzmann equation, the non-local, i.e. wave-vector-dependent,
charge and thermal conductivities as well as the non-local viscosity
were calculated. The longitudinal and transverse non-local charge
conductivities as functions of wave-vector $q$ and frequency $\omega$
are given by 
\begin{eqnarray}
\sigma_{\parallel} & = & \frac{\sigma_{0}}{1-i\tau_{c,1}\omega+\frac{1}{4}v^{2}\tau_{c,1}q^{2}\left(\frac{2i}{\omega}+\frac{1}{M_{c}\left(q,\omega\right)-i\omega}\right)},\nonumber \\
\sigma_{\bot} & = & \frac{\sigma_{0}}{1-i\tau_{c,1}\omega+\frac{\frac{1}{4}v^{2}\tau_{c,1}q^{2}}{M_{c}\left(q,\omega\right)-i\omega}},
\end{eqnarray}
where $\sigma_{0}=\frac{2e^{2}\log\left(2\right)k_{B}T\tau_{c,1}}{\pi\hbar^{2}}$
is the conductivity at vanishing wave-numbers and frequencies \cite{Fritz2008}.
$M_{c}$ is a memory function containing information on scattering
in high angular momentum channels $m\geq2$: 
\begin{equation}
M_{c}\left(q,\omega\right)=\tau_{c,2}^{-1}+\frac{1}{2}vq\frac{\textrm{I}_{3+\frac{\eta_{c}}{\gamma_{c}}-i\omega\tau_{c}}\left(\tau_{c}vq\right)}{\textrm{I}_{2+\frac{\eta_{c}}{\gamma_{c}}-i\omega\tau_{c}}\left(\tau_{c}vq\right)}.\label{eq:memory}
\end{equation}
This result is a direct consequence of the depence of the scattering
rate $\tau_{c,m}^{-1}\sim\alpha^{2}k_{B}T\left|m\right|$ on the angular
momentum state of the Dirac electron. A similar $\tau_{c,m}^{-1}\sim\left|m\right|$
behavior was found in Ref.\cite{Mirlin1997} for scattering off a
random magnetic field and gives rise to similar expressions for the
nonlocal conductivities, caused by rather different microscopic mechanism.
In Eq. (\ref{eq:memory}), $\tau_{c}$, $\gamma_{c}$ and $\eta_{c}$
determine the slopes and the offset in Eq. (\ref{eq:summary_scattering_rates})
(see Sec. \ref{subsec:Scattering-times}). The results for the non-local
thermal conductivity and viscosity are given in Eqs. (\ref{eq:thermal_conductivity_result}),
and (\ref{eq:viscosity_result}). The transport coefficients show
pronounced resonance features at $vq\approx\omega$ where $q$ and
$\omega$ are the wavenumber and frequency of the applied electric
field or thermal gradient (see Figs. \ref{fig:Fig_charge_conductivities},
\ref{fig:Fig_thermal_conductivities}) and $v$ is the electron group
velocity. The longitudinal charge conductivity can be measured in
experiments with surface acoustic waves (SAWs) \cite{Ingebrigtsen1969,Simon1996,Efros1990,Wixforth1989,Rotter1998,Govorov2000}.
The transverse conductivity determines the skin effect, which is however
not a feasible measurement for a two-dimensional graphene sheet. In
section \ref{sec:Surface-acoustic-waves} we consider a simple device
consisting of a graphene sheet laid on top of a piezoelectric crystal.
We calculate the velocity shift and damping of SAWs induced by the
graphene sheet and find that, while damping effects are small, a substantial
velocity shift can be expected. The damping and the velocity shift
measured as functions of temperature can give important insights into
the nature interaction effects in a Dirac fluid.

Non-local transport coefficients also determine in confined geometries.
The latter case is illustrated in Sec. \ref{sec:Poiseuille-profiles}
for the electric conductivity, using the Poiseuille geometry as an
example. The constitutive relation linking the electric current to
the electric field along the channel is interpreted as a differential
equation (Eq. (\ref{eq:pois_el_diff})) and solved with the appropriate
boundary conditions (Eq. (\ref{eq:boundary_condition_nl})). We find,
that the flow profiles strongly depend on the channel width $w$ as
compared to the electron-electron scattering lengths in the $m=1$
and $m=2$ channels: $l_{c,1}=v\tau_{c,1}$, $l_{c,2}=v\tau_{c,2}$.
While $l_{c,1}$ governs the decay of charge currents, $l_{c,2}$
determines the effectiveness of current transfer from regions with
high current density to regions with low current density. This latter
mechanism is analogous to viscous momentum transfer. The flow profiles
in dependence on $w$ can be separated into three regimes. For $w\gg l_{c,1}>l_{c,2}$,
the samples are in the Ohmic regime, where the current is dissipated
uniformly across the sample. The flow profile is flat. For $l_{c,1}<w<l_{c,2}$,
the profile curvature is maximal, since on the one hand the current
decay due to electron-electron scattering in the $m=1$ channel becomes
inefficient, on the other hand the current transfer to the boundaries
of the sample, where the flow is slowed down, is sufficiently strong.
For even smaller widths $w<l_{c,2}$, the profile turns flat again,
because the current transfer mechanism associated with $l_{c,2}$
ceases to be efficient. This characteristic pattern is shown in Fig.
\ref{fig:Poiseuille_slip}. Current profiles are accessible experimentally,
e.g. through the scanning single electron transistor technique of
Refs. \cite{Ella2019,Sulpizio2019}.

Finally, we calculated the dispersions of the collective modes of
a Dirac fluid. As do the transport coefficients, the collective modes
separate into a sector of charge excitations and a sector of energy
and imbalance excitations ($s=2$). These two sectors are decoupled
and can be studied separately. We find, that while the plasmon mode
is gapped out at small wave-numbers due to the interaction induced
resistivity (see Fig. \ref{fig:Fig_collective_charge_re}), a so-called
second sound mode, corresponding to a wavelike propagation of energy,
appears (Fig. \ref{fig:Fig_collective_energy_re}). Diffusive modes,
corresponding to the diffusion of charge, heat and quasiparticles
were found (see Figs. \ref{fig:Fig_collective_charge_im}, \ref{fig:Fig_collective_energy_im_Small_q}).
Their dispersion relations were calculated and showed to agree with
known results \cite{Briskot2015,Torre2019,Phan2013}. Besides these
well studied modes, an infinite set of damped modes connected to excitations
in higher angular harmonic channels was found (see Figs. \ref{fig:Fig_collective_charge_im},
\ref{fig:Fig_collective_energy_im}). The dispersions of these modes
are purely imaginary at vanishing wave-numbers and approach in the
long wavelength limit the values $\omega_{m}\left(q=0\right)=-i/\tau_{\varepsilon/c,m}$
for the $m$-th angular harmonic in the energy ($\varepsilon$) or
the charge ($c$) channels. At finite wave-numbers, these modes show
a complex structure of merging branches. Similar modes play an important
role in the equilibration of unitary fermi gases \cite{Brewer2015}
and the QCD plasma \cite{Romatschke2016,Romatschke2018,Heller2018}.
They also determine the unusual phase space dynamics of graphene electrons
which was the subject of an earlier work \cite{Kiselev2019b}.

\subsection*{Regime of validity}

Transport in graphene is of interest to researchers with diverse backgrounds.
Here we want to discuss the validity of our results in the context
of other graphene related research. Our paper is concerned with the
hydrodynmic regime, where electron tranport is governed by momentum
conserving electron-electron collisions and the electron-electron
mean free path is the smallest length scale \cite{Gurzhi1963}. In
particular, momentum relaxing scattering off impurities and phonons
must be weak. This demand sets serious limitations on sample sizes
and on the temperature range.

\subsubsection{The Dirac fluid of graphene at the charge neutrality point}

Throughout the paper we are interested in the low energy effective
behavior of graphene electrons near the Dirac point. Here, to a very
good approximation, the electron dispersion is given by the massless
two-dimensional Dirac Hamiltonian of Eq. (\ref{eq:Free_Dirac_Hamiltonian})
\cite{Castro2009}. At $T=0$, the lower Dirac cone is fully occupied
and the upper Dirac cone is empty. At finite temperatures, electrons
and holes in a region of size $k_{B}T$ around the Dirac cone are
created. These quasiparticles are carriers of electric and thermal
currents. Since their density is determined by temperature, $k_{B}T$
is the only energy scale in the system. We call this regime the Dirac
fluid regime. The chemical potential is vanishingly small: $\mu\ll k_{B}T$.
For the opposite case of a large chemical potential $\mu\gg k_{B}T$,
the system enters the Fermi liquid regime. Here, the scattering rate
is given by $\tau^{-1}\hbar\sim T^{2}/\mu$ \cite{Landau1957a,Landau1957b,Abrikosov1959}
(up to logarithmic corrections in 2D\cite{Hodges1971}). For the quantum
critical Dirac fluid, on the other hand, the electron-electron scattering
rate is determined by the temperature alone: 
\begin{equation}
\tau^{-1}\sim\alpha^{2}k_{B}T/\hbar\label{eq:approx_scattering_rate}
\end{equation}
where $\alpha=e^{2}/\left(\varepsilon v\hbar\right)$ is the graphene
fine structure constant. $v$ is the electron group velocity and $\varepsilon$
the dielectric constant. Higher order interaction effects can be treated
in terms of the renormalization group. Integrating out high energy
states above the thermal cut-off $k_{B}T$ results in a logarithmic
increase of the electron's group velocity \cite{Sheehy2007,Fritz2008}:
\begin{equation}
v=v_{0}\left(1+\frac{\alpha_{0}}{4}\log\left(\frac{\Lambda}{k_{B}T}\right)\right).\label{eq:velocity_renormalization}
\end{equation}
Here, $v_{0}\approx10^{6}\,\mathrm{m}/\mathrm{s}$ and $\alpha_{0}$
are the unrenormalized, bare electron velocity and the fine structure
constant. $\Lambda$ is an energy on the eV scale at which the electronic
bands begin to deviate from the linear Dirac-like shape. It is essential
to our theory, that the fine structure constant $\alpha\left(T\right)$
is renormalized to small values when the temperature is lowered. The
system is gradually approaching the free Dirac fermion fixed point,
thus ensuring the validity of the quasiparticle picture and the Boltzmann
approach chosen here to study the transport of electrons. Eq. (\ref{eq:velocity_renormalization})
is a perturbative result valid to lowest order in $\alpha$. However,
experiments show that the logarithmic increase of the Fermi velocity
at low energies is quite robust and holds even in the case of suspended
graphene where $\alpha_{0}\approx2$ as well as at intermediate temperatures
\cite{Elias2011}. Thus, there is good reason to believe that even
suspended graphene is located sufficiently near the free Dirac fermion
fixed point, such that weak coupling results are physically meaningful;
much more so for graphene grown on substrates with larger dielectric
constants.

\subsubsection{The quantum Boltzmann method}

The quantum Boltzmann method is well established for systems with
sharply defined quasiparticles \cite{KadanoffBaym,Mahan}, the prime
example being the Fermi liquid \cite{Abrikosov1959}. Here, thermally
excited quasiparticles have energies of the order of $\varepsilon_{\mathrm{qp}}=k_{B}T$,
such that the ratio $\varepsilon_{\mathrm{qp}}/\left(\hbar\tau^{-1}\right)\sim\mu/k_{B}T\gg1$
is large at temperatures below the Fermi temperature. This condition,
which is based on phase-space arguments rather than the interaction
strength, ensures the validity of the quasiparticle picture and the
Boltzmann equation.

In the case of the Dirac fluid, the ratio of the characteristic qusiparticle
energy and the scattering rate is
\begin{equation}
\frac{\varepsilon_{\mathrm{qp}}}{\tau^{-1}}\sim\alpha^{2}\left(T\right).
\end{equation}
Thus, the quasiparticle picture is valid only at small coupling strengths.
However, as discussed in the preceding section, for small temperatures
$\alpha\left(T\right)$ decreases, and the Dirac fluid asymptotically
approaches the free Dirac fermion limit. In this regime, the Boltzmann
equation provides a powerfull tool for the study of transport phenomena.
Coulomb interactions between electrons enter through a long-range
Vlasov term which describes electrostatic forces due to an inhomogeneous
charge distribution, as well as through the collision operator describing
short-range electron-electron collisions. We use the collision operator
derived in Ref. \cite{Fritz2008}, which includes all scattering processes
to second order in the fine sctructure constant (Born approximation).
While this approach is formally exact in the small $\alpha\left(T\right)$,
low temperature limit, we believe, as argued above, that it should
also provide reasonable results for larger values of the fine structure
constant.

In this paper, we consider the linear response of the Dirac fluid
to electric fields and thermal gradients at finite frequencies. The
Boltzmann approach limits our discussion to small frequencies: 
\begin{equation}
\omega\ll\frac{k_{B}T}{\hbar}.\label{eq:small_frequencies}
\end{equation}
At small frequencies, the system's response is governed by intra-band
processes which take place within one of the two Dirac cones. Inter-band
processes, on the other hand, involve the creation of electron-hole
pairs and therefore can only be excited at energies comparable to
$k_{B}T$ \cite{Link2016}. This means that the off-diagonal elements
of the density matrix $\left\langle \psi_{\lambda,\mathbf{k}}^{\dagger}\psi_{\lambda',\mathbf{k}}\right\rangle ,$where
$\psi_{\lambda,\mathbf{k}}^{\dagger}$, $\psi_{\lambda,\mathbf{k}}$
are electron creation and anihilation operators and the band index
$\lambda$ labes the two Dirac cones, are strongly suppressed. Allowing
us to interpret the diagonal components as a distribution function
\[
f_{\mathbf{k}\lambda}=\left\langle \psi_{\lambda,\mathbf{k}}^{\dagger}\psi_{\lambda,\mathbf{k}}\right\rangle 
\]
which can be found by solving the Boltzmann equation \cite{KadanoffBaym,Mahan}.
For further details on the quantum Boltzmann approach we refer to
Sec. \ref{sec:Theoretical-framework}, Appendix \ref{sec:App_Collision-operator}
and Ref. \cite{Fritz2008}.

\subsubsection{Impurities and Phonons}

At the temperature of $\sim50\,\mathrm{K}$, and assuming $\varepsilon\approx5$,
we estimate the electron-electron mean free path as $l_{ee}=v\tau\sim2\,\upmu\mathrm{m}$.
In clean graphene samples, impurity mean free paths of more than $10\upmu\mathrm{m}$
can be achieved \cite{Wang2013}, such that tranport indeed will be
dominated by electron-electron scattering. A major concern in experiments
with graphene near the charge neutrality point are small variations
of the local chemical potential $\mu\left(\mathbf{x}\right)$ which
have been dubbed electron-hole puddles \cite{Hwang2007,Martin2008}.
While the origins and properties of electron puddles and their influence
on transport are the subject of many studies (see e.g. \cite{Cheianov2007,Lucas2016,Gibertini2012,Sule2014}),
we choose not to include them in the present theory, which is concerned
with interaction effects in a clean Dirac fluid. Our results are relevant
for experiments with graphene sheets in the hydrodynamic regime. Here,
the dominance of electron-electron scattering over any impurity induced
effects was clearly demonstrated in Ref. \cite{Gallagher2019} by
showing that the electron scattering rates grow linearly in accordance
with Eq. (\ref{eq:approx_scattering_rate}) above a trashhold temperature.

Electron-phonon scattering is a significant disturbance for hydrodynamic
electron flows at high temperatures, unless one is in a regime governed
by phonon drag, see e.g. \cite{levchenko2020}. In graphene, the scattering
of electrons by 2D graphene lattice phonons is limited by the small
size of the Fermi-surface \cite{Efetov2010}, as well as by the high
Debye temperature which loweres the phonon density of states \cite{Efetov2010}.
These limitations are even more pronounced at the Dirac point, where
due to momentum conservation only phonons with momenta $k_{\mathrm{ph}}<k_{B}T/v$
participate in scattering events. However, scattering with surface
optical phonons of the substrate can lead to a significant increase
of the sheet resistance at higher temperatures. In Ref. \cite{Chen2008}
this mechanism was reported to set in above $150\,\mathrm{K}$ for
graphene grown on SiO$_{2}$. To a large extend, scattering on surface
acoustic photons determines the decay rates of graphene plasmons at
finite charge densities \cite{Hwang2010,karimi2017}. Experiments
on the hydrodynamics of Dirac fluids have been carried out with graphene
sheets encapsuled in hexagonal boron nitride \cite{Crossno2016,Gallagher2019}.
Here electron-phonon scattering is also reported to set in at the
relatively high temperatures of $70\,\mathrm{K}$ \cite{Crossno2016},
or even to be insignificant up to room temperatures \cite{Gallagher2019}.

\subsubsection{Sample sizes}

Currently, high quality graphene sheets have sizes on the order of
tenth of micrometers. On the one hand side this means that the effects
of boundary scattering can be important \cite{Kiselev2019a}. On the
other hand, it has been demonstrated that such samples are sufficiently
large to go well beyond the ballistic regime and to observe hydrodynamic
behavior \cite{Sulpizio2019,KrishnaKumar2017,Crossno2016,Gallagher2019,Bandurin2016,Bandurin2018,Berdyugin2019}. 

In graphene nanoribbons, gaps opening at the Dirac point can significantly
influence the behavior of collective modes \cite{Son2006,karimi2017}.
These gaps can be estimated as $\Delta\approx t/N$, where $t$ is
a characteristic tight-binding hopping amplitude on the $1$eV scale
and $N$ is the number of unit cells over which the ribbon extends.
For hydrodynamic samples $N\approx10^{5}$, and therefore the gaps
are much smaller than quasiparticle energies at experimental temperatures.

Boundary effects on collective mode propagation will give a larger
correction of order $l_{ee}/w$, where $w$ is the sample size (see
e.g. \cite{wild2012}).

\section{Theoretical framework\label{sec:Theoretical-framework}}

\subsection{Kinetic equation\label{subsec:Kinetic-equation_QB_Graphene}}

In order to clarify our notation, in this section we sketch the derivation
of the quantum Boltzmann formalism for the Dirac fluid, which was
developed in Ref. \cite{Fritz2008}. We begin with the Hamiltonian
of graphene electrons at the charge neutrality point: 
\begin{equation}
H=H_{0}+H_{\mathrm{int}},\label{eq:nl_Hamiltonian}
\end{equation}
where the free part is given by 
\begin{equation}
H_{0}=v\hbar\int_{\mathbf{k}}\sum_{a,b,i}\psi_{a,i}^{\dagger}\left(\mathbf{k}\right)\left(\mathbf{k\cdot\sigma}\right)_{ab}\psi_{b,i}\left(\mathbf{k}\right),\label{eq:Free_Dirac_Hamiltonian}
\end{equation}
and the interaction part reads 
\begin{equation}
H_{\mathrm{int}}=\frac{1}{2}\int_{\mathbf{k},\mathbf{k}',\mathbf{q}}\sum_{a,b,i,j}V\left(\mathbf{q}\right)\psi_{\mathbf{k+q},a,i}^{\dagger}\psi_{\mathbf{k}^{\prime}-\mathbf{q},b,j}^{\dagger}\psi_{\mathbf{k}^{\prime},b,j}\psi_{\mathbf{k,}a,i}.
\end{equation}
$V\left(\mathbf{q}\right)=\frac{2\pi e^{2}}{\varepsilon\left\vert \mathbf{q}\right\vert }$
is the 2D Coulomb potential. The indices $i,j=1,2\,...,N=4$ refer
to the spin and valley quantum numbers of an electron, whereas the
two sub-lattices are labelled by the indices $a,b$. The free particle
Hamiltonian $H_{0}$ is diagonalized by the unitary transformation
\begin{equation}
U_{\mathbf{k}}=\frac{1}{\sqrt{2}}\left[\begin{array}{cc}
1 & o_{\mathbf{k}}^{*}\\
1 & -o_{\mathbf{k}}^{*}
\end{array}\right],\label{eq:free_transform}
\end{equation}
where $o_{\mathbf{k}}=\left(k_{x}+ik_{y}\right)/\sqrt{k_{x}^{2}+k_{y}^{2}}$.

For the derivation of the quantum Boltzmann equation, it is convenient
to use the band representation of Dirac spinors $\psi_{\lambda,\mathbf{k}}=U_{\mathbf{k},\lambda a}\psi_{\mathbf{k},a}$
with $\lambda=\pm1$ labeling the upper and lower Dirac cones. In
this way, one can easily distinguish between processes that involve
the creation of particle-hole pairs and those which do not. The thermally
excited electron-hole pairs occupy states in a window of $k_{B}T$
around the Dirac point. Thus, if the applied fields have frequencies
$\omega<2k_{B}T/\hbar$, which is true in the hydrodynamic regime,
processes that create electron-hole pairs are unlikely and can be
neglected. This translates to neglecting the off-diagonal components
of the distribution function in the band representation, which is
then given by its diagonal elements: 
\[
f_{\mathbf{k}\lambda}=\left\langle \psi_{\lambda,\mathbf{k}}^{\dagger}\psi_{\lambda,\mathbf{k}}\right\rangle .
\]
The quantum Boltzmann equation then reads 
\begin{equation}
\left(\partial_{t}+\mathbf{v}_{\mathbf{k}\lambda}\cdot\nabla_{\mathbf{r}}-\left(e\nabla\varphi_{\mathrm{tot}}\right)\cdot\nabla_{\mathbf{k}}+C\right)f_{\mathbf{k}\lambda}\left(\mathbf{r},t\right)=0.\label{eq:Boltzmann}
\end{equation}
Here, $\mathbf{v}_{\mathbf{k}\lambda}=\partial\varepsilon_{\mathbf{k}\lambda}/\partial\mathbf{k}$
is the group velocity and 
\begin{equation}
\mathbf{\varphi_{\mathrm{tot}}}\left(\mathbf{r},t\right)=\varphi_{\mathrm{ext}}\left(\mathbf{r},t\right)+\varphi_{\mathrm{ind}}\left(\mathbf{r},t\right)\label{eq:potential_tot_ind_ext}
\end{equation}
is the sum of the external electrostatic potential and the induced
potential which is the result of an inhomogeneous distribution of
charges. The term associated with $\varphi_{\mathrm{tot}}$ was first
introduced by Vlasov\cite{Vlasov1938}. It will be dealt with at the
end of this section. $C$ represents the central part of the kinetic
theory - the Boltzmann collision operator describingelectron-electron
Coulomb scattering. Details on the derivation of $C$ are summarized
in Appendix \ref{sec:App_Collision-operator}, based on Refs. \cite{Fritz2008,Kiselev2019b}.

Studying the linear response to $\mathbf{\varphi_{\mathrm{tot}}}$,
we expand the distribution function around the local equilibrium distribution
$f_{k\lambda}^{\left(0\right)}$ 
\begin{equation}
f_{\mathbf{k}\lambda}\left(\mathbf{r},t\right)=f_{k\lambda}^{\left(0\right)}+w_{k}\psi_{\mathbf{k}\lambda}\left(\mathbf{r},t\right).\label{eq:Distribution_expanded}
\end{equation}
where $f_{k,\lambda}^{0}$ is given by 
\begin{equation}
f_{k\lambda}^{\left(0\right)}=\frac{1}{e^{\beta\left(\varepsilon_{\mathbf{k}\lambda}-\mathbf{u}\cdot\mathbf{k}\right)}+1}.\label{eq:Local_equilibrium}
\end{equation}
The product $w_{k}\equiv f_{k}^{\left(0\right)}\left(1-f_{k}^{\left(0\right)}\right)$
, that will soon play the role of a weight function in the scalar
product, does not depend on $\lambda$, and the corresponding index
is dropped in Eq. (\ref{eq:Distribution_expanded}) and in the following.

Performing a Fourier transformation $\psi{}_{\mathbf{k}\lambda}\left(\mathbf{r},t\right)\rightarrow\psi_{\mathbf{k}\lambda}\left(\mathbf{q},\omega\right)$
to frequency and momentum space, we obtain the linearized Boltzmann
equation 
\begin{equation}
\left({\cal L}+{\cal C}\right)\psi_{\mathbf{k}\lambda}\left(\mathbf{q},\omega\right)=S_{\mathbf{k}\lambda}\left(\mathbf{q},\omega\right).\label{eq:Kin_Eq}
\end{equation}
$\mathcal{L}$ is the Liouville operator and given by 
\begin{equation}
{\cal L}=-i\omega+i\mathbf{q}\cdot\mathbf{v}_{\mathbf{k}\lambda}
\end{equation}
The linearization of the collision operator can be expressed in the
form 
\begin{equation}
{\cal C}\psi_{\mathbf{k}\lambda}\approx\frac{1}{w_{k}}\sum_{\lambda'}\int_{\mathbf{k}'}\frac{\delta\left(C\psi\right)_{\mathbf{k}\lambda}}{\delta\psi_{\mathbf{k}'\lambda'}}\psi_{\mathbf{k}'\lambda'},
\end{equation}
where the weight function $w_{k}$ was introduced above.

Let the $\psi_{\boldsymbol{k}}$ be element of a function space with
inner product 
\begin{equation}
\left\langle \phi\mid\psi\right\rangle =\sum_{\lambda}\int_{\mathbf{k}}w_{k}\phi_{\mathbf{k}\lambda}^{*}\psi_{\mathbf{k}\lambda},\label{eq:Scalar_product}
\end{equation}
such that 
\begin{eqnarray}
\left\langle \phi\left|{\cal C}\right|\psi\right\rangle  & = & \sum_{\lambda}\int_{\mathbf{k}}w_{k}\phi_{\mathbf{k}\lambda}^{*}{\cal C}\psi_{\mathbf{k}\lambda}\nonumber \\
 & = & \sum_{\lambda\lambda'}\int_{\mathbf{k}\mathbf{k'}}\phi_{\mathbf{k}\lambda}^{*}\frac{\delta\left(C\psi\right)_{\mathbf{k}\lambda}}{\delta\psi_{\mathbf{k}'\lambda'}}\psi_{\mathbf{k}'\lambda'}.
\end{eqnarray}
One can show that the entropy production in the absence of external
driving terms is $\frac{\partial S}{\partial t}=k_{B}\left\langle \psi\left|{\cal C}\right|\psi\right\rangle $
which ensures that the collision operator is positive definite. In
fact, ${\cal C}$ is Hermitian under the above scalar product. Therefore
its eigenvalues are real and its eigenfunction form an orthonormal
basis of the function space.

The right hand side of Eq. (\ref{eq:Kin_Eq}) is determined by the
forces acting on the system. The three force terms studied here are
due to electric fields, thermal gradients and viscous forces. For
an electric field oriented along the $x$-axis, $\mathbf{E}=E_{0}\hat{\mathbf{e}}_{x}$,
the force term reads 
\begin{equation}
S_{E}=-eE_{0}\cos\theta\left(\lambda v\beta\right),\label{eq:Electric_force}
\end{equation}
where $\theta$ is the polar angle of the momentum $\mathbf{k}$.
It is important to notice, that
\[
\mathbf{E}=-\nabla\varphi_{\mathrm{tot}}.
\]
The corresponding term for a thermal gradient $\nabla T$ is given
by 
\begin{equation}
S_{T}=-k\left|\nabla T\right|\cos\theta k_{B}\left(v\beta\right)^{2}.\label{eq:Thermal_force}
\end{equation}
A viscous force is present if the drift velocity $\mathbf{u}$ in
the local equilibrium distribution function (\ref{eq:Local_equilibrium})
is a function of the coordinate $\mathbf{x}$. Then the drift term
of the Boltzmann equation (\ref{eq:Boltzmann}) can be thought of
as a force term 
\begin{align}
S_{S} & =-vkX_{0,\alpha\beta}\left(\frac{k_{\alpha}k_{\beta}}{k^{2}}-\frac{1}{2}\delta_{\alpha\beta}\right)\lambda\beta\nonumber \\
 & =-\frac{1}{2}kX_{0}\sin\left(2\theta\right)\left(\lambda v\beta\right),\label{eq:Shear_force}
\end{align}
where the stress tensor is given by 
\begin{align*}
X_{0,\alpha\beta} & =\frac{1}{2}\left(\frac{\partial u_{\alpha}}{\partial x_{\beta}}+\frac{\partial u_{\beta}}{\partial x_{\alpha}}-2\delta_{\alpha\beta}\nabla\cdot\mathbf{u}\right).
\end{align*}
In the following, we consider a flow with $\mathbf{u}\left(y\right)=u\left(y\right)\hat{\mathbf{e}}_{x}$
and therefore only include the component $X_{0,xy}$, which is relevant
for the calculation of the shear viscosity.

The collision operator is given by 
\begin{eqnarray}
\left(C\psi\right)_{\mathbf{k}\lambda} & = & \frac{2\pi}{\hbar}\int_{k^{\prime}q}\delta\left(k+k'-\left|\mathbf{k}+\mathbf{q}\right|-\left|\mathbf{k}'-\mathbf{q}\right|\right)\label{eq:Collision_operator}\\
 & \times & \left(1-f_{k}^{\left(0\right)}\right)\left(1-f_{k'}^{\left(0\right)}\right)f_{\left\vert \mathbf{k}+\mathbf{q}\right\vert }^{\left(0\right)}f_{\left\vert \mathbf{k}'-\mathbf{q}\right\vert }^{\left(0\right)}\nonumber \\
 & \times & \left\{ \gamma_{\mathbf{k},\mathbf{k}^{\prime},\mathbf{q}}^{\left(1\right)}\left(\psi_{\mathbf{k}+\mathbf{q},\lambda}+\psi_{\mathbf{k}'-\mathbf{q},\lambda}-\psi_{\mathbf{k}',\lambda}-\psi_{\mathbf{k},\lambda}\right)\right.\nonumber \\
 & + & \left.\gamma_{\mathbf{k},\mathbf{k}^{\prime},\mathbf{q}}^{\left(2\right)}\left(\psi_{\mathbf{k}+\mathbf{q},\lambda}-\psi_{-\mathbf{k}'+\mathbf{q},\bar{\lambda}}+\psi_{-\mathbf{k}',\bar{\lambda}}-\psi_{\mathbf{k},\lambda}\right)\right\} .\nonumber 
\end{eqnarray}
The matrix elements $\gamma_{\mathbf{k},\mathbf{k}^{\prime},\mathbf{q}}^{\left(1\right)}$,
$\gamma_{\mathbf{k},\mathbf{k}^{\prime},\mathbf{q}}^{\left(2\right)}$
can be found in Appendix \ref{sec:App_Collision-operator}.

Another important term in the kinetic equation describes the electrostatic
forces that arise due to an inhomogeneous distribution of charges.
These forces are mediated by a self consistent potential $\varphi_{\mathrm{ind}}$,
first introduced by Vlasov \cite{Vlasov1938}. It reads 
\begin{equation}
e\varphi_{\mathrm{ind}}\left(\mathbf{r},t\right)=\alpha vN\int d^{2}r'\sum_{\lambda}\int\frac{d^{2}k}{\left(2\pi\right)^{2}}\,\frac{\delta f_{\mathbf{k}\lambda}\left(\mathbf{r}',\omega\right)}{\left|\mathbf{r}-\mathbf{r}'\right|},\label{eq:Vlasov_field}
\end{equation}
where we have used the abbreviation $\delta f_{\mathbf{k}\lambda}\left(\mathbf{r},\omega\right)=w_{k}\psi_{\mathbf{k},\lambda}\left(\mathbf{r},\omega\right)$
and multiplied the potential by $e$ for notational convenience. A
derivation of the term can be found in Ref. \cite{KadanoffBaym} (Eqs.
(7-3) and (9-16)). Applying a Fourier transform to Eq. (\ref{eq:Vlasov_field})
one finds 
\begin{eqnarray*}
e\varphi_{\mathrm{ind}}\left(\mathbf{q},t\right) & = & \alpha vN\sum_{\lambda}\int\frac{d^{2}k}{\left(2\pi\right)^{2}}\,\frac{2\pi\delta f_{\mathbf{k}\lambda}\left(\mathbf{q},\omega\right)}{q}.
\end{eqnarray*}
In Sec. \ref{sec:Collective-modes} we will not be interested in the
response to the total electric field $\mathbf{E}=-i\mathbf{q}\varphi_{\mathrm{tot}}$,
but rather in solutions of the homogeneous Boltzmann equation
\[
\left({\cal L}+\mathcal{V}+{\cal C}\right)\psi_{\mathbf{k}\lambda}\left(\mathbf{q},\omega\right)=0.
\]
Here, the Vlasov term 
\begin{equation}
\mathcal{V}\psi_{\mathbf{k}\lambda}=-i\mathbf{q}\cdot\mathbf{v}_{\mathbf{k}\lambda}e\varphi_{\mathrm{ind}}.\label{eq:Vlasov_term}
\end{equation}
has to be included explicitely.

\subsection{Collinear zero modes\label{subsec:Collinear-zero-modes}}

In this section, we summarize how Eq. (\ref{eq:Kin_Eq}) is solved
in the limit of a small fine structure constant. A standard way to
deal with an integral equation like (\ref{eq:Kin_Eq}) is to expand
the function $\psi_{\mathbf{k},\lambda}$ into a set of suitable basis
functions. The choice of this basis is facilitated by the fact that
for small values of the graphene fine structure constant $\alpha$,
the collision operator (\ref{eq:Collision_operator}) logarithmically
diverges if the velocities of involved particles are parallel to each
other. This is a consequence of the linear single particle spectrum,
and the resulting momentum independent velocity of massless Dirac
particles. Intuitively speaking, the scattering is enhanced, because
particles traveling in the same direction interact with each other
over a particularly long period of time. A more mathematical picture
of this so-called collinear scattering anomaly is presented in Appendix
(\ref{sec:App_Collinear-scattering}). It is convenient to write the
collision operator as a sum of the collinear part $\mathcal{C}_{c}$
and the non-collinear part $\mathcal{C}_{nc}$: 
\begin{equation}
\mathcal{C}=\log{\left(1/\alpha\right)}\mathcal{C}_{c}+\mathcal{C}_{nc}.\label{eq:Collinear_non_collinear_separation}
\end{equation}
The factor $\log{\left(1/\alpha\right)}$ is large at small $\alpha$.
Both operators, $\mathcal{C}_{c}$ and $\mathcal{C}_{nc}$, are hermitian
with respect to the scalar product of Eq.\ref{eq:Scalar_product}.
Let $\varphi_{\mathbf{k},\lambda}^{n}$ be the orthogonal eigenfunctions
of $\mathcal{C}_{c}$ such that 
\begin{equation}
\left(\mathcal{C}_{c}\varphi^{n}\right)_{\mathbf{k},\lambda}=b_{n}\varphi_{\mathbf{k},\lambda}^{n}.
\end{equation}
$\psi_{\mathbf{k},\lambda}$ is expanded in terms of these functions:
\begin{equation}
\psi_{\mathbf{k},\lambda}=\sum_{n}\gamma_{n}\varphi_{\mathbf{k},\lambda}^{n}.\label{eq:Collinear_expansion}
\end{equation}
Suppose, some of the orthogonal basis functions $\varphi^{n}$, namely
those with $n<n_{0}$, set the collinear part of the collision operator
to zero, i.e. 
\begin{equation}
\mathcal{C}_{c}\varphi^{n<n_{0}}=0.
\end{equation}
Then, inserting the expansion (\ref{eq:Collinear_expansion}) into
Eq. (\ref{eq:Kin_Eq}) and projecting it onto the basis functions
$\varphi^{n'}$, one finds 
\begin{equation}
\gamma_{n'>n_{0}}=\frac{\Braket{\varphi^{n'}|S}-\Braket{\varphi^{n'}|\left(\mathcal{L}+\mathcal{C}_{nc}\right)\psi}}{b_{n'}\log{\left(1/\alpha\right)}}.\label{eq:Non_collinear_suppression}
\end{equation}
Hence, zero modes of $\mathcal{C}_{c}$ are enhanced by factor $\log{\left(1/\alpha\right)}$\cite{Fritz2008}.
These colinear zero modes can be found from the collision operator
given in Eq.\ref{eq:Collision_operator}: 
\begin{equation}
\chi_{\mathbf{k},\lambda}^{\left(m,s\right)}=\lambda^{m}e^{im\theta}\left\{ 1,\lambda,\lambda\beta v\hbar k\right\} .\label{eq:Collinear_modes}
\end{equation}
Here, $m$ labels the angular momentum, $s\in\left\{ 1,2,3\right\} $
the modes $\left\{ 1,\lambda,\lambda\beta vk\right\} $, and $\theta$
is the polar angle of the momentum vector $\mathbf{k}$. All modes
set the integral (\ref{eq:Collision_operator}) to zero for collinear
processes (see Appendix \ref{sec:App_Collinear-scattering}).

From Eq. (\ref{eq:Non_collinear_suppression}) follows that for small
values of $\alpha$, only the collinear zero modes have to be retained
in the expansion of the entire collision operator Eq. (\ref{eq:Collinear_expansion}),
i.e. the kinetic equation (\ref{eq:Boltzmann}) can be solved using
the restricted subspace of basis functions of Eq. (\ref{eq:Collinear_modes}).
The stronger colinear scattering processes give rise to a rapid equilibration
to the subset of modes given in Eq. (\ref{eq:Collinear_modes}) which
then dominate the long-time dynamics.

In order to proceed, the matrix elements of Eq. (\ref{eq:Kin_Eq})
in this basis must be calculated. The matrix elements of the Liouville
operator $\mathcal{L}$ are given by \begin{widetext} 
\begin{equation}
\left\langle \chi_{\mathbf{k},\lambda}^{\left(m,s\right)}\left|\mathcal{L}\right|\chi_{\mathbf{k},\lambda}^{\left(m',s'\right)}\right\rangle =\left(-i\omega\delta_{m,m'}+\frac{1}{2}ivq\left(e^{-i\vartheta_{\mathbf{q}}}\delta_{m,m'+1}+e^{i\vartheta_{\mathbf{q}}}\delta_{m,m'-1}\right)\right)\left(v\beta\hbar\right)^{-2}L_{s,s'},\label{eq:Liouville_projected}
\end{equation}
\end{widetext}where $\vartheta_{\mathbf{q}}$ is the polar angle
of the wave-vector $\mathbf{q}$ and 
\begin{equation}
L=\left[\begin{array}{ccc}
\frac{\log(2)}{\pi} & 0 & 0\\
0 & \frac{\log(2)}{\pi} & \frac{\pi}{6}\\
0 & \frac{\pi}{6} & \frac{9\zeta(3)}{2\pi}
\end{array}\right].
\end{equation}
The rows and columns of the matrix notation refer to the mode index
$s$ of Eq. (\ref{eq:Collinear_modes}).

We calculate the matrix elements of the collision operator $\mathcal{C}$
numerically (some values are given in Appendix \ref{sec:App_Matrix-elements}).
Due to the rotational invariance of the low-energy Dirac Hamiltonian
(\ref{eq:nl_Hamiltonian}), they are diagonal in the angular harmonic
representation. Most importantly, the matrix elements rapidly approach
a linear behavior for large $\left|m\right|$: 
\begin{equation}
\left\langle \chi_{\mathbf{k},\lambda}^{\left(m,s\right)}\left|\mathcal{C}\right|\chi_{\mathbf{k},\lambda}^{\left(m',s'\right)}\right\rangle =\frac{\delta_{m,m'}}{v^{2}\beta^{3}\hbar^{3}}\left(\left|m\right|\gamma_{s,s'}-\eta_{s,s'}\right).\label{eq:Linear_approximation}
\end{equation}
$\gamma_{s,s'}$ and $\eta_{s,s'}$ are numerical coefficients that
are listed below Eqs. (\ref{eq:linear_scattering_rates_charge}) and
(\ref{eq:linear_scattering_rates_energy}). This surprising result
is due to the linear Dirac spectrum of the system. It allows to solve
the Boltzmann equation exactly, as will be seen later. The linear
behavior of the scattering rates is also shown in Fig. \ref{fig:The-matrix-elements}.
To find closed expressions for the non-local transport coefficients,
the scattering rates are approximated by Eq. (\ref{eq:Linear_approximation})
for $m>2$. In principle, the numerically exact scattering rates up
to an arbitrary $m$ can be included. Here, the rates for $m>2$ will
be assumed to follow Eq.\ref{eq:Linear_approximation} in order to
keep the algebraic efforts at a minimum. The projections of the force
terms (\ref{eq:Electric_force})-(\ref{eq:Shear_force}) onto collinear
zero modes read 
\begin{equation}
\Braket{S_{E}|\chi_{\mathbf{k},\lambda}}=-\frac{eE_{0}}{2\hbar^{2}\beta v}\delta_{\left|m\right|,1}\left[\begin{array}{c}
\frac{\log(2)}{\pi}\\
0\\
0
\end{array}\right],\label{eq:electric_force_projected}
\end{equation}
\begin{equation}
\Braket{S_{T}|\chi_{\mathbf{k},\lambda}}=\frac{\left|\nabla T\right|k_{B}\pi^{4}}{v\beta\hbar^{2}}\delta_{\left|m\right|,1}\left[\begin{array}{c}
0\\
\frac{\pi}{6}\\
\frac{9\zeta(3)}{2\pi}
\end{array}\right],\label{eq:Thermal_force_projected}
\end{equation}
\begin{align}
\Braket{S_{S}|\chi_{\mathbf{k},\lambda}} & =-\frac{iX_{0}}{4\left(v\beta\hbar\right)^{2}}\textrm{sign}\left(m\right)\delta_{\left|m\right|,2}\left[\begin{array}{c}
0\\
\frac{\pi}{6}\\
\frac{9\zeta\left(3\right)}{2\pi}
\end{array}\right].\label{eq:viscous_projected}
\end{align}
For the Vlasov term (\ref{eq:Vlasov_term}) one finds 
\begin{eqnarray}
\left\langle \psi_{\mathbf{k},\lambda}^{\left(m,s\right)}|\mathcal{V}|\psi_{\mathbf{k},\lambda}^{\left(m',s'\right)}\right\rangle  & = & i\alpha N\left(e^{-i\vartheta_{\bm{q}}}\delta_{m,1}+e^{i\vartheta_{\bm{q}}}\delta_{m,-1}\right)\nonumber \\
 &  & \times\frac{\delta_{1,s}\delta_{1,s'}\delta_{m',0}}{2v^{2}\beta^{3}\hbar^{3}}\left[\begin{array}{c}
\frac{\log(2)^{2}}{\pi^{2}}\\
0\\
0
\end{array}\right].\label{eq:Vlasov_projected}
\end{eqnarray}
The non-equilibrium part of the distribution function expanded in
the subset of colinear zero modes becomes 
\begin{equation}
\psi_{\mathbf{k},\lambda}=\sum_{m=-\infty}^{\infty}\sum_{s=1}^{3}a_{m,s}\left(\omega,\mathbf{q}\right)\chi_{\mathbf{k},\lambda}^{\left(m,s\right)}.\label{eq:Distr_dunction_expanded}
\end{equation}
Together, the expressions (\ref{eq:Kin_Eq}), (\ref{eq:Liouville_projected}),
(\ref{eq:Linear_approximation}), (\ref{eq:electric_force_projected})-(\ref{eq:viscous_projected}),
(\ref{eq:Vlasov_projected}) and (\ref{eq:Distr_dunction_expanded})
provide a linearized kinetic equation restricted to the basis of collinear
zero modes that becomes exact for small values of the fine structure
constant $\alpha$. Since no assumptions on the spatial dependencies
were made, except that they are be within the limits of the applicability
of the kinetic equation, this expansion can be used to derive the
non-local transport coefficients in the linear-response regime, as
well as the dispersion relations of collective excitations.

\section{Non-local Transport\label{sec:Non-local-Transport} }

\subsection{Effects of electron-hole symmetry, momentum conservation and thermal
transport\label{subsec:Charge-and-energy}}

Within the kinetic approach, the charge current $\mathbf{j}_{c}$
and the heat current $\mathbf{j_{\varepsilon}}$ are given by 
\begin{align}
\mathbf{j}_{c} & =e\sum_{\lambda}\int_{\mathbf{k}}\lambda v\frac{\mathbf{k}}{k}f_{\mathbf{k},\lambda},\label{eq:charge_current_exp}\\
\mathbf{j_{\varepsilon}} & =\sum_{\lambda}\int_{\mathbf{k}}v^{2}\hbar\mathbf{k}f_{\mathbf{k},\lambda}.\label{eq:heat_current_exp}
\end{align}
In these expressions intra-band processes that create particle-hole
pairs are neglected (see Appendix \ref{sec:App_Collision-operator}).
It follows from Eqs. (\ref{eq:charge_current_exp}) (\ref{eq:heat_current_exp}),
that the even in $\lambda$ part of the distribution function $f_{\mathbf{k},\lambda}$
contains information about thermal transport, whereas the odd part
governs the transport of charge. Since the electric field contribution
to the kinetic equation (\ref{eq:Electric_force}) is odd in $\lambda$,
and the thermal gradient leads to a term that is even in $\lambda$
(Eq. (\ref{eq:Thermal_force})), the phenomena of thermal and charge
transport are decoupled to linear order in the external fields at
the neutrality point. This can be traced back to particle-hole symmetry
and is the ultimate reason why the Wiedemann-Franz law is dramatically
violated in a Dirac fluid\cite{Crossno2016}. The distribution function
shows a similar decoupling ocf charge and heat modes for higher $m$:
The collinear modes of Eq. (\ref{eq:Collinear_modes}) are proportional
to $\lambda^{m}$ for $s=1$ and to $\lambda^{m+1}$ for $s=2,3$.
Consequently the kinetic equation in the subspace of collinear zero
modes is block diagonal in the $s=1$ and $s=2,3$ modes, as can be
seen from Eqs. (\ref{eq:Collision_operator}), (\ref{eq:Liouville_projected}),
(\ref{eq:electric_force_projected})-(\ref{eq:viscous_projected}).
In the following this will further simplify the calculation of transport
coefficients.

Another important consequence of the linear graphene spectrum is that
the heat current $\mathbf{j}_{\varepsilon}$ is proportional to the
momentum density $\mathbf{g}=\sum_{\lambda}\int_{k}\hbar\mathbf{k}f_{\mathbf{k},\lambda}$
and is therefore conserved. The charge current, unlike in Galilean
invariant systems, is not conserved, and decays due to interactions,
giving rise to a finite restistvity in the clean system.

\subsection{Scattering times\label{subsec:Scattering-times}}

The matrix elements of the collision operator determine the scattering
rates of the three collinear zero modes in different angular harmonic
channels. In the absence of spatial inhomogeneities and external forces,
the kinetic equation in the basis of collinear zero modes (\ref{eq:Collinear_modes})
reads 
\begin{equation}
\sum_{s'}\left(\partial_{t}\delta_{s,s'}+\Gamma_{m}^{s,s'}\right)a_{m,s'}=0,\label{eq:Decay_eq}
\end{equation}
where the $a_{m,s}$ are the coefficients of the expansion (\ref{eq:Distr_dunction_expanded}).
Posed as an initial value problem, this equation describes the exponential
decay of collinear zero modes. This decay governs the behavior of
the system at long time scales, because modes that do not set the
collinear part of the collision integral to zero decay faster by a
factor $\log\left(1/\alpha\right)$ (see Eq. (\ref{eq:Collinear_non_collinear_separation})).

The scattering rates $\Gamma_{m}^{s,s'}$ are given by 
\begin{equation}
\Gamma_{m}^{s,s'}=\left(v\beta\hbar\right)^{2}L_{s,s'}^{-1}\left\langle \chi_{\mathbf{k},\lambda}^{\left(m,s\right)}\left|\mathcal{C}\right|\chi_{\mathbf{k},\lambda}^{\left(m',s'\right)}\right\rangle .\label{eq:Definition_scattering_rates}
\end{equation}
Because of the definition of the scalar product in Eq. (\ref{eq:Scalar_product}),
the matrix elements have dimension ${\rm length}$$^{2}/{\rm time}$.
Vanishing scattering rates indicate conservation laws, and the corresponding
modes are zero modes of the full collision operator as well as its
collinear part. These modes reflect the conservation of particle density,
imbalance density, energy density and momentum density: 
\[
\begin{array}{ccc}
\chi_{\mathbf{k},\lambda}^{\left(s=1,m=0\right)}=1, & \qquad & \chi_{\mathbf{k},\lambda}^{\left(s=2,m=0\right)}=\lambda,\end{array}
\]
\[
\chi_{\mathbf{k},\lambda}^{\left(s=3,m=0\right)}=\lambda\beta v\hbar k,\qquad\chi_{\mathbf{k},\lambda}^{\left(s=3,m=1\right)}=\lambda e^{i\theta}\beta v\hbar k.
\]
The imbalance density is conserved only to order $\alpha^{2}$, as
it decays due to higher order interaction processes. An important
simplification stems from the fact that all scattering rates, for
large $\left|m\right|$, share the asymptotic behavior $\Gamma_{m}\sim\left|m\right|$.
This becomes a reasonable approximation for the scattering rates with
$m\geq2$. In the next section it is shown, how this behavior allows
us to obtain closed form expressions for the non-local transport coefficients.
As discussed in the previous section, the matrix of scattering rates
$\Gamma_{m}^{s,s'}$ is block diagonal in the modes describing charge
($s=1$) and thermal excitations ($s=2,3$), i.e. $\Gamma_{m}^{1,2}=\Gamma_{m}^{2,1}=\Gamma_{m}^{1,3}=\Gamma_{m}^{3,1}=0$.
Therefore, the scattering times determining the non-local electric
conductivity are given by $\tau_{c,m}=1/\Gamma_{m}^{1,1}$: $\tau_{c,0}\rightarrow\infty$,
$\tau_{c,1}=\frac{1}{\alpha^{2}}\frac{\hbar}{k_{B}T}\frac{\log2}{0.804\pi}$,
$\tau_{c,2}=\frac{1}{\alpha^{2}}\frac{\hbar}{k_{B}T}\frac{\log2}{2.617\pi}$
as well as 
\begin{align}
\tau_{c,m} & \approx\frac{1}{\alpha^{2}}\frac{\hbar}{k_{B}T}\frac{\log2}{\pi}\left(\gamma_{c}\cdot\left|m\right|-\eta_{c}\right)^{-1}\,{\rm if}\ m>2,\label{eq:linear_scattering_rates_charge}
\end{align}
where $\gamma_{c}=2.57$ and $\eta_{c}=3.45$ (see Appendix \ref{sec:App_Matrix-elements}
for more numerical values). It is also convenient to define an effective
scattering time for the Vlasov term: 
\begin{equation}
\tau_{V}=\frac{2\pi^{2}\beta\hbar}{\alpha N\log\left(2\right)}.\label{eq:Vlasov_time}
\end{equation}
Notice, that $\tau_{V}/\tau_{c,m}\sim1/\alpha$ is large for small
$\alpha$.

In the thermal sector, there are two relevant modes. However, the
$s=3$ mode is physically more important, because the vanishing of
the corresponding scattering rates for the $m=0$ and $m=1$ channels
indicate the conservation of energy and momentum. In the following,
it is shown that the neglecting of the $s=2$ imbalance mode in the
calculation of the thermal conductivity and viscosity, while significantly
simplifying the analysis, does only result in a small numerical error.
Therefore, for the purpose of calculating the transport coefficients,
only the $s=3$ energy mode will be considered. The scattering times
are then given by $\tau_{\varepsilon,m}=1/\Gamma_{m}^{3,3}$. Because
of energy and momentum conservation, we have $\tau_{\varepsilon,m=0,1}\rightarrow\infty$,
and for $m=2$, it is $\tau_{\varepsilon,2}=\frac{1}{\alpha^{2}}\frac{\hbar}{k_{B}T}\frac{9\zeta(3)}{3.341\cdot2\pi}$.
For $m>2$ the linear approximation can be used: 
\begin{equation}
\tau_{\varepsilon,m}\approx\frac{1}{\alpha^{2}}\frac{\hbar}{k_{B}T}\frac{9\zeta(3)}{2\pi}\left(\gamma_{\varepsilon}\cdot\left|m\right|-\eta_{\varepsilon}\right)^{-1}\ m>2,\label{eq:linear_scattering_rates_energy}
\end{equation}
with $\gamma_{\varepsilon}=5.18$ and $\eta_{\varepsilon}=11.3$.

\subsection{Non-local transport coefficients\label{subsec:Non-local-transport-coefficients}}

The linear, non-local response of a system to external forces $\mathcal{F}\left(\mathbf{r}\right)$
is characterized by constitutive relations of the form 
\begin{equation}
\mathcal{J}\left(\mathbf{r},t\right)=\int d^{d}r'dt'\,\nu\left(\mathbf{r}-\mathbf{r}',t-t'\right)\mathcal{F}\left(\mathbf{r}',t'\right),\label{eq:non-local-constitutive}
\end{equation}
where $\mathcal{J}\left(\mathbf{r},t\right)$ is a current sourced
by the field $\mathcal{F}\left(\mathbf{r}',t'\right)$ and $\nu\left(\mathbf{r}-\mathbf{r}',t-t'\right)$
is the corresponding transport coefficient. $\mathcal{F}$ can be
a scalar potential, a vector field (an electric field or a thermal
gradient), or a tensor. Eq. (\ref{eq:non-local-constitutive}) takes
a much simpler form in Fourier space: 
\begin{equation}
\mathcal{J}\left(\mathbf{q},\omega\right)=\nu\left(\mathbf{q},\omega\right)\mathcal{F}\left(\mathbf{q},\omega\right).\label{constitutive_fourier}
\end{equation}
If the system is confined to a geometry of a characteristic size $l_{\mathrm{geo}}$,
the relevant wave vectors $\mathbf{q}$ in Eq. (\ref{constitutive_fourier})
will be of the order of $q_{\mathrm{geo}}\approx2\pi/l_{\mathrm{geo}}$.
On the other hand $\nu\left(\mathbf{q},\omega\right)$ varies on scales
of the inverse mean free path $q_{mf}\approx2\pi/l_{mf}$, where $l_{mf}=v\tau$
and $\tau$ is the relevant relaxation time. Thus if $l_{\mathrm{geo}}\gg l_{mf}$,
we can approximate $\nu\left(\mathbf{q}_{\mathrm{geo}},\omega\right)\approx\nu\left(\mathbf{q}=0,\omega\right)$.
We then have 
\begin{equation}
\nu\left(\mathbf{r}-\mathbf{r}',\omega\right)\approx\nu_{0}\left(\mathbf{q}=0,\omega\right)\delta\left(\mathbf{r}-\mathbf{r}'\right)
\end{equation}
and the constitutive relation (\ref{eq:non-local-constitutive}) reduces
to its local form $\mathcal{J}\left(\mathbf{r},\omega\right)=\nu_{0}\left(\omega\right)\mathcal{F}\left(\mathbf{r},\omega\right)$.
The non-locality of Eq. (\ref{eq:non-local-constitutive}) matters
if $l_{\mathrm{geo}}\lesssim l_{mf}$. On scales comparable to the
mean free path, transport is intrinsically non-local, because particles
loose their memory of previous events through collisions with other
particles or impurities - a mechanism that ceases to be efficient.
A good example is the Poiseuille flow through narrow channels described
in Sec. \ref{sec:Poiseuille-profiles}. We proceed with the calculation
of the non-local, i.e. wavenumber dependent electric conductivity,
thermal conductivity and viscosity using the kinetic equation (\ref{eq:Boltzmann})
and the collinear zero mode expansion summarized in Sec. \ref{subsec:Collinear-zero-modes}.

\subsubsection{Electric conductivity\label{subsec:Electrical-conductivity}}

As mentioned in Sec. \ref{subsec:Scattering-times}, only the first
collinear mode $s=1$ is involved in the calculation of the electric
conductivity. Inserting the expansion of the distribution function
in terms of collinear zero modes (\ref{eq:Distr_dunction_expanded})
into the kinetic equation (\ref{eq:Kin_Eq}) using its matrix representation
of Eqs. (\ref{eq:Liouville_projected}), (\ref{eq:Linear_approximation}),
(\ref{eq:electric_force_projected})-(\ref{eq:viscous_projected})
and (\ref{eq:Vlasov_projected}), the left hand side of (\ref{eq:Kin_Eq})
can be transformed into a recurrence relation for the coefficients
$a_{1,m}$, where, for the rest of this section, the $s=1$ index
is dropped. A similar analysis for electrons in a random magnetic
field was performed in Ref.\cite{Mirlin1997}. For $m>2$, Eq. (\ref{eq:linear_scattering_rates_charge})
can be used, and the recurrence relation reads 
\begin{equation}
a_{m+1}=\frac{2ie^{-i\vartheta_{\mathbf{q}}}}{vq}\left(i\omega-\tau_{c,m}^{-1}\right)a_{m}-e^{-2i\vartheta_{\mathbf{q}}}a_{m-1}.\label{eq:eq:Recurrence_charge}
\end{equation}
This recurrence relation has the form 
\begin{equation}
a_{m+1}=\left(\alpha'm+\beta'\right)a_{m}-e^{i\delta}a_{m-1}\label{eq:Recurrence_math}
\end{equation}
with $\alpha'=-\frac{2ie^{-i\vartheta_{\mathbf{q}}}}{vq}\frac{k_{B}T}{\hbar}\frac{\pi}{\log2}\gamma_{c}$,
$\beta'=\frac{2ie^{-i\vartheta_{\mathbf{q}}}}{vq}\left(i\omega-\eta_{c}\frac{k_{B}T}{\hbar}\frac{\pi}{\log2}\right)$
and $\delta=-2\vartheta_{\mathbf{q}}$. It has two solutions that
can be given in terms of modified Bessel functions. The physically
interesting solution is 
\begin{equation}
a_{m}=c\cdot e^{i\frac{\delta}{2}\left(m+\frac{\beta'}{\alpha'}\right)}\mathrm{I}_{m+\frac{\beta'}{\alpha'}}\left(-\frac{2e^{i\delta/2}}{\alpha'}\right),\label{eq:Recurrence_math_sol}
\end{equation}
where $\mathrm{I}_{\nu}\left(z\right)$ is the modified Bessel function
of the first kind. Another solution that diverges for $m\rightarrow\infty$
is given by 
\[
c_{m}=c\cdot e^{i\frac{\delta}{2}\left(m+\frac{\beta'}{\alpha'}\right)}\mathrm{K}_{m+\frac{\beta'}{\alpha'}}\left(\frac{2e^{i\delta/2}}{\alpha'}\right).
\]
$\mathrm{K}_{\nu}$ is the modified Bessel function of the second
kind. Making use of the coefficients $a_{m}$ for $m>2$ as given
by Eq. (\ref{eq:Recurrence_math_sol}), the kinetic equation can be
reduced to a $5\times5$ component matrix equation:\begin{widetext}
\begin{equation}
\left[\begin{array}{ccccc}
-i\omega+M_{c}\left(q,\omega\right) & \frac{1}{2}ivqe^{i\vartheta_{\mathbf{q}}} & 0 & 0 & 0\\
\frac{1}{2}ivqe^{-i\vartheta_{\mathbf{q}}} & -i\omega+\tau_{c,1}^{-1} & \frac{1}{2}ivqe^{i\vartheta_{\mathbf{q}}} & 0 & 0\\
0 & \frac{1}{2}ivqe^{-i\vartheta_{\mathbf{q}}} & -i\omega & \frac{1}{2}ivqe^{i\vartheta_{\mathbf{q}}} & 0\\
0 & 0 & \frac{1}{2}ivqe^{-i\vartheta_{\mathbf{q}}} & -i\omega+\tau_{c,1}^{-1} & \frac{1}{2}ivqe^{i\vartheta_{\mathbf{q}}}\\
0 & 0 & 0 & \frac{1}{2}ivqe^{-i\vartheta_{\mathbf{q}}} & -i\omega+M_{c}\left(q,\omega\right)
\end{array}\right]\left[\begin{array}{c}
a_{-2}\\
a_{-1}\\
a_{0}\\
a_{1}\\
a_{2}
\end{array}\right]=\left[\begin{array}{c}
0\\
\frac{eE_{0}\beta v}{2}\\
0\\
\frac{eE_{0}\beta v}{2}\\
0
\end{array}\right],\label{eq:5-Comp-Conductivity}
\end{equation}
\end{widetext}where $M_{c}\left(q,\omega\right)=\tau_{c,2}^{-1}+a_{3}\left(q,\omega\right)/a_{2}\left(q,\omega\right)$
is a memory function containing information on scattering channels
with higher angular momentum numbers. Using the Eqs. (\ref{eq:Recurrence_math})
and (\ref{eq:Recurrence_math_sol}), the memory function is written
\begin{equation}
M_{c}\left(q,\omega\right)=\tau_{c,2}^{-1}+\frac{1}{2}vq\frac{\textrm{I}_{3+\frac{\eta_{c}}{\gamma_{c}}-i\omega\tau_{c}}\left(\tau_{c}vq\right)}{\textrm{I}_{2+\frac{\eta_{c}}{\gamma_{c}}-i\omega\tau_{c}}\left(\tau_{c}vq\right)},\label{eq:Memory_charge}
\end{equation}
with the abbreviation $\tau_{c}=\frac{\hbar}{k_{B}T}\frac{\log2}{\pi}\gamma_{c}^{-1}$.
It is now straightforward to calculate the electric conductivity from
the relation 
\begin{equation}
j_{c,x}\left(\mathbf{q},\omega\right)=\sigma_{xx}\left(\mathbf{q},\omega\right)E_{x}\left(\mathbf{q},\omega\right).
\end{equation}
The non-local conductivity can be decomposed into a longitudinal part
$\sigma_{\parallel}\left(\omega,q\right)$ and a transverse part $\sigma_{\bot}\left(\omega,q\right)$,
both depending on the modulus of $\mathbf{q}$. The longitudinal and
transverse parts describe currents that flow in the direction of $\mathbf{q}$,
or orthogonal to $\mathbf{q}$, respectively: 
\begin{equation}
\sigma_{\alpha\beta}=\frac{q_{\alpha}q_{\beta}}{q^{2}}\sigma_{\parallel}\left(q,\omega\right)+\left(\delta_{\alpha\beta}-\frac{q_{\alpha}q_{\beta}}{q^{2}}\right)\sigma_{\bot}\left(q,\omega\right).\label{eq:Decomposition_electr_explicit}
\end{equation}
We assumed that the electric field is parallel to the $x$-axis. According
to Eq. (\ref{eq:Decomposition_electr_explicit}), $\sigma_{\parallel}\left(q,\omega\right)$
can be read off from the $x$-component of the current density $j_{c,x}$
by letting $\mathbf{q}$ be parallel to $\mathbf{e}_{x}$, and $\sigma_{\bot}\left(q,\omega\right)$
by considering the case $\mathbf{q}\parallel\mathbf{e}_{y}$. The
conductivities are then given by 
\begin{eqnarray}
\sigma_{\parallel} & = & \frac{\sigma_{0}}{1-i\tau_{1,c}\omega+\frac{1}{4}v^{2}\tau_{c,1}q^{2}\left(\frac{2i}{\omega}+\frac{1}{M_{c}\left(q,\omega\right)-i\omega}\right)},\nonumber \\
\sigma_{\bot} & = & \frac{\sigma_{0}}{1-i\tau_{c,1}\omega+\frac{\frac{1}{4}v^{2}\tau_{c,1}q^{2}}{M_{c}\left(q,\omega\right)-i\omega}},\label{eq:charge_conductivity_result}
\end{eqnarray}
where $\sigma_{0}=N\frac{e^{2}\log\left(2\right)\tau_{c,1}}{2\pi\beta\hbar^{2}}$
is the quantum critical conductivity calculated in Ref. \cite{Fritz2008}.
Note that $\sigma_{\parallel}\left(q\neq0,\omega=0\right)=0$ holds,
which also follows from formula (\ref{eq:Density_Density_Conductivity}).
If this was not the case, static currents with a finite wave-vector
$\mathbf{q}$ would lead to an infinite accumulation of charge at
certain points, which is forbidden by the conservation of charge.
In Fig. \ref{fig:Fig_charge_conductivities} the charge conductivities
are plotted as functions of $\omega$ for different values of $q$.\onecolumngrid 

\begin{center}
\begin{figure}
\begin{centering}
\includegraphics[scale=0.35]{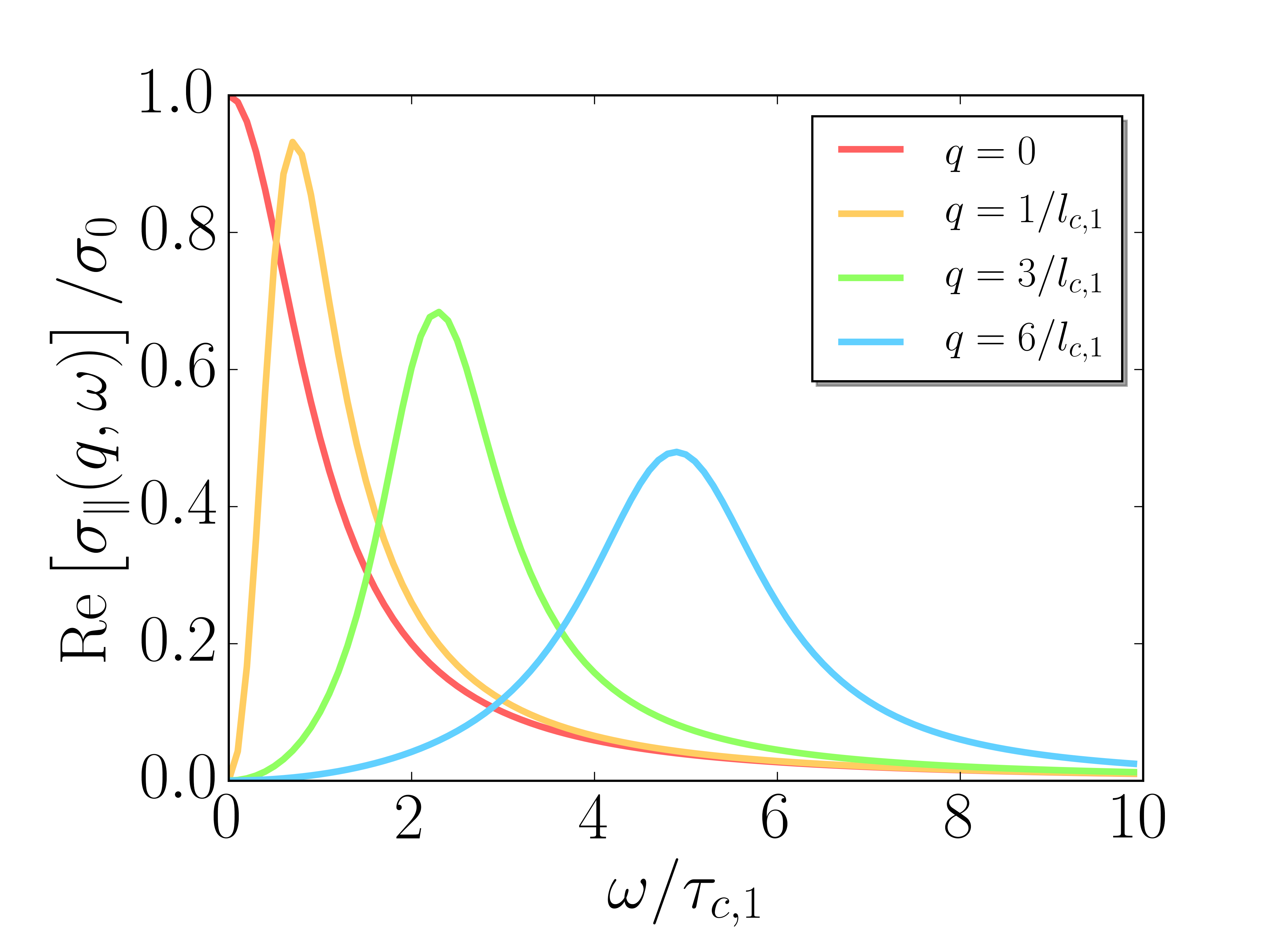}\includegraphics[scale=0.35]{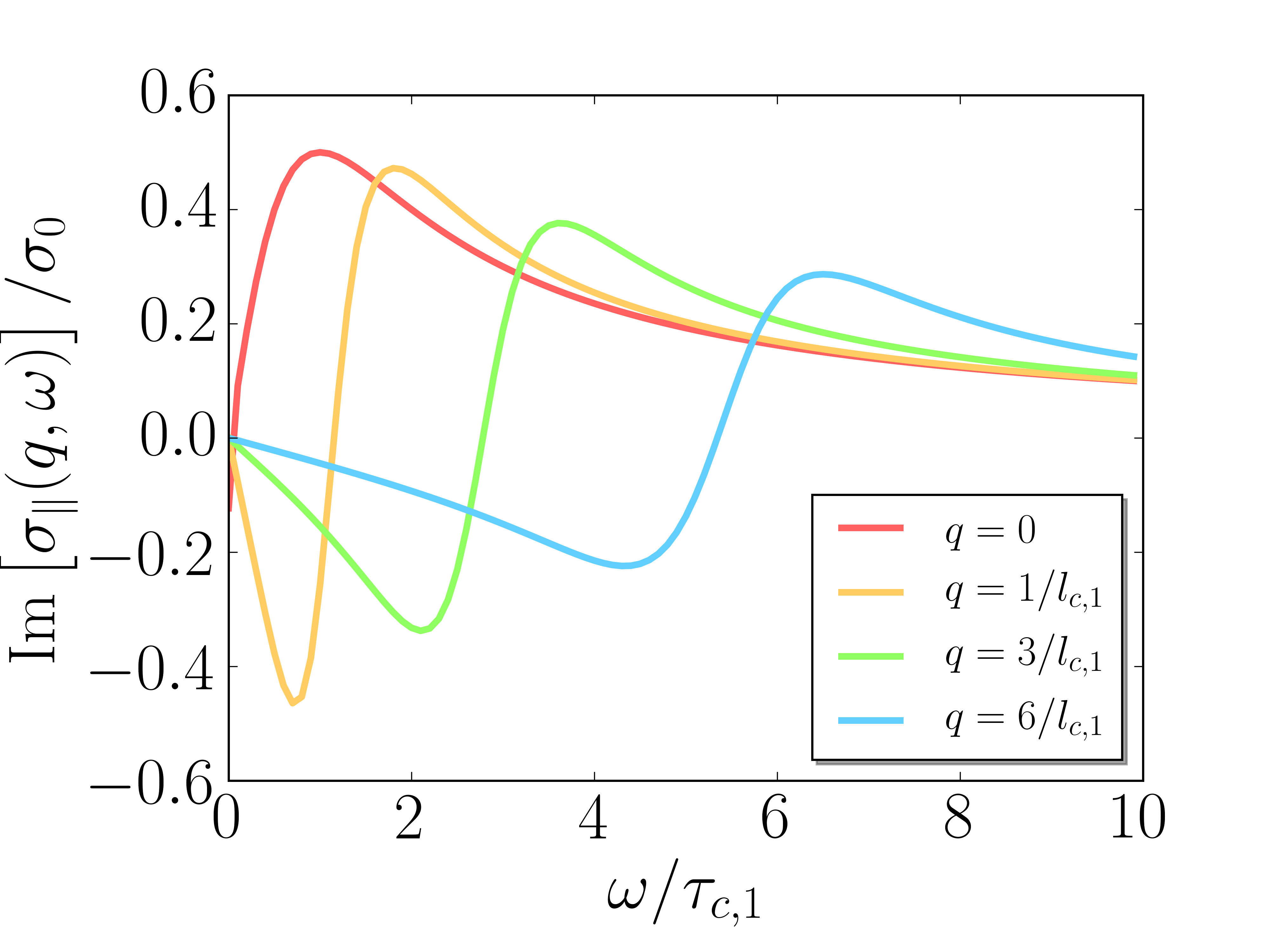} 
\par\end{centering}

\begin{centering}
\includegraphics[scale=0.35]{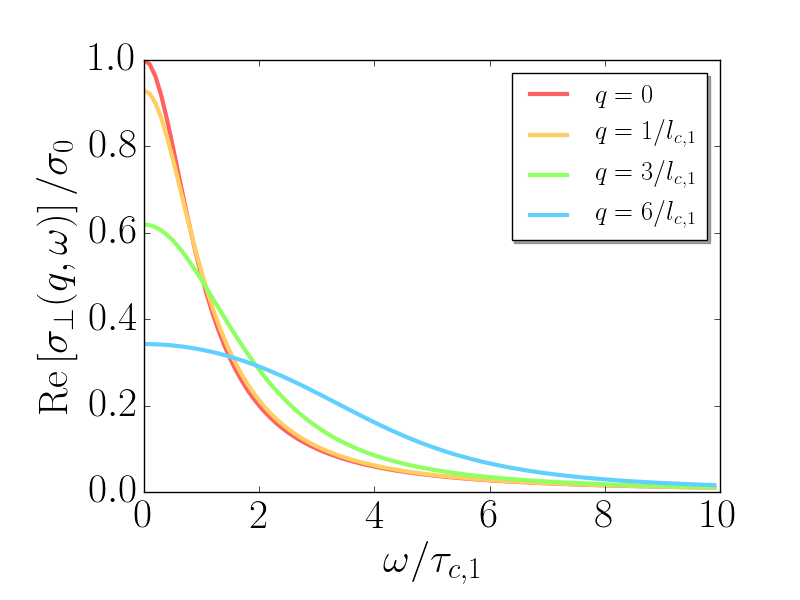}\includegraphics[scale=0.35]{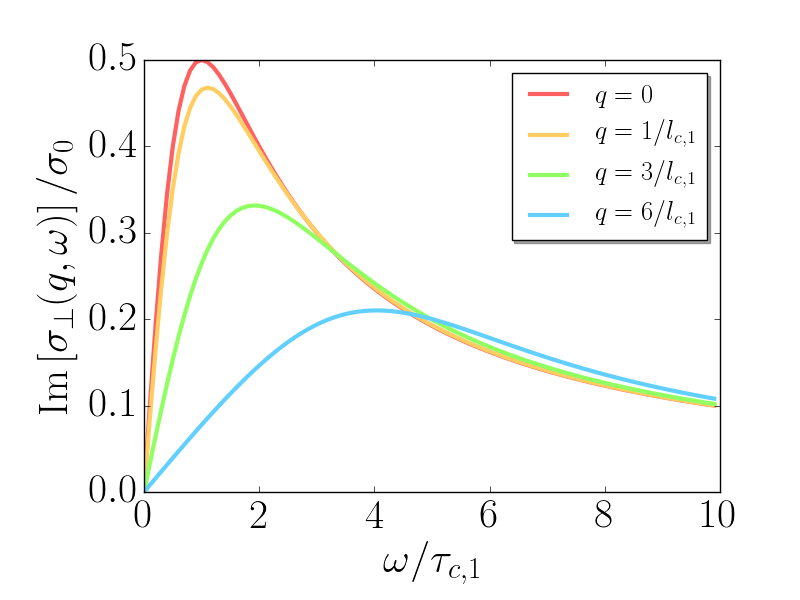} 
\par\end{centering}

\centering{}\caption{Longitudinal (upper row) and transverse (lower row) electric conductivities
of charge neutral graphene as functions of the electric field frequency
$\omega$ as given by Eqs (\ref{eq:charge_conductivity_result}).
Different colors indicate different values of the wavenumber $q$.
Frequencies and wave-numbers are normalized to the characteristic
scattering times and lengths $\tau_{c,1}$, $l_{c,1}=v\tau_{c,1}$.
$\sigma_{0}$ is the interaction induced conductivity at the neutrality
point \cite{Fritz2008,Kashuba2008}. The graphs show distinct resonant
features at frequencies $\omega\sim q/v$, where $v$ is the electron
group velocity. Whereas the real part of the longitudinal and the
imaginary part of the transverse conductivities are peaked around
$\omega\sim q/v$, the imaginary part of the longitudinal conductivity
exhibits a sign change indicating an abrubt phase change of the current
response. The real parts approach $\sigma_{0}$ for $q\rightarrow0$,
$\omega\rightarrow0$. For $q\protect\neq0$, $\omega=0$ the longitudinal
conductivity vanishes. This general property of the charge conductivity
follows from the conservation of charge (see Eq. (\ref{eq:Density_Density_Conductivity})).
\label{fig:Fig_charge_conductivities}}
\end{figure}

\par\end{center}

\twocolumngrid

The electric conductivity tensor $\sigma_{\alpha\beta}\left(\mathbf{q},\omega\right)$
of Eq. (\ref{eq:charge_conductivity_result}) gives access to different
electric response functions. The current-current correlation function
is given by 
\begin{equation}
\chi_{J_{\alpha}J_{\beta}}\left(\mathbf{q},\omega\right)=-i\omega\sigma_{\alpha\beta}\left(\mathbf{q},\omega\right)\,,\label{eq:current-current-correlation}
\end{equation}
where $\alpha$, $\beta$ denote the components of the current vector
(see e.g. Ref. \cite{Link2016}). With the help of the continuity
equation, the charge density-density correlation function is obtained
from Eq. (\ref{eq:current-current-correlation}): 
\begin{align}
\chi_{\rho\rho}\left(q,\omega\right) & =\frac{q_{\alpha}q_{\beta}}{\omega^{2}}\chi_{J_{\alpha}J_{\beta}}\left(\mathbf{q},\omega\right).\nonumber \\
 & =\frac{q^{2}}{i\omega}\sigma_{\parallel}\left(q,\omega\right).\label{eq:Density_Density_Conductivity}
\end{align}
The non-local conductivity is related to the dielectric constant $\varepsilon\left(\mathbf{q},\omega\right)$
which is defined as (see Eq. (\ref{eq:potential_tot_ind_ext}))
\begin{equation}
\varepsilon=\frac{\varphi_{\mathrm{ext}}}{\varphi_{\mathrm{tot}}}.
\end{equation}
Observing that $\varphi_{\mathrm{ind}}\left(\mathbf{q},\omega\right)=V\left(q\right)\delta\rho\left(\mathbf{q},\omega\right)$,
where $\delta\rho$ is the induced charge density, we find 
\begin{equation}
\varepsilon=1-V\left(q\right)\frac{\delta\rho}{\varphi_{\mathrm{tot}}}.\label{eq:dielectric_density}
\end{equation}
In linear response it is $\delta\rho=\chi_{\rho\rho}\left(q,\omega\right)\varphi_{\mathrm{tot}}$,
so that we can write
\begin{equation}
\varepsilon=1-V\left(q\right)\chi_{\rho\rho}.
\end{equation}
Taking the divergence of Ohm's law $j_{\alpha}\left(\mathbf{q},\omega\right)=\sigma_{\alpha\beta}\left(\mathbf{q},\omega\right)E_{\beta}\left(\mathbf{q},\omega\right)$,
and using the continuity equation $i\omega\delta\rho=iq_{\alpha}j_{\alpha}$
to express the electric current in terms of the induced charge density,
we obtain 
\[
\varphi_{\mathrm{tot}}=\frac{i\omega\delta\rho}{q^{2}\sigma_{\parallel}}.
\]
Inserting in Eq (\ref{eq:dielectric_density}) we have
\[
\varepsilon\left(\mathbf{q},\omega\right)=1-V\left(q\right)\frac{iq^{2}}{i\omega}\sigma_{\parallel}\left(\mathbf{q},\omega\right),
\]
which is in accordance with Eq. (\ref{eq:Density_Density_Conductivity}).
Notice, that both the longitudinal conductivity $\sigma_{\parallel}$
and the charge susceptibility $\chi_{\rho\rho}$ describe the response
to the total potential $\varphi_{\mathrm{tot}}$. Hence the Vlasov
term does not enter these quantities explicitely (for an in-depth
discussion see Ref. \cite{PinesNozieres1}, Chapter 3, in particular
Eq. (3.56)) Finally, the charge compressibility $K=\partial\rho/\partial\mu$
is given by 
\begin{equation}
K\left(q\right)=\chi_{\rho\rho}\left(\omega=0\right).
\end{equation}
The role of interaction effects for the compressibility were discussed
in Ref.\cite{Sheehy2007}.

\subsubsection{Thermal conductivity}

Next we presemt our analysis for the non-local thermal conductivity.
Since momentum conservation implies for a Dirac fluid the conservation
of the heat current, thermal transport is expected to display classical
hydrodynamic behavior, i.e. one expects non-local effects to be even
more important than for charge transport.\cite{Foster2009,Link2016}.

As pointed out in Sec. \ref{subsec:Scattering-times}, the $s=3$
energy mode must be kept in the calculation of the thermal conductivity,
whereas the $s=2$ imbalance mode can be neglected, contributing only
a small correction to the overall result. With only a single mode
involved, the calculation is formally analogous to the calculation
of the electrical conductivity in Sec. \ref{subsec:Electrical-conductivity},
even though there are crucial differences in the actual result, given
the distinct role of momentum conservation. The relaxation time $\tau_{c,m}$
must be replaced by $\tau_{\varepsilon,m}$ as given by Eq. (\ref{eq:linear_scattering_rates_energy}).
The conservation of momentum is incorporated via $\tau_{\varepsilon,1}\rightarrow\infty$,
whivch follows from the Boltzmann approach. The resulting longitudinal
and transverse thermal conductivities read 
\begin{eqnarray}
\kappa_{\parallel}\left(q,\omega\right) & = & \frac{\kappa_{0}}{i\omega\tau_{\varepsilon,2}-\frac{1}{4}v^{2}q^{2}\tau_{\varepsilon,2}\left(\frac{2i}{\omega}-\frac{1}{M_{\varepsilon}\left(q,\omega\right)+i\omega}\right)}\nonumber \\
\kappa_{\bot}\left(q,\omega\right) & = & \frac{\kappa_{0}}{i\omega\tau_{\varepsilon,2}+\frac{\frac{1}{4}v^{2}q^{2}\tau_{\varepsilon,2}}{M_{\varepsilon,2}\left(q,\omega\right)+i\omega}},\label{eq:thermal_conductivity_result}
\end{eqnarray}
with the memory function 
\[
M_{\varepsilon}\left(q,\omega\right)=\tau_{\varepsilon,2}^{-1}+\frac{1}{2}vq\frac{\textrm{I}_{3+\frac{\eta_{\varepsilon}}{\gamma_{\varepsilon}}+i\omega\tau_{\varepsilon}}\left(\tau_{\varepsilon}vq\right)}{\textrm{I}_{2+\frac{\eta_{\varepsilon}}{\gamma_{\varepsilon}}+i\omega\tau_{\varepsilon}}\left(\tau_{\varepsilon}vq\right)}.
\]
The abbreviation $\tau_{\varepsilon,2}=\frac{1}{\alpha^{2}}\frac{\hbar}{k_{B}T}\frac{9\zeta(3)}{3.341\cdot2\pi}$
is used. For convenience $\kappa_{\parallel/\bot}$ is given in units
of a thermal conductivity $\kappa_{0}=9N\pi^{3}k_{B}\zeta(3)\tau_{\varepsilon,2}/2\beta^{2}\hbar^{2}$,
however, $\tau_{\varepsilon,2}$ is the relaxation time in the $\left|m\right|=2$
channel, and should not be confused with an alleged relaxation time
of the energy current, which is infinite due to the conservation of
momentum.

In Fig. \ref{fig:Fig_charge_conductivities} the thermal conductivities
are plotted as functions of $\omega$ for different values of $q$.
The fact that thermal currents are protected by momentum conservation
leads to a divergence of the thermal conductivity at small frequencies:
for $q=0$, $\kappa$ is purely imaginary and shows the characteristic
$1/\omega$ Drude behavior.\onecolumngrid 

\begin{center}
\begin{figure}
\begin{centering}
\includegraphics[scale=0.35]{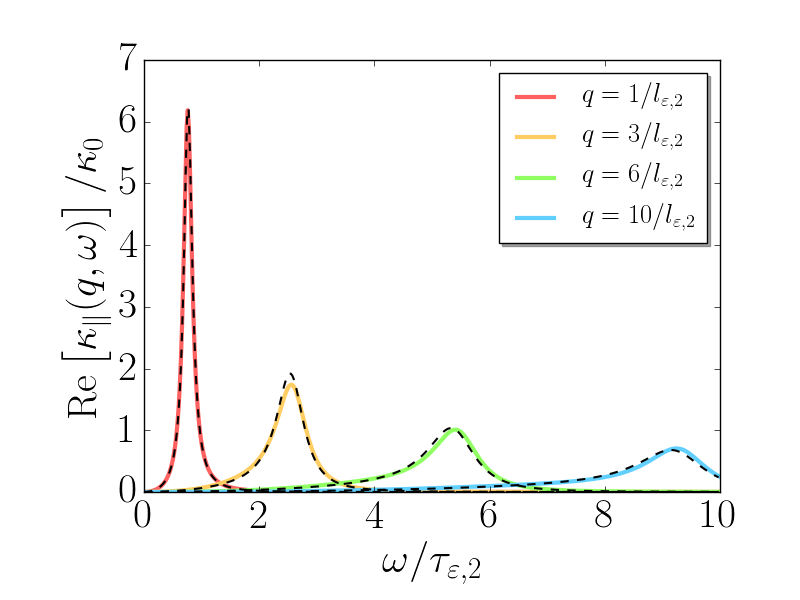}\includegraphics[scale=0.35]{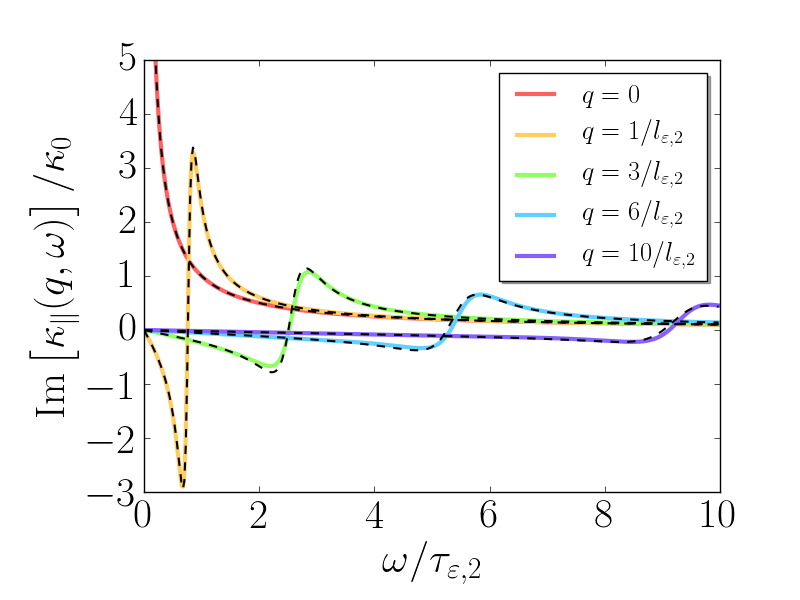} 
\par\end{centering}

\begin{centering}
\includegraphics[scale=0.35]{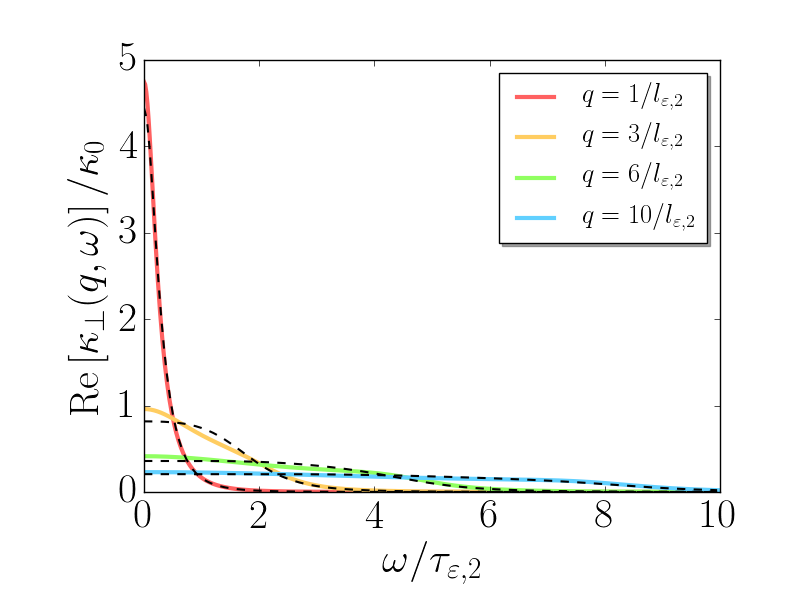}\includegraphics[scale=0.35]{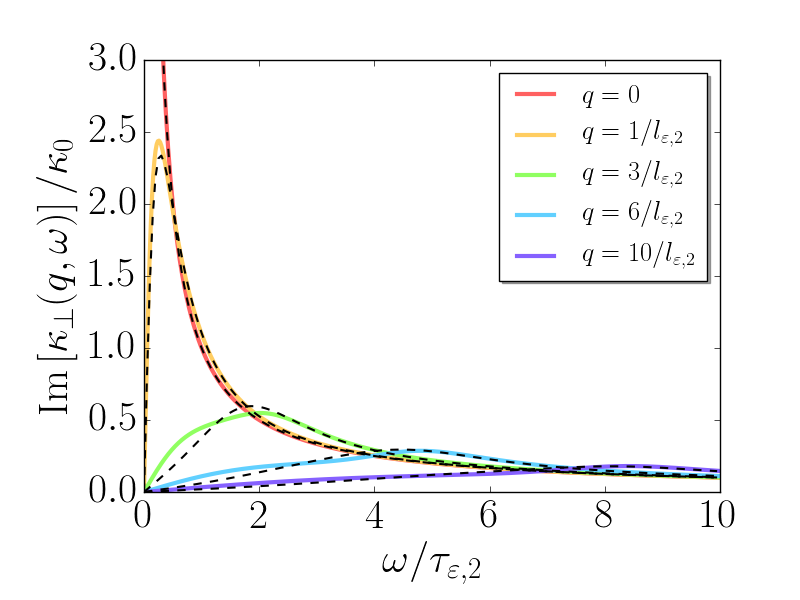} 
\par\end{centering}

\centering{}\caption{The figure shows the longitudinal (upper row) and transverse (lower
row) thermal conductivities (\ref{eq:charge_conductivity_result})
as functions of the electric field frequency $\omega$. Different
colors indicate different values of the wavenumber $q$. The conductivities
are normalized to $\kappa_{0}=9N\pi^{3}k_{B}\zeta(3)\tau_{\varepsilon,2}/2\beta^{2}\hbar^{2}$.
For small $\omega$ and vanishing $q$, the imaginary part of $\kappa_{\parallel/\perp}$
diverges as $1/\omega$, whereas the real part vanishes - a behavior
indicating that thermal transport in the system is ballistic. The
solid lines show the analytical result of Eq. (\ref{eq:thermal_conductivity_result}),
the dashed lines show the full numerical result including all modes
and the exact scattering times. \label{fig:Fig_thermal_conductivities}}
\end{figure}

\par\end{center}

\twocolumngrid

\subsubsection{Non-local shear viscosity\label{subsec:Non-local-shear-viscosity}}

The non-local viscosity is defined through a constitutive relation
of the form of Eq. (\ref{constitutive_fourier}), linking the shear
force $X_{0,\alpha\beta}\left(\mathbf{r}'\right)$ to the momentum-current
tensor $\tau_{\alpha\beta}$:

\begin{equation}
\tau_{\alpha\beta}\left(\mathbf{r},t\right)=\int d^{2}r'\int dt'\,\eta_{\alpha\beta\gamma\delta}\left(\mathbf{r}-\mathbf{r}',t-t'\right)X_{0,\gamma\delta}\left(\mathbf{r}',t'\right).\label{eq:constitutive_viscosity}
\end{equation}
Since the system is isotropic, the shear force can be chosen such
that the flow velocity is aligned with the $x$-axis, and its gradient
shows in the $y$ direction. It is assumed that the shear force is
wavelike: $X_{0,xy}\left(\mathbf{r}\right)=X_{0,xy}e^{i\mathbf{q}\cdot\mathbf{r}-i\omega t}$.
The wave-vector $\mathbf{q}$ can have an arbitrary direction in the
$xy$-plane, introducing a preference direction to the system's response.
In addition to $\tau_{xy}$, this gives rise to nonzero components
$\tau_{xx}$, $\tau_{yy}$, if $\mathbf{q}$ does not align with the
$x$ or the $y$-axes. The viscosity tensor $\eta_{\alpha\beta xy}$
can be decomposed into transverse and longitudinal parts (see Eq.
(\ref{eq:Decomposition_electr_explicit})) analogously to the electric
and charge conductivities. Because $\eta_{\alpha\beta xy}$ is a fourth
rank tensor the decomposition is slightly more involved and the reader
is referred to Appendix \ref{sec:Decomposition_visc} for details.
The general $\mathbf{q}$-dependent viscosity tensor can be constructed
with the help of three rank two tensors: 
\begin{eqnarray}
e_{\alpha\beta}^{\left(1\right)} & = & \frac{q_{\alpha}q_{\beta}}{q^{2}}\nonumber \\
e_{\alpha\beta}^{\left(2\right)} & = & \delta_{\alpha\beta}-\frac{q_{\alpha}q_{\beta}}{q^{2}}\nonumber \\
e_{\alpha\beta}^{\left(3\right)} & = & \frac{1}{\sqrt{2}}\left(q_{\alpha}p_{\beta}+p_{\alpha}q_{\beta}\right)/\left(pq\right),\label{eq:Tensor_basis}
\end{eqnarray}
where 
\begin{equation}
p_{\alpha}=q_{\gamma}\varepsilon_{\gamma\alpha}.
\end{equation}
The viscosity tensor is parameterized by two frequency and momentum
dependent functions, $\eta_{\parallel}\left(\mathbf{q},\omega\right)$
and $\eta_{\bot}\left(\mathbf{q},\omega\right)$, which we will call
longitudinal and transverse viscosities: 
\begin{align*}
\eta_{\alpha\beta\gamma\delta}\left(\mathbf{q},\omega\right) & =\eta_{1}\left(\mathbf{q},\omega\right)\left(e_{\alpha\beta}^{\left(1\right)}e_{\gamma\delta}^{\left(1\right)}+e_{\alpha\beta}^{\left(2\right)}e_{\gamma\delta}^{\left(2\right)}\right)\\
 & \qquad+\eta_{2}\left(\mathbf{q},\omega\right)e_{\alpha\beta}^{\left(3\right)}e_{\gamma\delta}^{\left(3\right)}.
\end{align*}
Let the flow be in x-direction: $\mathbf{u}\left(y\right)=u\left(y\right)\mathbf{\hat{\mathbf{e}}}_{x}$,
and let the wave-vector be parameterized by $\mathbf{q}=q\left(\cos\left(\vartheta_{\mathbf{q}}\right),\sin\left(\vartheta_{\mathbf{q}}\right)\right)^{T}$,
where $\theta$ is measured with respect to the $x$-axis. For $\vartheta_{\mathbf{q}}=0$
or $\vartheta_{\mathbf{q}}=\pi/2$ follows $e_{\alpha\beta}^{\left(1,2\right)}=0$,
$\eta_{xxxy}=\eta_{xxyx}=0$ and $\eta_{xyxy}=\eta_{2}/2$. This corresponds
to the familiar shear flow in e.g. a Poiseuille geometry where $\tau_{xx}=\tau_{yy}=0$.
The momentum current flows orthogonal to the direction of the momentum
density. For $\vartheta_{\mathbf{q}}=\pi/4$, the viscosity is determined
by $\eta_{1}$: $\eta_{xyxy}=\eta_{1}/2$.

As in the case of thermal conductivity, dropping the $s=2$ imbalance
mode produces only a small numerical correction in the final result
for the viscosity. With an external shear force of the form of Eqs.
(\ref{eq:Shear_force}), (\ref{eq:viscous_projected}) applied to
the system, the kinetic equation can be written as $5\times5$ component
matrix equation, similar to the case of an applied electric field
(see Eq. (\ref{eq:5-Comp-Conductivity})). The force acts in the $\left|m\right|=2$
channels, and the equation reads\begin{widetext} 
\begin{equation}
\left[\begin{array}{ccccc}
-i\omega+M_{\varepsilon}\left(q,\omega\right) & \frac{1}{2}ivqe^{i\theta} & 0 & 0 & 0\\
\frac{1}{2}ivqe^{-i\theta} & -i\omega & \frac{1}{2}ivqe^{i\theta} & 0 & 0\\
0 & \frac{1}{2}ivqe^{-i\theta} & -i\omega & \frac{1}{2}ivqe^{i\theta} & 0\\
0 & 0 & \frac{1}{2}ivqe^{-i\theta} & -i\omega & \frac{1}{2}ivqe^{i\theta}\\
0 & 0 & 0 & \frac{1}{2}ivqe^{-i\theta} & -i\omega+M_{\varepsilon}\left(q,\omega\right)
\end{array}\right]\left[\begin{array}{c}
a_{-2}\\
a_{-1}\\
a_{0}\\
a_{1}\\
a_{2}
\end{array}\right]=\left[\begin{array}{c}
-\frac{iX_{0}}{4}\\
0\\
0\\
0\\
\frac{iX_{0}}{4}
\end{array}\right].\label{eq:5-Comp_Visc}
\end{equation}
\end{widetext}Solving the matrix equation (\ref{eq:5-Comp_Visc})
for $a_{\pm2}$, the viscosity is calculated with the help of Eq.
(\ref{eq:constitutive_viscosity}) which takes the form $\tau_{xy}=N\sum_{\lambda}\int_{k}v_{x}k_{y}f_{\mathbf{k},\lambda}=\eta_{xyxy}X_{0,xy}$.
As explained above, the viscosity components $\eta_{1}$ and $\eta_{2}$
can be read off from the general result $\eta_{xyxy}\left(\mathbf{q}=q\left(\cos\left(\vartheta_{\mathbf{q}}\right),\sin\left(\vartheta_{\mathbf{q}}\right)\right)^{T},\omega\right)$
by setting $\vartheta_{\mathbf{q}}=0$ and $\vartheta_{\mathbf{q}}=\pi/2$:
\begin{eqnarray}
\eta_{1}\left(q,\omega\right) & = & \frac{2\eta_{0}}{-i\tau_{\varepsilon,2}\omega-q^{2}v^{2}\frac{i\tau_{\varepsilon,2}\omega}{2q^{2}v^{2}-4\omega^{2}}+\tau_{\varepsilon,2}M_{\varepsilon}\left(q,\omega\right)},\nonumber \\
\eta_{2}\left(q,\omega\right) & = & \frac{2\eta_{0}}{-i\tau_{\varepsilon,2}\omega-\frac{q^{2}v^{2}\tau_{\varepsilon,2}}{4i\omega}+\tau_{\varepsilon,2}M_{\varepsilon}\left(q,\omega\right)}.\nonumber \\
\label{eq:viscosity_result}
\end{eqnarray}
Here, $\eta_{0}$ is the viscosity at $q=0$, $\omega=0$, $\eta_{0}=N\left(k_{B}T\right)^{3}\tau_{\varepsilon,2}/\left(8\hbar^{2}v^{2}\right)$,
as it was first calculated in Ref. \cite{Mueller2009} including both
modes, $s=2$ and $s=3$.

\section{Collective modes\label{sec:Collective-modes}}

Collective modes are solutions to the homogeneous part of the kinetic
equation (\ref{eq:Boltzmann}), (\ref{eq:Kin_Eq}) (see e.g. \cite{Lucas2018}).
Consider Eq. (\ref{eq:Kin_Eq}). With the force terms set to zero
it holds 
\[
\left(\mathcal{L}+\mathcal{V}+\mathcal{C}\right)\psi=0.
\]
Here, $\mathcal{L}$ and $\mathcal{C}$ have are the matrix operators
of Eqs. (\ref{eq:Liouville_projected}) and (\ref{eq:Linear_approximation}).
Solutions to this equation exist only if 
\begin{equation}
\det\left(\mathcal{L}+\mathcal{V}+\mathcal{C}\right)=0\label{eq:Collective_mode_condition}
\end{equation}
holds. This is only the case for certain values of the variable pairs
$\omega$, $q$. Eq. (\ref{eq:Collective_mode_condition}) is an eigenvalue
problem where the eigenvalues $\omega\left(q\right)$ determine the
dispersion relations of the collective modes. On the other hand, collective
modes can be found from poles of response functions for an external
force $S$. The two methods are equivalent. Within the kinetic equation
formalism, response functions are calculated as averages over the
distribution function $\psi=\left(\mathcal{L}+\mathcal{V}+\mathcal{C}\right)^{-1}S$.
If the condition (\ref{eq:Collective_mode_condition}) is fulfilled,
the operator $\left(\mathcal{L}+\mathcal{V}+\mathcal{C}\right)^{-1}$
is singular and thus singularities in the response to $S$ appear.
We will use Eq. (\ref{eq:Collective_mode_condition}) to study the
collective modes of a Dirac fluid on an infinite domain.

As in the previous sections, the kinetic equation will be expanded
in terms of collinear zero modes (\ref{eq:Collinear_modes}): $\chi_{\mathbf{k},\lambda}^{\left(m,s\right)}=\lambda^{m}e^{im\theta}\left\{ 1,\lambda,\lambda\beta v\hbar k\right\} $.
For $m=0$ these modes correspond to excitations of the charge, imbalance
and energy densities; for $\left|m\right|=1$ they correspond to the
associated currents. At the end of this section it will be shown that
including non-collinear zero modes in the calculation does not change
the result as long as the fine structure constant $\alpha$ is kept
small.

To get a feeling for the structure of collective modes in the system,
it is useful to begin with the case $q=0$. In the subspace of collinear
zero modes, the kinetic equation reduces to Eq. (\ref{eq:Decay_eq})
and the condition (\ref{eq:Collective_mode_condition}) reads 
\begin{equation}
\det\left(-i\omega\delta_{s,s'}+\Gamma_{m}^{s,s'}\right)=0.\label{eq:q=00003D00003D0_modes_eiegnvalues}
\end{equation}
This is an eigenvalue equation for the frequencies of collective modes
that can be solved independently for any $m$. Since, as pointed out
in Sec. \ref{subsec:Charge-and-energy}, $\Gamma_{m}^{s,s'}$ is block-diagonal
in the subspaces of electric ($s=1$) and imbalance/energy ($s=2,3$)
excitations, the above equation, as well as its extension to $q\neq0$,
can be solved independently in these two sectors. For $s=1$, the
eigenfrequencies are $\omega_{m}\left(q=0\right)=-i/\tau_{c,m}$.
Since in this scenario the time evolution of the modes is given by
the factor $e^{-i\omega_{m}t}$, all but the $m=0$ mode, which is
protected by charge conservation, exponentially decay at a rate inversely
proportional to their scattering time. The $m=0$ zero mode corresponds
to the charge density, which is conserved, and therefore does not
decay. In the following two sections, the collective charge, as well
as energy and imbalance excitations will be described at finite $\mathbf{q}$.
Figs. \ref{fig:Fig_collective_charge_im}, \ref{fig:Fig_collective_charge_re},
\ref{fig:Fig_collective_energy_im_Small_q}, \ref{fig:Fig_collective_energy_im},
\ref{fig:Fig_collective_energy_re} show the dispersion relations
of these modes.

\subsection{Collective charge excitations}

\begin{figure}
\centering{}\includegraphics[scale=0.35]{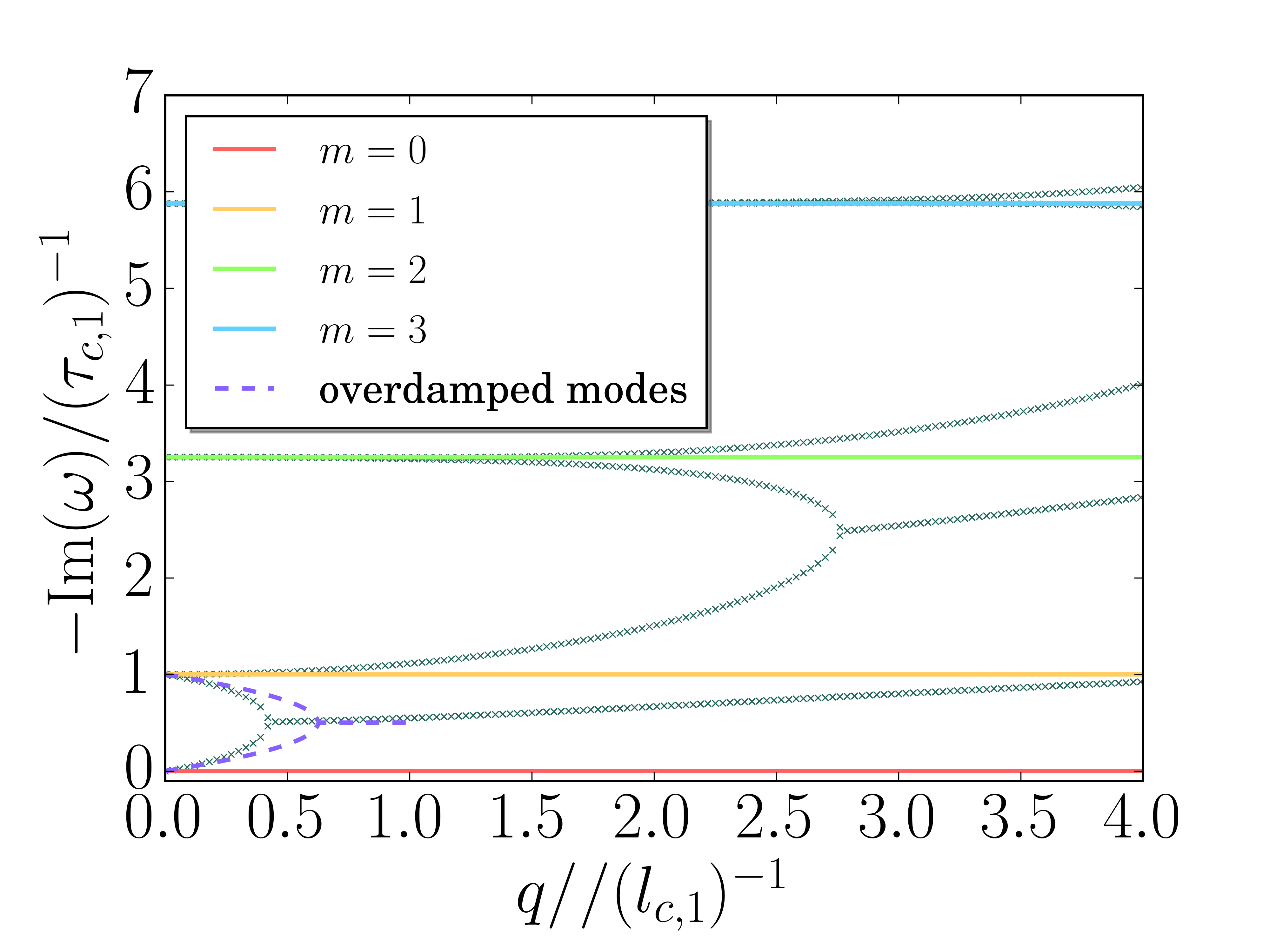}\caption{\label{fig:Fig_collective_charge_im} The imaginary parts of the dispersion
relations of collective charge excitations in different angular harmonic
channels $m$ are shown. The wave-vector $q$ is given in units of
the inverse scattering length $v\tau_{c,1}^{-1}$. The grey symbols
correspond to the numerical solution of Eq. (\ref{eq:Collective_mode_condition}).
The purely imaginary $m=0$ diffusive mode is the only mode approaching
zero for small $q$ - a behavior necessitated by charge conservation.
Modes with a higher $m$ are damped and approach the values $-i/\tau_{c,m}$
for $q\rightarrow0$. The corresponding excitations decay even in
the absence of spatial inhomogeneities. At a value $q=q_{pl}^{*}$
(Eq. \ref{eq:crit_q}), the dispersions of the diffusive mode and
the $m=1$ excitation merge, giving rise to a plasmon mode, which
has a finite real part (see Fig \ref{fig:Fig_collective_charge_re}).
This value is slightly overestimated by the simplified expression
of Eq. (\ref{eq:crit_q}).}
\end{figure}

\begin{figure}
\centering{}\includegraphics[scale=0.35]{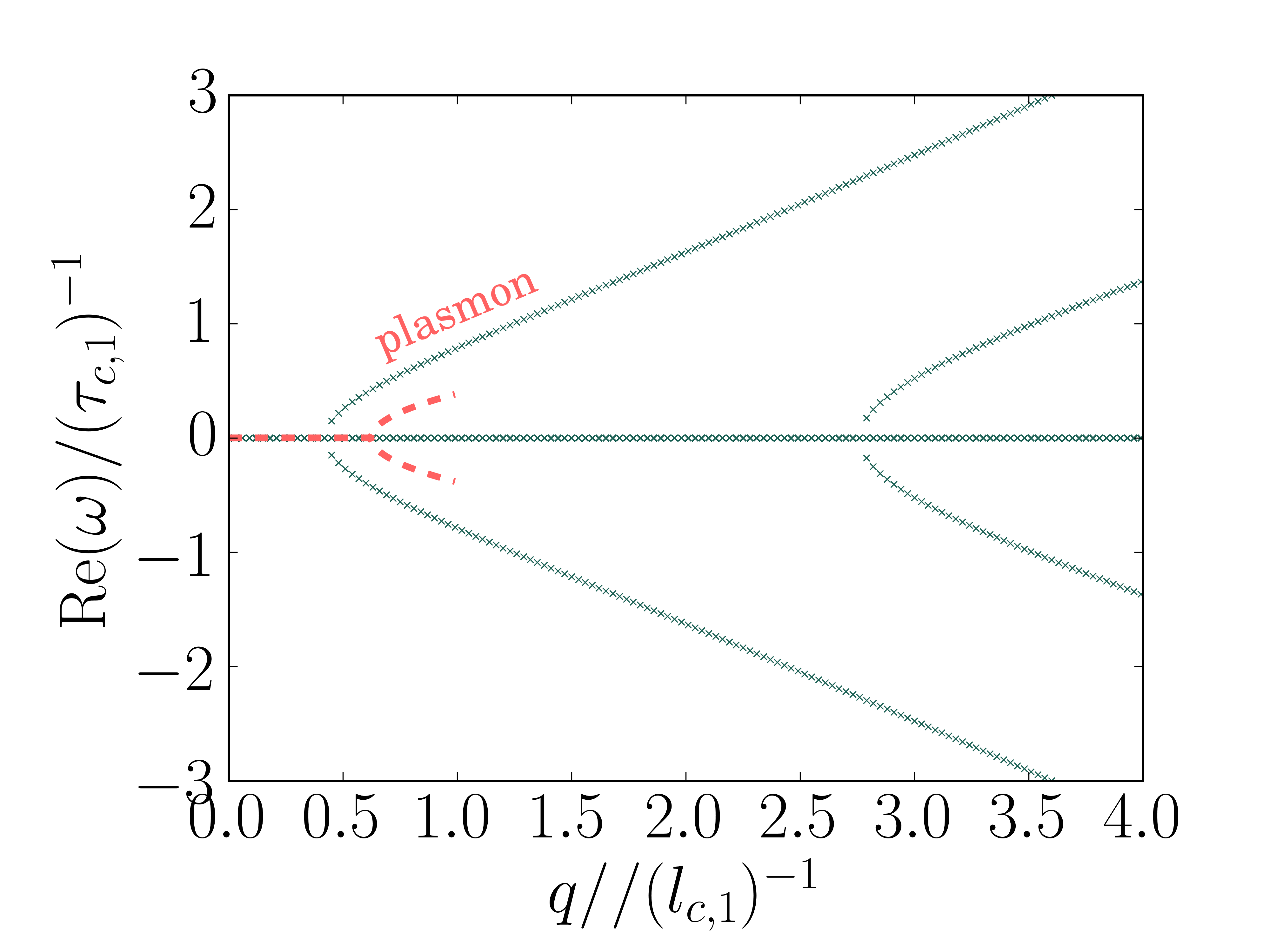}\caption{\label{fig:Fig_collective_charge_re}The figure shows the real parts
of the dispersion relations of collective charge excitations in different
angular harmonic channels $m$. The wave-vector $q$ is given in units
of the inverse scattering length $v\tau_{c,1}^{-1}$. The grey symbols
correspond to the numerical solution of Eq. (\ref{eq:Collective_mode_condition}).
The plasmon mode is gapped out by the interaction induced conductivity
and only obtains a finite real part around $q=q_{pl}^{*}$ (the simplified
value of $q_{pl}^{*}$ given in Eq. (\ref{eq:crit_q}) (red dashed
line) overestimates the branching point). At higher $q$, other, strongly
damped modes corresponding to higher angular harmonics appear. The
dampings of these modes are given by the $m>1$ modes of Fig. \ref{fig:Fig_collective_charge_im}.}
\end{figure}

In general, conserved modes do not decay at $q=0$, and therefore
their dispersion relations must vanish in a spatially homogeneous
system. The only conserved mode in the charge sector is the charge
density mode $\chi_{\mathbf{k},\lambda}^{\left(m=0,s=1\right)}=1$.
In the limit $q\ll v\tau_{c,1}$, the memory matrix (\ref{eq:Memory_charge})
reduces to $M_{c}\left(q,\omega\right)\approx\tau_{c,2}^{-1}$ and
Eq. (\ref{eq:Collective_mode_condition}) can be solved analytically.
The dispersions of the two lowest modes are 
\begin{equation}
\omega_{\mathrm{charge\,diff.}}\approx\omega_{\pm}=-\frac{i}{2\tau_{c,1}}\pm\sqrt{\frac{vq}{\tau_{V}}-\frac{1}{4\tau_{c,1}^{2}}}.\label{eq:Charge_lowest_modes_expansion}
\end{equation}
The conserved charge density mode is described by $\omega_{-}$. The
dispersion relations of Eq. (\ref{eq:Charge_lowest_modes_expansion})
have a non-vanishing real part for 
\begin{equation}
q>q_{pl}^{*}=\frac{\tau_{V}}{4v\tau_{c,1}^{2}}.\label{eq:crit_q}
\end{equation}
For wave-vectors below $q_{pl}^{*}$, the plasmon is over-damped (see
Fig. \ref{fig:Fig_collective_charge_im}). However, we have $vq_{pl}^{*}\sim\alpha^{3}k_{B}T/\hbar$
such that the plasmon mode becomes more and more pronounced at low
temperatures.

The plasmon mode is gapped out due to the intrinsic interaction induced
resistivity. At $q=0$ it has a vanishing real part and its decay
rate is given by the scattering rate in the $m=1$ channel: 
\begin{equation}
\omega_{pl}\left(q\rightarrow0\right)=-i/\tau_{c,1}
\end{equation}
(see also \cite{Briskot2015}). It is the most weakly damped of an
infinite set of modes corresponding to higher angular harmonics (see
Fig. \ref{fig:Fig_collective_charge_im}). It is clearly seen, that
the modes relate to different angular harmonic channels $m$. For
$q=0$ their dispersions approach $\omega_{m}\left(q=0\right)=-i/\tau_{c,m}$.
Such modes play a crucial role in the relaxation mechanism of focused
current beams in graphene \cite{Kiselev2019b}. Similar collective
modes have been argued to influence the relaxation behavior of unitary
fermi gases \cite{Brewer2015} and QCD plasmas \cite{Romatschke2016,Romatschke2018,Heller2018}.

\subsection{Collective energy and imbalance excitations}

In the energy sector spanned by the modes $s=2,3$, the Eqs. (\ref{eq:Collective_mode_condition})
and (\ref{eq:q=00003D00003D0_modes_eiegnvalues}) give rise three
zero eigenvalues. These correspond to the conserved energy ($\chi_{\mathbf{k},\lambda}^{\left(m=0,s=3\right)}=\lambda\beta v\hbar k$)
and quasiparticle (imbalance) densities ($\chi_{\mathbf{k},\lambda}^{\left(m=0,s=2\right)}=\lambda$),
as well as momentum ($\chi_{\mathbf{k},\lambda}^{\left(m=1,s=3\right)}+\left(-\right)\chi_{\mathbf{k},\lambda}^{\left(m=-1,s=3\right)}=2\left(i\right)\beta v\hbar k_{x\left(y\right)}$).
The first two conservation laws lead to two diffusive modes. The conservation
of momentum gives rise to second sound - ballistic thermal waves propagating
through the two dimensional graphene plane \cite{Phan2013}. This
mode is the analogue of the density modes of a clean neutral Galilean
invariant system.

\begin{figure}
\centering{}\includegraphics[scale=0.35]{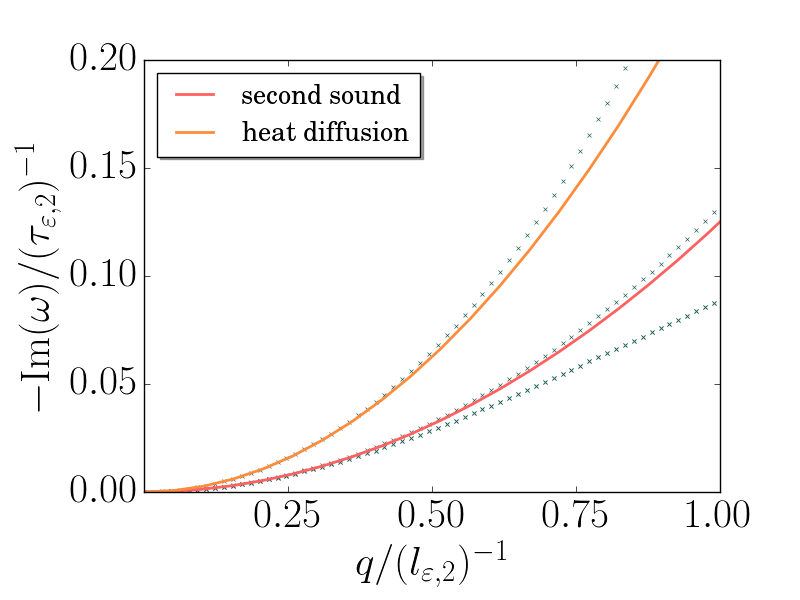}\caption{\label{fig:Fig_collective_energy_im_Small_q}The figure shows the
imaginary part of the dispersion relations of second sound, heat diffusion,
and quasiparticle (imbalance) diffusion excitations. The wavevector
$q$ is given in units of the inverse scattering length $v\tau_{c,2}^{-1}$.
The grey symbols correspond to the numerical solution of Eq. (\ref{eq:Collective_mode_condition}).
The damping of second sound is due to scattering in the $m=2$ channel
and follows the dispersion $\mathrm{Im}\left(\omega_{\mathrm{sec.\,sound}}\right)\approx\frac{1}{8}v^{2}q^{2}\tau_{\varepsilon,2}$
(red curve). For small $q$ the imaginary part of the second sound
dispersion and the dispersion of the quasiparticle diffusion mode
merge. A third diffusive mode corresponds to the diffusion of heat
(orange curve).}
\end{figure}

\begin{figure}
\centering{}\includegraphics[scale=0.35]{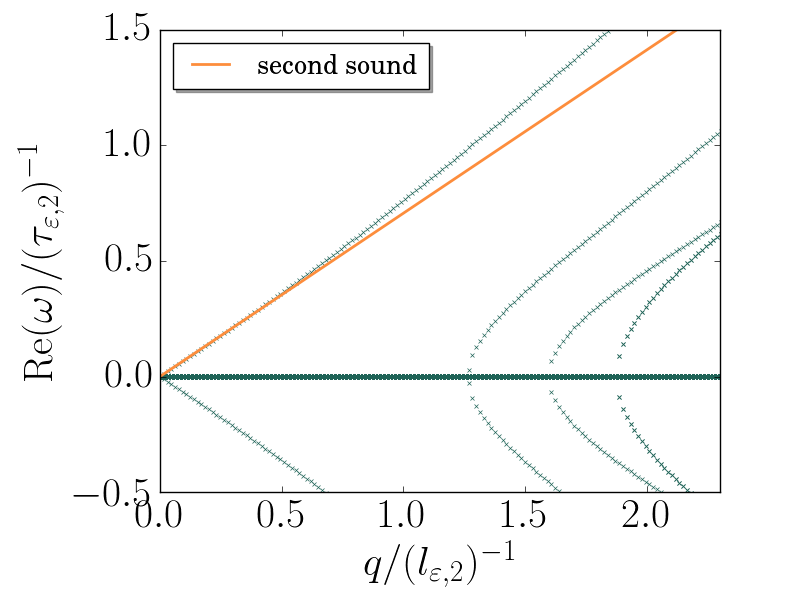}\caption{\label{fig:Fig_collective_energy_re}The real parts of the dispersion
relations of collective energy and imbalance excitations are depicted.
The grey symbols correspond to the numerical solution of Eq. (\ref{eq:Collective_mode_condition}).
The linear dispersion of the second sound mode given by $vq/\sqrt{2}$
for small $q$ is shown in orange color. The wave-vector $q$ is given
in units of the inverse scattering length $v\tau_{c,2}^{-1}$.}
\end{figure}

Truncating the mode expansion of Eq. (\ref{eq:Collective_mode_condition})
at $m=2$, which is a good approximation for low wave-numbers, yields
the dispersions 
\begin{align}
\omega_{\mathrm{heat\,diff.}} & \approx\frac{1}{4}v^{2}q^{2}\tau_{\varepsilon,2},\nonumber \\
\omega_{\mathrm{qp\,diff.}} & \approx\frac{1}{8}v^{2}q^{2}\tau_{\varepsilon,2},
\end{align}
for the heat and quasiparticle (imbalance) diffusion modes, respectively.
The second sound dispersion is given by 
\begin{equation}
\omega_{\mathrm{sec.\,sound}}\approx\frac{vq}{\sqrt{2}}+i\tau_{\varepsilon,2}\frac{v^{2}q^{2}}{8}.
\end{equation}
Second sound mediated by phonons has been previously observed in solids
\cite{Narayanamurti1972} and had a velocity comparable to the velocity
of sound. Here, the second sound is carried by electrons and propagates
with a velocity $v_{0}/\sqrt{2}$. The above dispersion relations
are shown in Figs. \ref{fig:Fig_collective_energy_im_Small_q} and
\ref{fig:Fig_collective_energy_re}. 

\begin{figure}
\centering{}\includegraphics[scale=0.35]{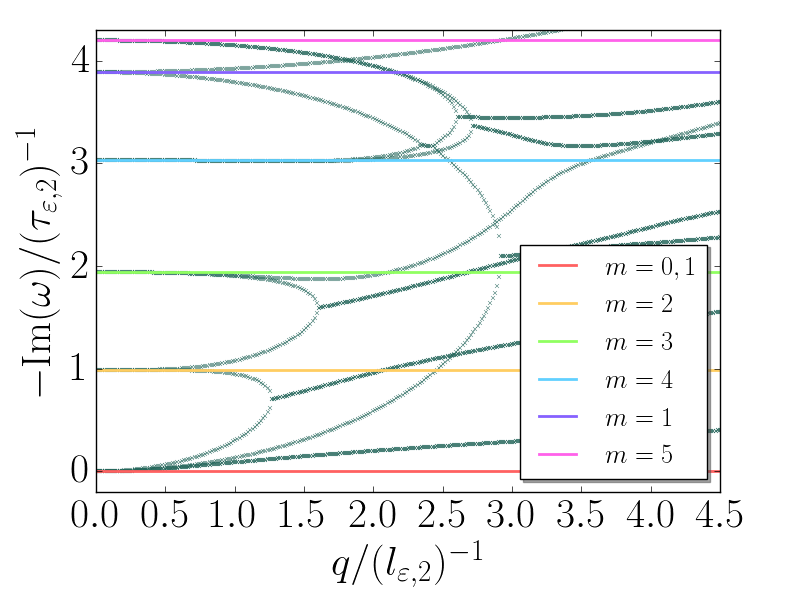}\caption{\label{fig:Fig_collective_energy_im}The imaginary part of the dispersion
relations of collective charge excitations in different angular harmonic
channels $m$ are shown. The wave-vector $q$ is given in units of
the inverse scattering length $v\tau_{c,2}^{-1}$. The grey symbols
correspond to the numerical solution of Eq. (\ref{eq:Collective_mode_condition}).
For small $q$, the modes approach values given by the scattering
rates $-i/\tau_{c,m}$ and are thus strongly damped. At larger values
of $q$, the dispersions tend to merge in a complex fashion. Fig.
\ref{fig:Fig_collective_energy_im_Small_q} shows the weakly damped
modes (second sound and diffusive modes) for small values of $q$.}
\end{figure}

The dispersion of the quasiparticle diffusion mode and the imaginary
part of the second sound dispersion merge at low wave-numbers. As
in the case of charge excitations, there exists an infinite number
of damped modes associated with scattering in higher angular harmonic
channels. These modes are depicted in Figs. \ref{fig:Fig_collective_energy_im}
and \ref{fig:Fig_collective_energy_re}. Note, that modes associated
with imbalance excitations ($s=2$) are damped stronger by an order
of magnitude as compared to energy excitations ($s=3$). 

\subsection{Validity of the collinear zero mode approximation for collective
modes\label{subsec:Validity-Restriction}}

The discussion so far was carried out in the restricted subspace of
collinear zero modes. In this section it is shown that the results
for collective excitations obtained within the restricted subspace
remain valid, if this restriction is lifted, and non-collinear zero
modes are added. These modes introduce large corrections to the matrix
of scattering rates $\Gamma_{m}^{s,s'}$, and it is not obvious that
they can be neglected. It is sufficient to consider the $\mathbf{q}=0$
case. The extension to finite wave-numbers is straightforward.

The scattering rate matrix $\Gamma_{m}^{s,s'}$ of Eq. (\ref{eq:Definition_scattering_rates})
is extended to include modes that are not collinear zero modes, which
are labeled with indices $s>3$. It is useful to define the following
matrices 
\begin{align*}
S & =\left(v\beta\hbar\right)^{2}\left\langle \chi_{\mathbf{k},\lambda}^{\left(s<3\right)}\left|\mathcal{C}\right|\chi_{\mathbf{k},\lambda}^{\left(s'<3\right)}\right\rangle \\
P & =\left(v\beta\hbar\right)^{2}\left\langle \chi_{\mathbf{k},\lambda}^{\left(s>3\right)}\left|\mathcal{C}\right|\chi_{\mathbf{k},\lambda}^{\left(s'<3\right)}\right\rangle \\
Q & =\left(v\beta\hbar\right)^{2}\left\langle \chi_{\mathbf{k},\lambda}^{\left(s<3\right)}\left|\mathcal{C}\right|\chi_{\mathbf{k},\lambda}^{\left(s'>3\right)}\right\rangle \\
R & =\left(v\beta\hbar\right)^{2}\left\langle \chi_{\mathbf{k},\lambda}^{\left(s>3\right)}\left|\mathcal{C}\right|\chi_{\mathbf{k},\lambda}^{\left(s'>3\right)}\right\rangle .
\end{align*}
Here, $\chi_{\mathbf{k},\lambda}^{\left(s<3\right)}$ are the familiar
collinear zero modes (\ref{eq:Collinear_modes}). $\chi_{\mathbf{k},\lambda}^{\left(s>3\right)}$
are modes with a different $\left|\mathbf{k}\right|$-dependence,
such that the full set of modes forms a complete basis. Since $\mathcal{C}$
is Hermitian, we have $Q=P^{T}$. The mode expansion of the Liouville
operator $L_{s,s'}$ of Eq. (\ref{eq:Liouville_projected}) also has
to be enlarged by the $s>3$ modes. However, we do not need to know
the precise values of the corresponding elements of $L$. The eigenvalue
equation (\ref{eq:q=00003D00003D0_modes_eiegnvalues}) reads 
\begin{equation}
\det\left(-i\omega L-F\right)=0,\label{eq:modes_equation_full_space}
\end{equation}
where $F$ is the composite matrix 
\[
F=\left[\begin{array}{cc}
S & P\\
P^{T} & R
\end{array}\right].
\]
In the following, the Liouville matrix $L$ will also be separated
into blocks corresponding the same subspaces: $L=\left(\left(L_{S},L_{P}\right),\left(L_{P}^{T},L_{S}\right)\right)$.
It follows from Eq. (\ref{eq:Collinear_non_collinear_separation})
and the Hermiticity of the collinear part of the collision operator
$\mathcal{C}_{c}$ that 
\begin{align*}
S & \sim P\sim1\\
R & \sim\log\left(1/\alpha\right),
\end{align*}
meaning that non collinear zero modes are scattered faster by a factor
of $\log\left(1/\alpha\right)$. The determinant can be found using
the block matrix identity 
\begin{equation}
\det\left[\begin{array}{cc}
A & B\\
C & D
\end{array}\right]=\det\left(D\right)\det\left(A-BD^{-1}C\right).\label{eq:block_matrix_identity}
\end{equation}
Applying this identity to Eq. (\ref{eq:modes_equation_full_space})
and noticing that for $\alpha\rightarrow0$ the inverse matrix in
the last determinant vanishes, one has 
\[
\det\left(i\omega L+F\right)\approx\det\left(i\omega L_{R}+R\right)\det\left(i\omega L_{S}+S\right).
\]
Eq. (\ref{eq:modes_equation_full_space}) therefore separates into
two independent parts: $\det\left(i\omega L_{R}+R\right)=0$ and $\det\left(i\omega L_{S}+S\right)=0$.
The second equation is equivalent to the eigenvalue equation (\ref{eq:q=00003D00003D0_modes_eiegnvalues}).
In the limit of a small fine structure constant, the weakly damped
collective modes can therefore be found by solving the kinetic equation
in the restricted subspace of collinear zero modes, even if there
is significant coupling between all modes.

\section{Surface acoustic waves\label{sec:Surface-acoustic-waves}}

The longitudinal electrical conductivity $\sigma_{\parallel}$ is
accessible through experiments with surface acoustic waves (SAWs)
\cite{Wixforth1989}. The simplest setup to measure $\sigma_{\parallel}$
is a sheet of graphene placed on top of a piezoelectric material.
Using interdigital transducers, SAWs are induced in the piezoelectric.
The real part of $\sigma_{\parallel}$ then determines the damping
of the SAWs, while the imaginary part changes the SAW velocity $v_{s}$.
Overall, for a small piezoelectric coupling the change of the SAW
velocity $\Delta v_{s}$, where the imaginary part describes the damping,
can be written as \cite{Ingebrigtsen1969,Simon1996} 
\begin{equation}
\frac{\Delta v_{s}}{v_{s,0}}=p_{e}\frac{1}{1+i\frac{\sigma_{\parallel}}{\sigma_{M}}}.\label{eq:SAW_velocity_shift_complex}
\end{equation}
Here $p_{e}<1$ is an effective coupling constant, and a $\sigma_{M}$
a reference conductivity. Both $p_{e}$ and $\sigma_{M}$ depend on
material parameters of the piecoelectric. A rough estimate for $\sigma_{M}$
is given by $\sigma_{M}\approx v_{s}\epsilon_{\mathrm{eff}}$ \cite{Ingebrigtsen1969,Simon1996},
where $\epsilon_{\mathrm{eff}}$ is the effective permittivity at
the surface of the piezoelectric. There has been experimental work
on the coupling between SAWs and graphene \cite{Bandhu2013,Miseikis2012}.
LiNbO$_{3}$ seems to be a suitable piezoelectric for such experiments\cite{Bandhu2013},
because it provides a relatively large coupling parameter $p_{e}\approx0.03$
\cite{Rotter1998}. While there might be better choices for the piezoelectric
material, here we consider LiNbO$_{3}$, since the feasibility of
a graphene-LiNbO$_{3}$ device has been demonstrated in Ref. \cite{Bandhu2013}.
The SAW velocity is $v_{s}\approx4\cdot10^{3}\,\mathrm{m/s}$ and
the effective dielectric constant is given by $\epsilon_{\mathrm{eff}}\approx0.5\epsilon_{0}\left(\sqrt{\epsilon_{xx}^{T}\epsilon_{zz}^{T}}+2\right)\approx24\epsilon_{0}$
(assuming that the dielectric constant above the graphene sheet is
$\epsilon_{0}$). One then has 
\[
\sigma_{M}\approx10^{-6}\,\mathrm{S}.
\]
The fine structure constant is small due to the large dielectric constant
and renormalization effects. We estimate $\alpha\approx0.1.$ Here
and in the following estimations, we assume a temperature of $50\,\mathrm{K}$.

Interdigital transducers induce SAWs with sharply defined wave-vectors
$q_{0}$. The frequency of the SAW $\omega_{0}$ is given by 
\[
\omega_{0}=v_{s}q_{0}.
\]
$\omega_{0}$ is much smaller than the characteristic hydrodynamic
frequency for a wave-vector of the same magnitude $\omega_{\mathrm{hydro}}\approx vq_{0}$,
where $v\approx10^{6}\,\mathrm{m}/\mathrm{s}$. It is 
\begin{equation}
\frac{\omega_{0}}{\omega_{\mathrm{hydro}}}\approx0.005.\label{eq:Ratio_SAW_Hydro_Freq}
\end{equation}
As shown in Fig. \ref{fig:Fig_charge_conductivities} the longitudinal
conductivity $\sigma_{\parallel}$ is peaked around $\omega_{\mathrm{hydro}}$
and vanishes in the limit $\omega=0$, $q\rightarrow0$. Therefore,
SAW experiments are confined to a highly ``off-resonant'' regime
due to the small ratio (\ref{eq:Ratio_SAW_Hydro_Freq}) and therefore
cannot be large. 

The damping coefficient is given by 
\begin{equation}
\Gamma=-\omega\mathrm{Im}\left(\frac{\Delta v_{s}}{v_{s}}\right)=\omega p_{e}\frac{\mathrm{\mathrm{Re}}\left(\sigma_{\parallel}\right)/\sigma_{M}}{1+\left|\frac{\sigma_{\parallel}}{\sigma_{M}}\right|^{2}}.\label{eq:dampling_coefficient}
\end{equation}
The relative velocity shift is 
\begin{equation}
\mathrm{Re}\left(\frac{\Delta v_{s}}{v_{s}}\right)=p_{e}\frac{1+\mathrm{\mathrm{Im}}\left(\sigma_{\parallel}\right)/\sigma_{M}}{1+\left|\frac{\sigma_{\parallel}}{\sigma_{M}}\right|^{2}}.\label{eq:SAW_velocity_shift}
\end{equation}

Since interdigital transducers excite SAWs of a fixed wavelength,
altering $q_{0}$ is difficult. Instead, the $q$ dependence of $\sigma_{\parallel}$
can be tested by varying the temperature, and thus the product of
the wave vector and the scatterng length and $q_{0}l_{c,m}$. Fig
\ref{fig:SAW_results} shows the damping and the velocity shift induced
by the graphene sheet as a function of temperature, according to Eqs.
(\ref{eq:dampling_coefficient}) and (\ref{eq:SAW_velocity_shift}).
As expected, the damping coefficients are very small, on the order
of $10^{5}\,\mathrm{Hz}$, corresponding to damping lengths of $1/\mathrm{cm}$.
Such small damping are measurable in GaAs 2DEG structures \cite{Govorov2000},
however they might be hard to observe with the more unconventional
LiNbO$_{3}$ device. On the other hand the low temperature (large
$q$) behavior of the conductivity sensitively depends on the scattering
rates in the higher angular harmonic channels (see lower left panel
of Fig. \ref{fig:SAW_results}), although the specific dependence
$\tau_{m>c,2}^{-1}\sim\left|m\right|$ will be very hard to distinguish
from e.g. constant scattering rates.. Finally we note, that here we
considered the SAW response in the hydrodynamic regime $l_{c,1}\ll w$,
where $w$ is the sample size. For small sample sizes, the results
will differ due to boundary scattering.\onecolumngrid 

\begin{center}
\begin{figure}
\includegraphics[scale=0.28]{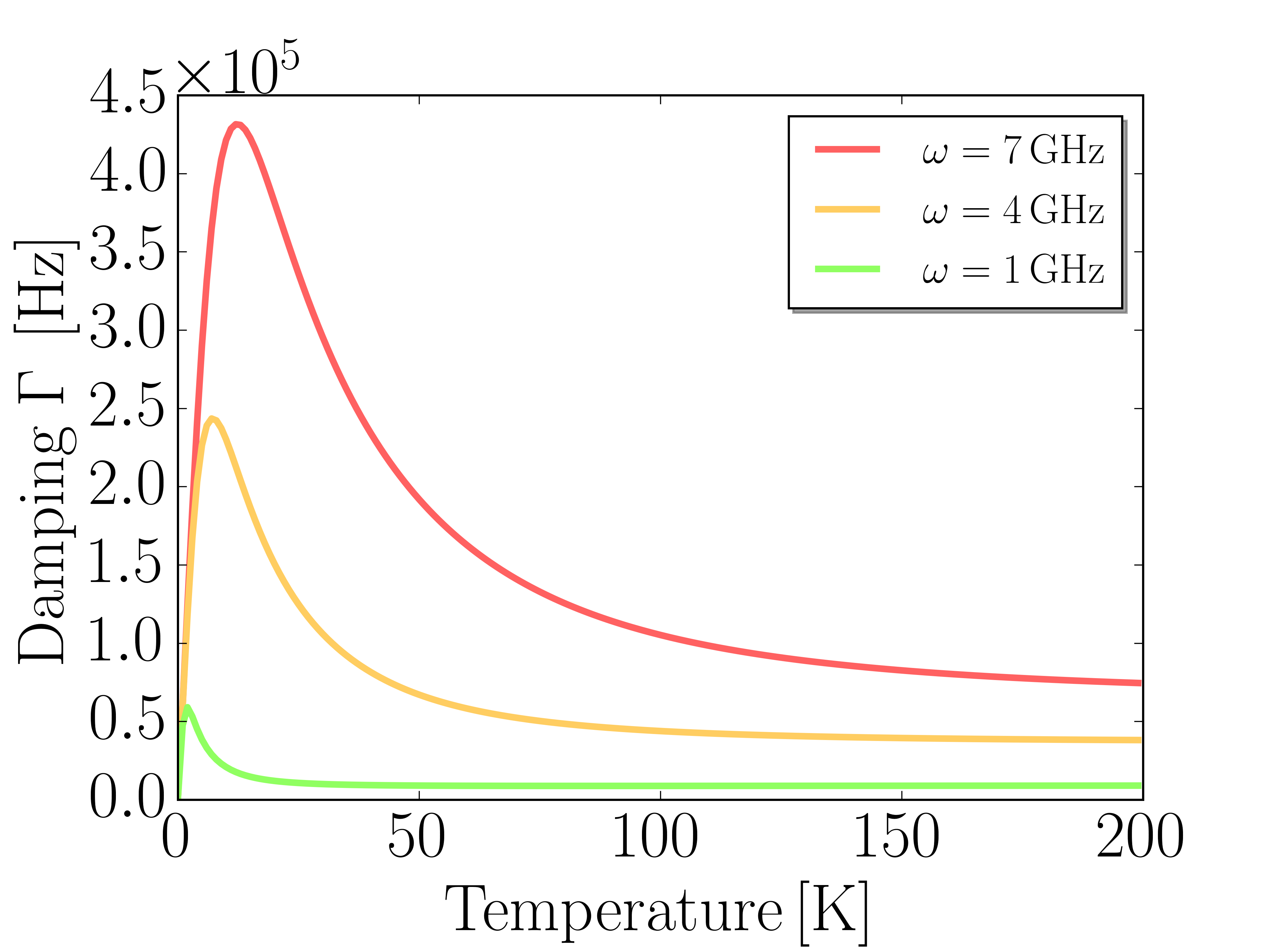}\includegraphics[scale=0.28]{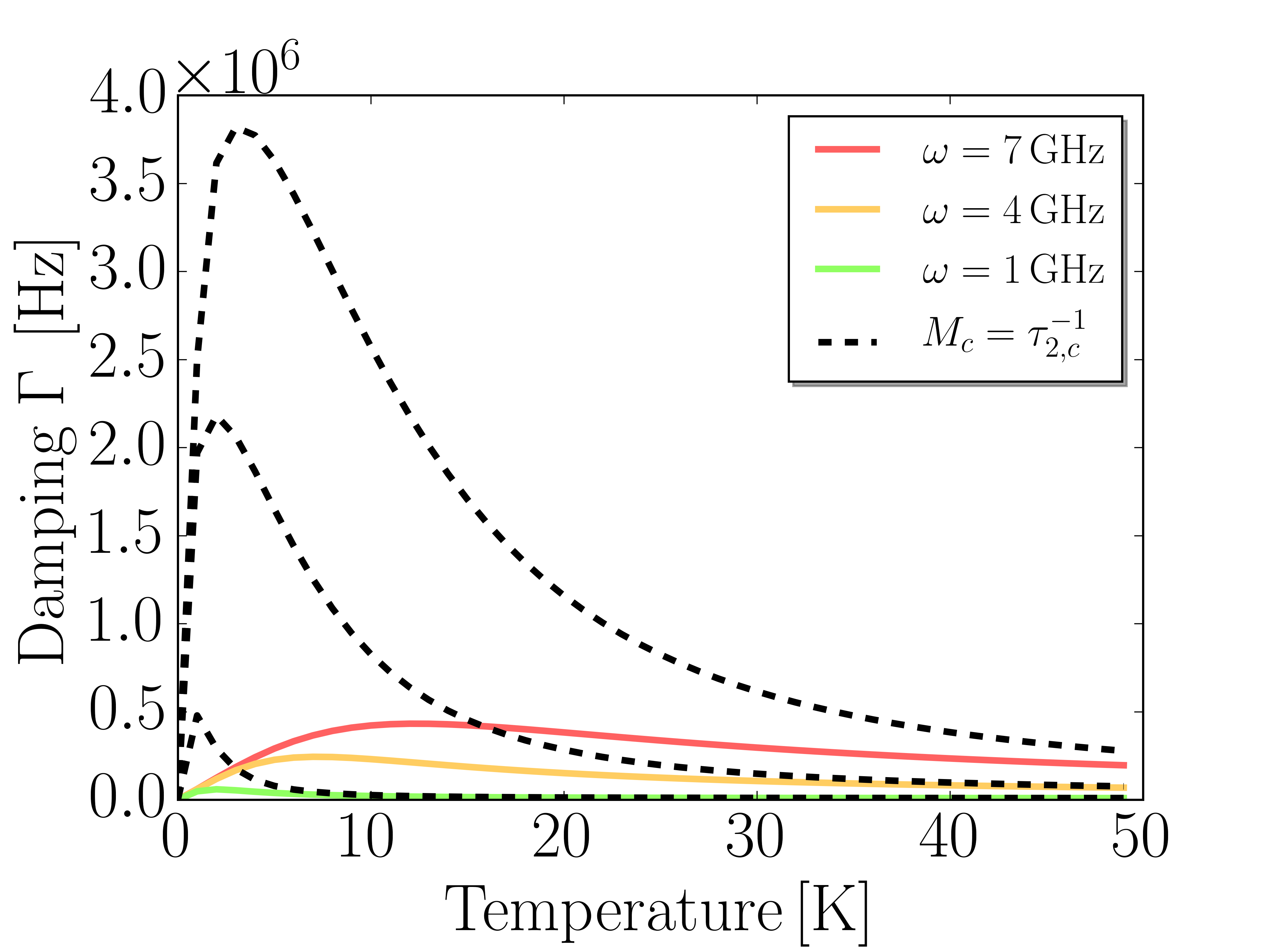}\includegraphics[scale=0.28]{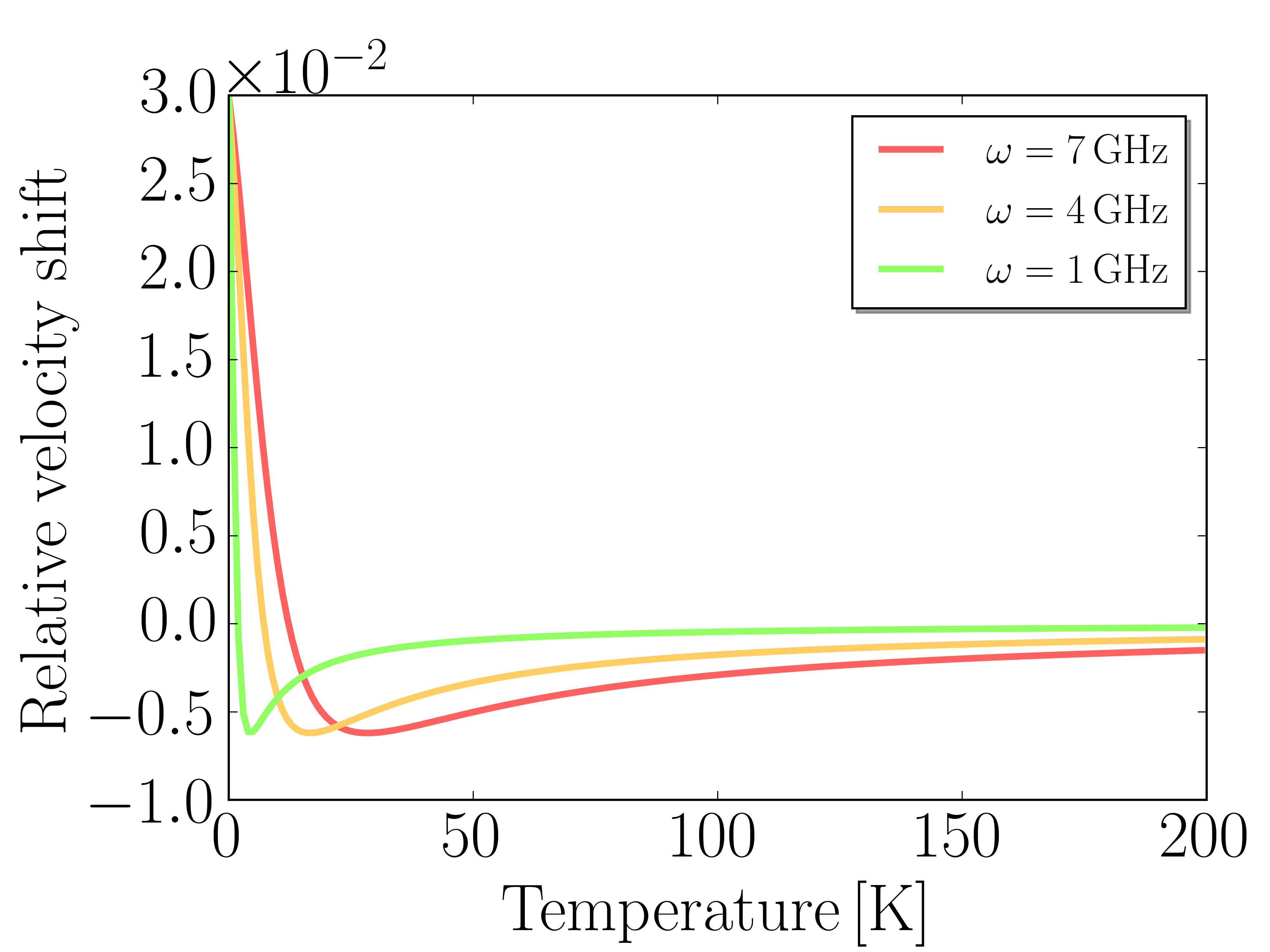}

\centering{}\caption{The Figure shows the damping coefficients and velocity shifts of LiNbO$_{3}$
surface acoustic waves induced by a graphene sheet laying on top of
the crystal. Due to the $\sim1/T$ dependence of the scattering times
$\tau_{c,m}$, changing the temperature alters the quantity $v\tau_{c,m}q$,
where $q$ is the SAW wave-vector, such that the functional dependence
of $\sigma_{\parallel}\left(q,\omega=v_{s}q\right)$ can be investigated
without switching the SAW frequencies. Here, $v_{s}$ is the SAW velocity.
Left figure: Damping coefficients of SAWs for three distinct frequencies.
The induced damping is small (of the order of $1/\mathrm{cm}$). Middle
figure: The damping coefficients at low temperatures depend sensitively
on the scattering in higher angular harmonic channels. Setting $M_{c}=\tau_{c,2}^{-1}$,
thus neglecting the scattering times $\tau_{c,m>2}$, raises the damping
by an order of magnitude. Right figure: Upper right pannel: Relative
velocity shifts $\Delta v_{s}/v_{s}$. The velocity shifts are large
(on the order of $1\%$). This is a consequence of the mainly capacitive
behavior of the graphene sheet at small frequencies (see main text).
\label{fig:SAW_results}}
\end{figure}

\par\end{center}

\twocolumngrid

\section{Poiseuille profiles\label{sec:Poiseuille-profiles}}

The wave-vector-dependence of transport coefficients is of importance
when the currents in a system are spatially inhomogeneous, either
because the applied fields are inhomogeneous, or because the inhomogeneity
is imposed by the geometry of the system. The simplest example for
the latter case is the Poiseuille flow. In undoped graphene, the energy
current is conserved due to the conservation of momentum, however
it is dissipated by the uneven boundaries of the sample \cite{Kiselev2019b}.
In a Poiseuille geometry, which consists of an infinitely long, straight
sample of width $w$, the boundaries slow down the current flow. The
current profile becomes parabolic across the sample. On the other
hand, charge currents decay in the bulk of undoped graphene due to
the interaction induced resistivity. In this case, there exists a
crossover from an almost flat current profile if $w\gg v\tau_{c,1}$
to a more parabola-like shape at $w<v\tau_{c,1}$. However, as shown
in Ref. \cite{Kiselev2019b}, the slowing down of the flow by the
boundaries becomes inefficient when $w\lesssim v\tau_{c,2}$, again
changing the profile. In this section we investigate the Poiseuille
profiles of charge currents in undoped graphene using the full non-local
conductivity (\ref{eq:charge_conductivity_result}).

\subsection{Flow equations and boundary conditions}

The thermal and charge flow is governed by the constitutive relations
\begin{equation}
\kappa^{-1}\left(\mathbf{q},\omega\right)_{\alpha\beta}j_{\varepsilon,\beta}=-\partial_{\alpha}T
\end{equation}
and 
\begin{equation}
\sigma^{-1}\left(\mathbf{q},\omega\right)_{\alpha\beta}j_{c,\beta}=E_{\alpha},
\end{equation}
where $j_{\varepsilon,\beta}$ is the thermal current and $j_{c,\beta}$
the electric current. With the thermal and electric conductivities
$\kappa$ and $\sigma$ depending on the wave vector $\mathbf{q}$,
these equations can be seen as Fourier transforms of differential
equations. Similar equations have been studied to describe non-localities
induced by vortices in type II superconductors \cite{Huse1993}. The
temperature gradient $-\partial_{\alpha}T$ and the electric field
$E_{\alpha}$ act as source terms. In a Poiseuille geometry, the force
fields act perpendicular to the gradient of the flow velocity, i.e.
it is $\mathbf{E}\perp\mathbf{q}$, $\nabla T\perp\mathbf{q}$. Therefore,
the currents are determined by the transverse conductivities. Let
the sample be oriented in $y$-direction and centered around $x=0$.
The equations then read 
\begin{align}
\kappa_{T}^{-1}\left(q_{x},\omega\right)j_{\varepsilon,y}\left(q_{x},\omega\right) & =-\partial_{y}T,\label{eq:pois_therm_diff}\\
\sigma_{T}^{-1}\left(q_{x},\omega\right)j_{c,y}\left(q_{x},\omega\right) & =E_{y}.\label{eq:pois_el_diff}
\end{align}
To solve the above equations, boundary conditions at the sample boundaries
at $\pm w/2$ are needed. As discussed in Ref. \cite{Kiselev2019a},
partial slip boundary conditions are appropriate: 
\begin{equation}
j_{\varepsilon/c,y}\left(x=\pm w/2,\omega\right)=\mp\zeta\left.\frac{\partial j_{\varepsilon/c,y}}{\partial x}\right|_{x=\pm w/2}.\label{eq:boundary_condition_nl}
\end{equation}
$\zeta$ is the so called slip length parameterizing the momentum
charge (current) dissipation at the sample boundaries. If the boundaries
are sufficiently rough, $\zeta$ is of the order of the mean free
path associated with the $m=2$ scattering time: $\zeta\sim v\tau_{\varepsilon/c,2}$.
In principle, the Eqs. (\ref{eq:pois_therm_diff}), (\ref{eq:pois_el_diff})
represent infinite order differential equations and require infinitely
many boundary conditions. However, this problem does not appear explicitly
in the calculation. The finite width of the sample $w$ sets a natural
cut-off for the wave-numbers $q$, and therefore only the low powers
of $q$ are relevant on the right hand side of Eqs. (\ref{eq:pois_therm_diff}),
(\ref{eq:pois_el_diff}). For simplicity, the boundary condition (\ref{eq:boundary_condition_nl})
is used, which is reasonable for not too small widths.

The Eqs. (\ref{eq:pois_therm_diff}), (\ref{eq:pois_el_diff}) now
can be solved by performing a Fourier transform. To fix the boundary
conditions two point-like delta-function inhomogeneities are positioned
at $\pm w$. In real space the equations take the form 
\begin{align}
\kappa_{T}^{-1}\left(\partial_{x},\omega\right)j_{\varepsilon,y}\left(x,\omega\right) & =-\partial_{y}T-\alpha\delta\left(x-w\right)-\beta\delta\left(x+w\right)\label{eq:pois_therm_real_sources}\\
\sigma_{T}^{-1}\left(\partial_{x},\omega\right)j_{c,y}\left(x,\omega\right) & =E_{y}-\alpha\delta\left(x-w\right)-\beta\delta\left(x+w\right).\label{eq:pois_el_real_sources}
\end{align}
If the constants $\alpha$, $\beta$ are chosen such that Eq. (\ref{eq:boundary_condition_nl})
is satisfied, the solution inside the sample will be identical to
the solution of the homogeneous equations with the matching boundary
conditions.

Here, the profiles of electric current flows through samples of different
widths will be calculated. Solving the Eq. (\ref{eq:pois_therm_real_sources})
in Fourier space one obtains 
\begin{align}
 & j_{c,y}\left(q_{x},\omega\right)=\nonumber \\
 & \qquad\left(2\pi E_{y}\delta\left(q_{x}\right)-\alpha e^{-iwq_{x}}-\beta e^{iwq_{x}}\right)\sigma_{T}\left(q_{x},\omega\right).\label{eq:Fourier_solution_charge}
\end{align}
Inserting this result into Eq. (\ref{eq:boundary_condition_nl}) gives
two algebraic equations, from which $\alpha$ and $\beta$ can be
determined: 
\begin{align*}
 & \zeta\int\frac{dq_{x}}{2\pi}\left(iq_{x}\right)\left(\alpha e^{-iq_{x}\frac{3w}{2}}+\beta e^{iq_{x}\frac{w}{2}}\right)\sigma_{T}\left(q_{x},\omega\right)=\\
 & \qquad\int\frac{dq_{x}}{2\pi}\left(\alpha e^{-iq_{x}\frac{3w}{2}}+\beta e^{iq_{x}\frac{w}{2}}\right)\sigma_{T}\left(q_{x},\omega\right)-\\
 & \qquad\qquad E_{y}\sigma_{T}\left(0,\omega\right)
\end{align*}
\begin{align*}
 & \zeta\int\frac{dq_{x}}{2\pi}\left(iq_{x}\right)\left(\alpha e^{-iq_{x}\frac{w}{2}}+\beta e^{iq_{x}\frac{3w}{2}}\right)\sigma_{T}\left(q_{x},\omega\right)=\\
 & \qquad-\int\frac{dq_{x}}{2\pi}\left(\alpha e^{-iq_{x}\frac{w}{2}}+\beta e^{iq_{x}\frac{3w}{2}}\right)\sigma_{T}\left(q_{x},\omega\right)+\\
 & \qquad\qquad E_{y}\sigma_{T}\left(0,\omega\right).
\end{align*}
The above integrals are calculated with the FFT algorithm. Once $\alpha$,
$\beta$ are found, a Fourier transform the of the solution (\ref{eq:Fourier_solution_charge})
gives the desired flow profiles.

\begin{figure}
\centering{}\includegraphics[scale=0.35]{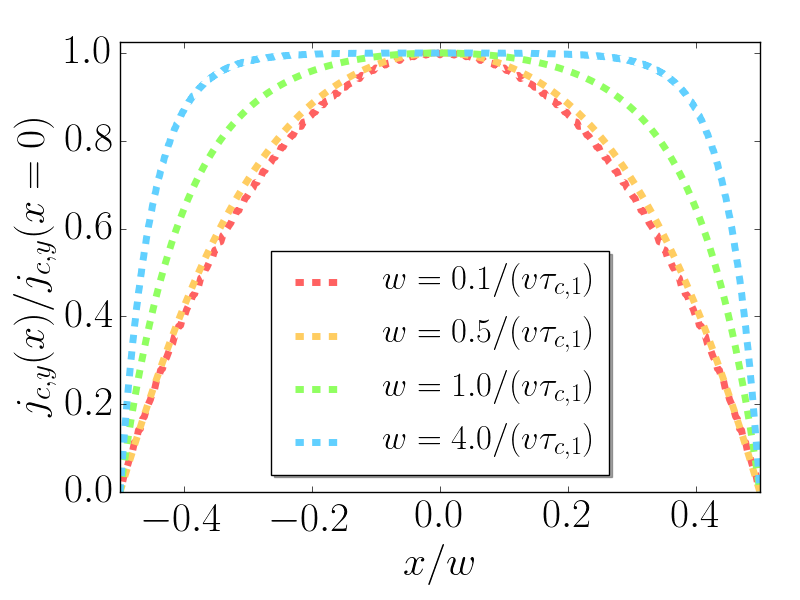}\caption{\label{fig:Poiseuille_no_slip}Poiseuille profiles of charge currents
in undoped graphene samples of different widths $w$. Although physically
incorrect, no-slip boundary conditions were assumed for clarity. The
profiles are normalized to the current at $x=0$. At large widths
$w>v\tau_{c,1}$, the flow profiles turn flat. In the bulk they resemble
Ohmic flow. For small widths $w<v\tau_{c,1}$, the momentum non-conserving
scattering becomes inefficient. The electrons travel a distance corresponding
to several width before loosing their momentum. Consequently, the
profiles take a parabolic form, resembling classical Poiseuille flow.
The profiles were calculated from Eq. (\ref{eq:Fourier_solution_charge}).}
\end{figure}

\begin{figure}
\centering{}\includegraphics[scale=0.35]{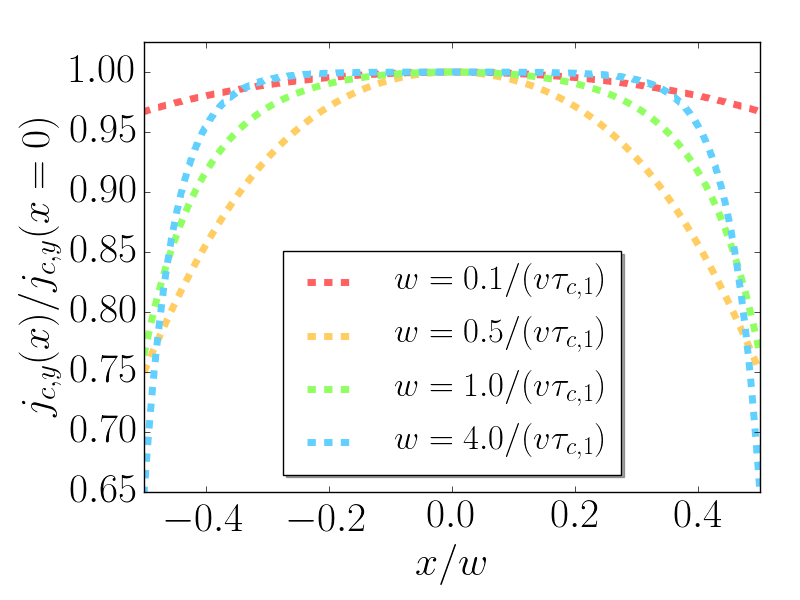}\caption{\label{fig:Poiseuille_slip}Poiseuille profiles of charge currents
in undoped graphene samples of different widths $w$, normalized to
the current at $x=0$. Partial slip boundary conditions with a slip
length $\zeta=v\tau_{c,2}$ were applied. At very small widths $w\ll\zeta$,
boundary scattering ceases to be an efficient mechanism for the dissipation
of electric current. The profiles turn flat, as they do in the nearly
Ohmic regime $w>v\tau_{c,1}$. In the crossover regime at widths $w\sim0.5v\tau_{c,1}$,
profile curvature is most pronounced. The profiles were calculated
from Eq. (\ref{eq:Fourier_solution_charge}).}
\end{figure}

Figs. \ref{fig:Poiseuille_no_slip} and \ref{fig:Poiseuille_slip}
show the results for different widths $w$. For demonstration purposes
no-slip boundary conditions ($\zeta=0$) were assumed in Fig. \ref{fig:Poiseuille_no_slip}.
Here, for $w>v\tau_{c,1}$ the flow profile turns flat in the middle
of the sample and steeply descends to zero at the boundaries (as necessitated
by the no-slip boundary conditions). This behavior is due to the interaction-induced
conductivity that dissipates current uniformly across the sample -
at a distance $d>v\tau_{c,1}$ away from the boundary, a uniform flow
is restored. On the other hand, for $w<v\tau_{c,1}$ the current-relaxing
scattering processes in the $m=1$ channel become less and less important.
The scattering in the $m=2$ channel dominates. It acts in the same
way viscous forces act in ordinary flows. Current is transported from
the middle of the sample, where it is maximal, to the sample edges,
where it is dissipated. A finite slip length (as discussed, $\zeta=v\tau_{c,2}$
was chosen for simplicity) alters these results (see Fig. \ref{fig:Poiseuille_slip}):
Whereas for widths $w>v\tau_{c,1}$ the finite slip gives the current
a non-negligible velocity at the sample boundary, for small widths
$w<v\tau_{c,2}$, the flow profiles are rendered flatter, and the
boundary effects become negligible. In the crossover region $w\sim v\tau_{c,1}$,
the profiles are curved and resemble a parabola. This takes place
around $w\sim0.5v\tau_{c,2}$ and is in accordance with the general
expectations \cite{Kiselev2019a}: for $w<v\tau_{c,2}$ the quasi-viscous
transport of currents from the middle of the sample towards the boundaries
becomes inefficient, and the boundary does efficiently dissipate the
current. 

An interesting question is how the collective modes investigated in
Sec. \ref{sec:Collective-modes} are changed when the Dirac fluid
is confined to a Poiseuille type sample with the boundary conditions
of Eq. (\ref{eq:boundary_condition_nl}). For large sample sizes one
can expect that e.g. the charge modes will exhibit a small correction
of the order of $l_{c,1}/w$. The effects for small $w$ should be
more interesting. They are, however, beyond the scope of the present
study.

\section{Conclusion}

In conclusion we have developed a kinetic theory of non-local charge
and thermal transport in a clean Dirac fluid in the hydrodynamic regime.
We obtained closed analytic expressions for the frequency and wave-vector-dependent,
charge and thermal conductivities as well as the non-local viscosity
due to electron-electron Coulomb interactions. Our solution is possible
due to the dominance of so-called colinear zero modes. In the limit
os a small fine-structure constant of graphene, all other mode relax
more rapidly, limiting the phase space of the collective excitations
that dominate the long-time dynamics. One aspect of the same physics,
that was discussed previously by us in Ref.\cite{Kiselev2019b}, is
the onset of superdiffusion in phase space, where Lévy-flight behavior
on the Dirac cone emerges. Frequent small angle scattering events
are interrupted by rate large-angle scattering processes. We made
specific predictions for measurements such as the velocity shift of
surface acoustic waves and for inhomogeneous flow pattern. Those become
identical to the one that follow from the solution of the Navier-Stokes
equations in the long wavelength limit, but include higher order gradients
that come into play as the sample geometry becomes smaller. In particular,
we have demonstrated how the non-local transport coefficients determine
the profiles of a hydrodynamic flow through narrow channels. In addition
we determined the collective mode spectrum of the system including
plasma waves and second sound like thermal waves. We find a complex
structure of damped collective excitations. These excitations are
similar to the so-called ``non-hydrodynamic'' modes that were shown
to be relevant for the equilibration of other collission-domuinated
quantum fluids \cite{Brewer2015} \cite{Romatschke2016,Romatschke2018,Heller2018}. 
\begin{acknowledgments}
This work was supported by the European Commission’s Horizon 2020
RISE program Hydrotronics (Grant No. 873028). We thank L. Levitov,
A. Lucas, and J. F. Karcher for interesting discussions and P. Witkowski
for drawing our attention to Refs. \cite{Brewer2015,Romatschke2016,Romatschke2018,Heller2018}.
We are grateful to B. Jeevanesan for pointing out that the identity
(\ref{eq:block_matrix_identity}) provides a simple proof of the dominance
of collinear zero modes in Sec. \ref{subsec:Validity-Restriction}
and to I. V. Gornyi for clarifying to us the role of the Vlasov term
in the conductivity.
\end{acknowledgments}

\appendix

\onecolumngrid

\section{The collision operator\label{sec:App_Collision-operator}}

Transformed to the band basis, the interaction part of the Hamilton
operator (\ref{eq:nl_Hamiltonian}) reads 
\begin{equation}
H_{{\rm int}}=\frac{1}{2}\int_{k,k^{\prime},q}\sum_{\alpha\beta}T_{\lambda\mu\mu^{\prime}\lambda^{\prime}}\left(\mathbf{k,k}^{\prime},\mathbf{q}\right)\psi_{\lambda^{\prime}}^{\dagger}\left(\mathbf{k+q},t\right)\psi_{\mu}^{\dagger}\left(\mathbf{k}^{\prime}-\mathbf{q},t\right)\psi_{\mu^{\prime}}\left(\mathbf{k}^{\prime},t\right)\psi_{\lambda}\left(\mathbf{k,}t\right)
\end{equation}
where the matrix elements $T_{\lambda\mu\mu^{\prime}\lambda^{\prime}}\left(\mathbf{k,k}^{\prime},\mathbf{q}\right)$
\begin{equation}
T_{\lambda\mu\mu^{\prime}\lambda^{\prime}}\left(\mathbf{k,k}^{\prime},\mathbf{q}\right)=V\left(q\right)\left(U_{\mathbf{k+q}}U_{\mathbf{k}}^{-1}\right)_{\lambda^{\prime}\lambda}\left(U_{\mathbf{k}^{\prime}-\mathbf{q}}U_{\mathbf{k}^{\prime}}^{-1}\right)_{\mu\mu^{\prime}}.
\end{equation}
$U$ is the usual transformation from sub-lattice space to the band
space (see Eq. \ref{eq:free_transform}). For the derivation of the
quantum Boltzmann equation, the self energies $\Sigma{}_{\lambda}^{\gtrless}$
and the Green's functions $g_{\lambda^{\prime}}^{\gtrless}$ are of
interest (the small $g$ is used for the Green's function transformed
to the band basis $g^{\gtrless}\left(\mathbf{X},T;\mathbf{k},\omega\right)=U_{\mathbf{k}}G^{\gtrless}\left(\mathbf{X},T;\mathbf{k},\omega\right)U_{\mathbf{k}}^{\dagger}$,
where $\left(\mathbf{X},T\right)$ are the center of mass coordinates,
and $\left(\mathbf{k},\omega\right)$ are the relative coordinates
after the Wigner transform). For details on the Wigner transform and
the definitions of $G^{\gtrless}$, $\Sigma{}^{\gtrless}$ see e.g.
\cite{Kita2010,KadanoffBaym,Mahan}). The off diagonal elements of
greens functions in band space can be neglected if the frequencies
of interest are smaller than the energies of thermally excited particles:
$\omega\ll k_{B}T$. In the following, only the weak space and time
dependencies induced by external forces and represented by the center
of mass coordinates will be of interest. For simplicity, the dependence
on $\left(\mathbf{X},T\right)$ will be suppressed. The Green's functions
$g_{\lambda^{\prime}}^{\gtrless}\left(\mathbf{k},\omega\right)$ can
be related to the distribution function: 
\begin{eqnarray}
g_{\lambda}^{>}\left(\mathbf{k},\omega\right) & = & -i2\pi\delta\left(\omega-\varepsilon_{\lambda}\left(\mathbf{k}\right)-U_{\mathrm{pot}}\right)\left(1-f_{\lambda,\mathbf{k}}\left(\omega\right)\right)\nonumber \\
g_{\lambda}^{<}\left(\mathbf{k},\omega\right) & = & i2\pi\delta\left(\omega-\varepsilon_{\lambda}\left(\mathbf{k}\right)-U_{\mathrm{pot}}\right)f_{\lambda,\mathbf{k}}\left(\omega\right).\label{eq:gs_in_distr_funct}
\end{eqnarray}
To second order in perturbation theory, for the self-energies 
\begin{eqnarray}
\Sigma{}_{\lambda}^{\gtrless}\left(\mathbf{k,}\omega\right) & = & N\sum_{\mu\mu^{\prime}\lambda^{\prime}}\int\frac{d^{2}qd^{2}k^{\prime}d\omega_{1}d\omega_{2}}{\left(2\pi\right)^{6}}\left\vert T_{\lambda\mu\mu^{\prime}\lambda^{\prime}}\left(\mathbf{k,k}^{\prime},\mathbf{q}\right)\right\vert ^{2}\nonumber \\
 & \times & g_{\lambda^{\prime}}^{\gtrless}\left(\mathbf{k+q,}\omega_{1}\right)g_{\mu}^{\gtrless}\left(\mathbf{k}^{\prime}-\mathbf{q,}\omega_{2}\right)g_{\mu^{\prime}}^{\lessgtr}\left(\mathbf{k}^{\prime}\mathbf{,}\omega_{1}+\omega_{2}-\omega\right)\nonumber \\
 & - & \sum_{\mu\mu^{\prime}\lambda^{\prime}}\int\frac{d^{2}qd^{2}k^{\prime}}{\left(2\pi\right)^{4}}\int\frac{d\omega_{1}d\omega_{2}}{\left(2\pi\right)^{2}}T_{\lambda\lambda^{\prime}\mu^{\prime}\mu}\left(\mathbf{k,k}^{\prime},\mathbf{k}^{\prime}-\mathbf{q-k}\right)T_{\lambda\mu\mu^{\prime}\lambda^{\prime}}\left(\mathbf{k,k}^{\prime},\mathbf{q}\right)^{\ast}\nonumber \\
 & \times & g_{\lambda^{\prime}}^{\gtrless}\left(\mathbf{k+q,}\omega_{1}\right)g_{\mu}^{\gtrless}\left(\mathbf{k}^{\prime}-\mathbf{q,}\omega_{2}\right)g_{\mu^{\prime}}^{\lessgtr}\left(\mathbf{k}^{\prime}\mathbf{,}\omega_{1}+\omega_{2}-\omega\right)
\end{eqnarray}
holds. $N=4$ accounts for the spin-valley degeneracy.

The collision operator, as it appears in Eq. (\ref{eq:Boltzmann}),
can now be determined from the self energies $\Sigma^{<}$ and $\Sigma^{>}$.
It can then be written in terms of the distribution function $f_{\lambda}\left(\mathbf{k}\right)$:
\begin{equation}
{\cal C}_{\lambda}\left(\mathbf{k}\right)=-i\Sigma_{\lambda}^{<}\left(\mathbf{k},\varepsilon_{\lambda}\left(\mathbf{k}\right)\right)\left(1-f_{\lambda}\left(\mathbf{k}\right)\right)-i\Sigma_{\lambda}^{>}\left(\mathbf{k},\varepsilon_{\lambda}\left(\mathbf{k}\right)\right)f_{\lambda}\left(\mathbf{k}\right).
\end{equation}
The delta function $\delta\left(\omega-\varepsilon_{\lambda}\left(\mathbf{k}\right)-U_{\mathrm{pot}}\left(\mathbf{x}\right)\right)$
sets the left hand side of the quantum Boltzmann equation to zero
and therefore cancels out. Inserting Eqs. (\ref{eq:gs_in_distr_funct})
into the self energies, parameterizing the deviations of $f_{\lambda}\left(\mathbf{k}\right)$
from the equilibrium distribution function as shown in Eq. (\ref{eq:Distribution_expanded}),
and linearizing in $\psi_{\mathbf{k}\lambda}\left(\mathbf{x},t\right)$
leads to the collision operator of Eq. (\ref{eq:Collision_operator}).
The matrix elements $\gamma_{\mathbf{k},\mathbf{k}',\mathbf{q}}^{\left(1,2\right)}$
of Eq. (\ref{eq:Collision_operator}) are given by:

\begin{eqnarray}
\gamma_{1}\left(\mathbf{k,k}^{\prime},\mathbf{q}\right) & = & \left(N-1\right)\left\vert T_{A}\left(\mathbf{k,k}^{\prime},\mathbf{q}\right)\right\vert ^{2}+\frac{1}{2}\left\vert T_{A}\left(\mathbf{k,k}^{\prime},\mathbf{k}^{\prime}-\mathbf{q-k}\right)-T_{A}\left(\mathbf{k,k}^{\prime},\mathbf{q}\right)\right\vert ^{2}\nonumber \\
 &  & -\left\vert T_{A}\left(\mathbf{k,k}^{\prime},\mathbf{k}^{\prime}-\mathbf{q-k}\right)\right\vert ^{2}\nonumber \\
\gamma_{2}\left(\mathbf{k,k}^{\prime},\mathbf{q}\right) & = & \left(N-1\right)\left\vert T_{B}\left(\mathbf{k,\mathbf{k}^{\prime}},\mathbf{\mathbf{k}^{\prime}-k-\mathbf{q}}\right)\right\vert ^{2}+\left(N-1\right)\left\vert T_{A}\left(\mathbf{k,\mathbf{k}^{\prime}},\mathbf{q}\right)\right\vert ^{2}\nonumber \\
 &  & +\left\vert T_{A}\left(\mathbf{k,\mathbf{k}^{\prime}},\mathbf{q}\right)-T_{B}\left(\mathbf{k,\mathbf{k}^{\prime}},\mathbf{\mathbf{k}^{\prime}-\mathbf{q}}\mathbf{-k}\right)\right\vert ^{2},\label{eq:app_gammas}
\end{eqnarray}
with 
\begin{eqnarray*}
T_{A}\left(\mathbf{k,k}^{\prime},\mathbf{q}\right) & = & T_{++++}\left(\mathbf{k,k}^{\prime},\mathbf{q}\right)=T_{----}\left(\mathbf{k,k}^{\prime},\mathbf{q}\right)\\
 & = & T_{+--+}\left(\mathbf{k,k}^{\prime},\mathbf{q}\right)=T_{-++-}\left(\mathbf{k,k}^{\prime},\mathbf{q}\right)\\
 & = & \frac{V\left(q\right)}{4}\left(1+\frac{\left(K+Q\right)K^{\ast}}{\left\vert \mathbf{k+q}\right\vert k}\right)\left(1+\frac{\left(K^{\prime}-Q\right)K^{\prime\ast}}{\left\vert \mathbf{k}^{\prime}\mathbf{-q}\right\vert k^{\prime}}\right)
\end{eqnarray*}
and 
\begin{eqnarray}
T_{B}\left(\mathbf{k,k}^{\prime},\mathbf{q}\right) & = & T_{++--}\left(\mathbf{k,k}^{\prime},\mathbf{q}\right)=T_{--++}\left(\mathbf{k,k}^{\prime},\mathbf{q}\right)\nonumber \\
 & = & \frac{V\left(q\right)}{4}\left(1-\frac{\left(K+Q\right)K^{\ast}}{\left\vert \mathbf{k+q}\right\vert k}\right)\left(1-\frac{\left(K^{\prime}-Q\right)K^{\prime\ast}}{\left\vert \mathbf{k}^{\prime}\mathbf{-q}\right\vert k^{\prime}}\right)\label{eq:app_Ts}
\end{eqnarray}
Upper-case letters like $K=k_{x}+ik_{y}$ etc. combine the two components
of the momentum vector onto a complex variable.

Since the quantum Boltzmann equation only accounts for the diagonal
in $\lambda$ components of the distribution function, the currents
also have to be decomposed into contributions that involve particle-hole
pair creation $\mbox{\ensuremath{\left(\mathbf{j}_{\mathrm{inter}}\right)}}$
and those who do not $\left(\mathbf{j}_{\mathrm{intra}}\right)$.
Here, the identity 
\begin{equation}
U_{\mathbf{k}}\mathbf{\sigma}U_{\mathbf{k}}^{-1}=\frac{\mathbf{k}}{k}\sigma_{z}-\frac{\mathbf{k\times e}_{z}}{k}\sigma_{y}
\end{equation}
is useful. The charge current 
\begin{equation}
\mathbf{j}_{c}=ev\int_{\mathbf{k}}\psi^{\dagger}\left(\mathbf{k}\right)\mathbf{\sigma}\psi\left(\mathbf{k}\right)
\end{equation}
can be written as 
\begin{equation}
\mathbf{j}_{c}=\mathbf{j}_{c,\mathrm{intra}}+\mathbf{j}_{c,\mathrm{inter}},
\end{equation}
where the two contributions are given by 
\begin{eqnarray}
\mathbf{j}_{c,\mathrm{intra}} & = & ev\int_{\mathbf{k}}\sum_{\lambda=\pm}\frac{\lambda\mathbf{k}}{k}\gamma_{\mathbf{k},\lambda}^{\dagger}\gamma_{\mathbf{k},\lambda}\nonumber \\
\mathbf{j}_{c,\mathrm{inter}} & = & iev\int_{\mathbf{k}}\frac{\mathbf{k\times e}_{z}}{k}\left(\gamma_{\mathbf{k},+}^{\dagger}\gamma_{\mathbf{k},-}-\gamma_{\mathbf{k},-}^{\dagger}\gamma_{\mathbf{k},+}\right).
\end{eqnarray}
The energy current $\mathbf{j}_{\varepsilon}$ and the momentum current
tensor $\tau_{xy}$ can be decomposed in a similar manner. This leads
to the expressions (\ref{eq:charge_current_exp}) and (\ref{eq:heat_current_exp})
of the main text and the expression that is used for $\tau_{xy}$
in Sec. \ref{subsec:Non-local-shear-viscosity}. As discussed above,
in the hydrodynamic regime, it is legitimate to focus on the intra-band
contributions, which dominate the transport behavior of the system.

\section{Collinear scattering and collinear zero modes\label{sec:App_Collinear-scattering}}

Here, the logarithmic divergence of the collision operator for collinear
processes is demonstrated following Ref. \cite{Fritz2008}. We then
show, that the $m$-dependent collinear zero modes are those given
in Eq. (\ref{eq:Collinear_modes}).

The essential mathematics behind the divergence is contained in phase
space density available for two particle collisions. The phase space
is restricted by the delta function ensuring energy conservation:
$\delta\left(k+k_{1}-\left|\mathbf{k}+\mathbf{q}\right|-\left|\mathbf{k}_{1}-\mathbf{q}\right|\right)$.
This can be seen from power counting in Eq. (\ref{eq:Collision_operator})
using Eqs. (\ref{eq:app_gammas}), (\ref{eq:app_Ts}).

Choosing $\mathbf{k}=\left(k,0\right)$ with $k>0$, and writing $\mathbf{k}_{1}=\left(k_{1},k_{\bot}\right),$
$\mathbf{q}=\left(q,q_{\bot}\right)$, collinear scattering occurs
when $k_{1}>0$, $k+q>0$, $k_{1}-q>0$ and $q_{\bot}\approx0$, $k_{\bot}\approx0$.
For small $q_{\bot}$, $k_{\bot}$ the argument of the delta function
can be approximated as 
\begin{equation}
k+k_{1}-\left|\mathbf{k}+\mathbf{q}\right|-\left|\mathbf{k}_{1}-\mathbf{q}\right|\approx\frac{k_{\bot}^{2}}{2k_{1}}-\frac{q_{\bot}^{2}}{2\left(k+q\right)}-\frac{\left(k_{\bot}-q_{\bot}\right)^{2}}{2\left(k_{1}-q\right)}.\label{eq:app_delta_argument}
\end{equation}
The right hand side of this equation is a polynomial in $q_{\bot}$,
and can be written in terms of linear factors as 
\[
\frac{k_{\bot}^{2}}{2k_{1}}-\frac{q_{\bot}^{2}}{2\left(k+q\right)}-\frac{\left(k_{\bot}-q_{\bot}\right)^{2}}{2\left(k_{1}-q\right)}=-\frac{k_{1}+k}{2\left(k+q\right)\left(k_{1}-q\right)}\left(q_{\bot}-\zeta_{1}k_{\bot}\right)\left(q_{\bot}-\zeta_{2}k_{\bot}\right).
\]
It is then easy to see by performing the $q_{\bot}$ integration that
\[
\int dk_{\bot}dq_{\bot}\delta\left(-\frac{k_{1}+k}{2\left(k+q\right)\left(k_{1}-q\right)}\left(q_{\bot}-\zeta_{1}k_{\bot}\right)\left(q_{\bot}-\zeta_{2}k_{\bot}\right)\right)\propto\int\frac{dk_{\bot}}{k_{\bot}}.
\]
This behavior leads to a logarithmic divergence. The divergence is
however cut off by the screening of the Coulomb potential \cite{Mueller2008}
\[
V\left(\left|\mathbf{q}\right|\right)\rightarrow V\left(\left|\mathbf{q}\right|+q_{TF}\right),
\]
where $q_{TF}$ is the Thomas Fermi screening length. In the case
of charge neutral graphene $q_{FT}=\alpha k_{B}T/v$. If the screening
is included, the integral of (\ref{eq:Collision_operator}) vanishes
in the infrared. Thus, the contribution of collinear processes to
the scattering rates is enhanced by the large factor 
\[
\log\left(1/\alpha\right).
\]

It was demonstrated in sec. \ref{subsec:Collinear-zero-modes} of
the main text, that relaxation processes in the hydrodynamic regime
are dominated by collinear zero modes. As demonstrated above, these
modes describe scattering events in which all particle velocities
show in the same direction. Examining the delta function responsible
for energy conservation $\delta\left(k+k_{1}-\left|\mathbf{k}+\mathbf{q}\right|-\left|\mathbf{k}_{1}-\mathbf{q}\right|\right)$,
we see that, if all momenta are parallel to each other, energy is
only conserved, if the above conditions $k>0$, $k_{1}>0$, $k+q>0$,
$k_{1}-q>0$ apply (except for unimportant isolated points in phase
space). The exchange momentum $q$, however, can be positive or negative.
To find those $\psi_{\mathbf{k}\lambda}$ that correspond to collinear
zero modes, two terms in the collision operator Eq. (\ref{eq:Collision_operator})
have to be considered: 
\begin{eqnarray}
A_{\mathbf{k},\mathbf{k}_{1},\mathbf{q},\lambda}^{\left(1\right)} & = & \psi_{\mathbf{k}+\mathbf{q}\lambda}+\psi_{\mathbf{k}_{1}-\mathbf{q}\lambda}-\psi_{\mathbf{k}_{1}\lambda}-\psi_{\mathbf{k}\lambda}\nonumber \\
A_{\mathbf{k},\mathbf{k}_{1},\mathbf{q},\lambda}^{\left(2\right)} & = & \psi_{\mathbf{k}+\mathbf{q}\lambda}-\psi_{-\mathbf{k}_{1}+\mathbf{q}\bar{\lambda}}+\psi_{-\mathbf{k}_{1}\bar{\lambda}}-\psi_{\mathbf{k}\lambda}.
\end{eqnarray}
Using the parameterization 
\begin{equation}
\psi_{\mathbf{k},\lambda}=a_{\lambda,m}\left(k\right)e^{im\theta_{\mathbf{k}}}
\end{equation}
yields 
\begin{eqnarray}
A_{\mathbf{k},\mathbf{k}',\mathbf{q},\lambda}^{\left(1\right)} & = & \left(a_{\lambda,m}\left(k+q\right)+a_{\lambda,m}\left(k_{1}-q\right)-a_{\lambda,m}\left(k_{1}\right)-a_{\lambda,m}\left(k\right)\right)e^{im\theta_{\mathbf{k}}}\nonumber \\
A_{\mathbf{k},\mathbf{k}',\mathbf{q},\lambda}^{\left(2\right)} & = & \left(a_{\lambda,m}\left(k+q\right)-\left(-1\right)^{m}a_{\bar{\lambda},m}\left(k_{1}-q\right)+\left(-1\right)^{m}a_{\bar{\lambda},m}\left(k_{1}\right)-a_{\lambda,m}\left(k\right)\right)e^{im\theta_{\mathbf{k}}}.
\end{eqnarray}
For collinear zero modes 
\begin{eqnarray*}
A_{\mathbf{k},\mathbf{k}',\mathbf{q},\lambda}^{\left(1\right)} & = & 0\\
A_{\mathbf{k},\mathbf{k}',\mathbf{q},\lambda}^{\left(2\right)} & = & 0
\end{eqnarray*}
has to hold. $A_{\mathbf{k},\mathbf{k}',\mathbf{q},\lambda}^{\left(1\right)}$
is set to zero by $a_{\lambda,m}\left(k\right)=\left\{ 1,\lambda,\beta v\hbar k,\lambda\beta v\hbar k\right\} $.
$A_{\mathbf{k},\mathbf{k}',\mathbf{q},\lambda}^{\left(2\right)}$
is more restrictive. For even $m$ its zero modes are given by $a_{\lambda,m}\left(k\right)=\left\{ 1,\lambda,\lambda\beta v\hbar k\right\} $,
for odd $m$ the zero modes are $a_{\lambda,m}\left(k\right)=\left\{ 1,\lambda,\beta v\hbar k\right\} $.
Summing up, the collinear zero modes are given by 
\[
a_{\lambda,m}=\lambda^{m}\left\{ 1,\lambda,\lambda\beta v\hbar k\right\} e^{im\theta_{\mathbf{k}}}.
\]

\section{Matrix elements of the collision operator\label{sec:App_Matrix-elements}}

The values of some matrix elements are shown in Table \ref{tab:Matrix-elements_Table}.
For $m\geq2$ the values can be approximated by 
\begin{eqnarray}
\left\langle \chi_{\mathbf{k},\lambda}^{\left(m,s=1\right)}\left|\mathcal{C}\right|\chi_{\mathbf{k},\lambda}^{\left(m,s=1\right)}\right\rangle  & = & 2.574\cdot\left|m\right|-3.456\nonumber \\
\left\langle \chi_{\mathbf{k},\lambda}^{\left(m,s=2\right)}\left|\mathcal{C}\right|\chi_{\mathbf{k},\lambda}^{\left(m,s=2\right)}\right\rangle  & = & 1.825\cdot\left|m\right|-2.741\nonumber \\
\left\langle \chi_{\mathbf{k},\lambda}^{\left(m,s=3\right)}\left|\mathcal{C}\right|\chi_{\mathbf{k},\lambda}^{\left(m,s=3\right)}\right\rangle  & = & 5.184\cdot\left|m\right|-11.37\nonumber \\
\left\langle \chi_{\mathbf{k},\lambda}^{\left(m,s=2\right)}\left|\mathcal{C}\right|\chi_{\mathbf{k},\lambda}^{\left(m,s=3\right)}\right\rangle  & = & 2.042\cdot\left|m\right|-4.398.\nonumber \\
\end{eqnarray}
All values are given in units of $\frac{1}{v^{2}\beta^{3}\hbar^{3}}.$

\begin{table}
\begin{centering}
\begin{tabular}{|c|c|c|c||c|c|c|c||c|c|c|c|}
\hline 
$m$  & $s$  & $s'$  & $\left\langle \chi_{\mathbf{k},\lambda}^{\left(m,s\right)}\left|\mathcal{C}\right|\chi_{\mathbf{k},\lambda}^{\left(m,s'\right)}\right\rangle $  & $m$  & $s$  & $s$'  & $\left\langle \chi_{\mathbf{k},\lambda}^{\left(m,s\right)}\left|\mathcal{C}\right|\chi_{\mathbf{k},\lambda}^{\left(m,s'\right)}\right\rangle $  & $m$  & $s$  & $s$'  & $\left\langle \chi_{\mathbf{k},\lambda}^{\left(m,s\right)}\left|\mathcal{C}\right|\chi_{\mathbf{k},\lambda}^{\left(m,s'\right)}\right\rangle $\tabularnewline
\hline 
\hline 
0  & 1  & 1  & 0  & 2  & 1  & 1  & 2.617  & 4  & 1  & 1  & 6.988\tabularnewline
\hline 
0  & 1  & 2  & 0  & 2  & 1  & 2  & 0  & 4  & 1  & 2  & 0\tabularnewline
\hline 
0  & 1  & 3  & 0  & 2  & 1  & 3  & 0  & 4  & 1  & 3  & 0\tabularnewline
\hline 
0  & 2  & 2  & 0  & 2  & 2  & 2  & 1.745  & 4  & 2  & 2  & 4.722\tabularnewline
\hline 
0  & 2  & 3  & 0  & 2  & 2  & 3  & 1.243  & 4  & 2  & 3  & 4.122\tabularnewline
\hline 
0  & 3  & 3  & 0  & 2  & 3  & 3  & 3.341  & 4  & 3  & 3  & 10.456\tabularnewline
\hline 
1  & 1  & 1  & 0.804  & 3  & 1  & 1  & 4.728  & 5  & 1  & 1  & 9.345\tabularnewline
\hline 
1  & 1  & 2  & 0  & 3  & 1  & 2  & 0  & 5  & 1  & 2  & 0\tabularnewline
\hline 
1  & 1  & 3  & 0  & 3  & 1  & 3  & 0  & 5  & 1  & 3  & 0\tabularnewline
\hline 
1  & 2  & 2  & 0.463  & 3  & 2  & 2  & 3.167  & 5  & 2  & 2  & 6.351\tabularnewline
\hline 
1  & 2  & 3  & 0  & 3  & 2  & 3  & 2.573  & 5  & 2  & 3  & 5.800\tabularnewline
\hline 
1  & 3  & 3  & 0  & 3  & 3  & 3  & 6.647  & 5  & 3  & 3  & 14.610\tabularnewline
\hline 
\end{tabular}
\par\end{centering}

\caption{Matrix elements of the collision operator (\ref{eq:Collision_operator})
with respect to the collinear zero modes $\chi_{\mathbf{k},\lambda}^{\left(m,s\right)}=\lambda^{m}e^{im\theta}\left\{ 1,\lambda,\lambda\beta v\hbar k\right\} $.
The index $m$ labels the angular harmonic and $s$ one of the modes
in curved brackets.\label{tab:Matrix-elements_Table}}
\end{table}

\section{Decomposition of the viscosity tensor into longitudinal and transverse
parts\label{sec:Decomposition_visc}}

Consider a system with a preference direction introduced by the wave-vector
$\mathbf{q}$. It is useful to define the orthogonal tensor basis
\begin{eqnarray}
e_{\alpha\beta}^{\left(1\right)} & = & \frac{q_{\alpha}q_{\beta}}{q^{2}}\nonumber \\
e_{\alpha\beta}^{\left(2\right)} & = & \delta_{\alpha\beta}-\frac{q_{\alpha}q_{\beta}}{q^{2}}\nonumber \\
e_{\alpha\beta}^{\left(3\right)} & = & \frac{1}{\sqrt{2}}\left(q_{\alpha}p_{\beta}+p_{\alpha}q_{\beta}\right)/\left(pq\right),\label{eq:Tensor_basis-appendix}
\end{eqnarray}
which is normalized according according to 
\[
\sum_{\alpha\beta}e_{\alpha\beta}^{\left(i\right)}e_{\alpha\beta}^{\left(j\right)}=\delta_{ij}.
\]
Here it is 
\[
p_{\alpha}=q_{\gamma}\varepsilon_{\gamma\alpha}.
\]
In this basis, the symmetric shear force tensor $X_{0,\alpha\beta}$
can be written 
\begin{equation}
X_{0,\alpha\beta}=X^{\left(1\right)}e_{\alpha\beta}^{\left(1\right)}+X^{\left(2\right)}e_{\alpha\beta}^{\left(2\right)}+X^{\left(3\right)}e_{\alpha\beta}^{\left(3\right)}.\label{eq:Appendix_shear_force_decomposition}
\end{equation}
The same holds for the momentum current (stress) tensor 
\begin{equation}
\tau_{\alpha\beta}=\tau^{\left(1\right)}e_{\alpha\beta}^{\left(1\right)}+\tau^{\left(2\right)}e_{\alpha\beta}^{\left(2\right)}+\tau^{\left(3\right)}e_{\alpha\beta}^{\left(3\right)}.\label{eq:Appendix_stress_tensor_decomposition}
\end{equation}
Since the system is fully isotropic, except for the preference direction
set by $\mathbf{q}$, the response of the system to different components
of $X_{0,\alpha\beta}$ can only be distinct as far as these components
relate differently to the direction of $\mathbf{q}$. Eqs (\ref{eq:Appendix_shear_force_decomposition})
and (\ref{eq:Appendix_stress_tensor_decomposition}) are decompositions
of the shear force and momentum current tensors into such components.
The fourth rank viscosity tensor $\eta_{\alpha\beta\gamma\delta}$
is defined through the constitutive relation 
\[
\tau_{\alpha\beta}=\eta_{\alpha\beta\gamma\delta}X_{0,\gamma\delta}.
\]
In general, such a tensor connecting the quantities $\tau_{\alpha\beta}$
and $X_{0,\alpha\beta}$ as given by Eqs. (\ref{eq:Appendix_shear_force_decomposition}),
(\ref{eq:Appendix_stress_tensor_decomposition}) can be written as
$\eta_{\alpha\beta\gamma\delta}=\sum_{ij}e_{\alpha\beta}^{\left(i\right)}e_{\gamma\delta}^{\left(j\right)}\eta^{\left(ij\right)}$.
However it follows from an Onsager reciprocity relation that $\eta_{\alpha\beta\gamma\delta}$
has to be symmetric with respect to an interchange of the first and
last pairs of indices: 
\[
\eta_{\left(\alpha\beta\right)\left(\gamma\delta\right)}=\eta_{\left(\gamma\delta\right)\left(\alpha\beta\right)}.
\]
This condition further restricts the form of $\eta_{\alpha\beta\gamma\delta}$
to 
\begin{equation}
\eta_{\alpha\beta\gamma\delta}=\sum_{i}e_{\alpha\beta}^{\left(i\right)}e_{\gamma\delta}^{\left(i\right)}\eta^{\left(i\right)}.\label{eq:appendix_visc_tensor_form}
\end{equation}
Calculating the scalars $\eta^{\left(i\right)}$ using the quantum
Boltzmann equation, one finds $\eta^{\left(1\right)}=\eta^{\left(2\right)}\neq\eta^{\left(3\right)}$.
For reasons explained in the main text, we call $\eta^{\left(1\right)}=\eta^{\left(2\right)}=\eta_{\perp}$
the transverse, and $\eta^{\left(3\right)}=\eta_{\parallel}$ the
longitudinal viscosity. In the sense that $\eta_{\alpha\beta\gamma\delta}$
is spanned by projection operators onto the tensorial subspaces which
span the force and current tensors and are given in Eqs. (\ref{eq:Tensor_basis-appendix}),
the decomposition (\ref{eq:appendix_visc_tensor_form}) is completely
analogous to the decomposition of a conductivity tensor into transverse
and longitudinal parts (see Eq. (\ref{eq:Decomposition_electr_explicit})).

\twocolumngrid

\bibliographystyle{thesis-bib}
\bibliography{References}

\end{document}